\documentclass[aps,pra,reprint,showpacs]{revtex4-1}
\usepackage{graphics}
\usepackage{epsfig}
\usepackage{amsmath}
\usepackage{amssymb}
\usepackage{rotating} 

\usepackage{float}
\allowdisplaybreaks

\usepackage{hyperref}

\hypersetup{
 bookmarks=true,		
 unicode=false,			
 pdftoolbar=true,		
 pdfmenubar=true,		
 pdffitwindow=false,		
 pdfstartview={FitH},		
 pdftitle={Multipartite entanglement measures},    
 pdfauthor={Szil{\'a}rd Szalay},	
 pdfsubject={research article on multipartite entanglement},	
 pdfcreator={pdflatex},		
 pdfproducer={vim},		
 pdfkeywords={quantum} {entanglement} {multipartite} {mixed states}, 
 pdfnewwindow=true,		
 colorlinks=true,		
 linktoc=page,			
 linkcolor=blue,		
 citecolor=blue,		
 filecolor=blue,		
 urlcolor=blue			
}

\newcommand{\field}[1]{\mathbb{#1}}
\newcommand{\ve}[1]{\mathbf{#1}}
\newcommand{\vs}[1]{\boldsymbol{#1}}
\newcommand{\vvs}[1]{\underline{\boldsymbol{#1}}}

\newcommand{\cket}[1]{\vert #1 \rangle}
\newcommand{\bra}[1]{\langle #1 \vert}
\newcommand{\bracket}[1]{\langle #1 \rangle}
\newcommand{\ee}{\mathrm{e}}
\newcommand{\Id}{\mathrm{I}}

\newcommand{\defn}{\overset{\text{def.}}{\Longleftrightarrow}}
\providecommand{\abs}[1]{{\lvert#1\rvert}}
\providecommand{\norm}[1]{{\lVert#1\rVert}}
\newcommand{\cmpl}[1]{\overline{#1}}
\newcommand{\pcmpl}[1]{\overline{#1}}

\newcommand{\LieGrp}[1]{\mathrm{#1}}

\DeclareMathOperator{\rk}{rk}
\DeclareMathOperator{\dom}{dom}
\DeclareMathOperator{\tr}{tr}
\DeclareMathOperator{\Lin}{Lin}
\DeclareMathOperator{\Conv}{Conv}
\DeclareMathOperator{\Span}{Span}
\DeclareMathOperator{\Extr}{Extr}

\DeclareMathOperator*{\argmin}{argmin}

\DeclareMathOperator{\downset}{\downarrow}
\DeclareMathOperator{\upset}{\uparrow}

\newcommand{\cnvroof}[1]{{#1}^{\cup}}
\newcommand{\nsubset}{\not\subset}

\providecommand{\abs}[1]{\lvert#1\rvert}
\providecommand{\norm}[1]{\lVert#1\rVert}

\newcommand{\set}[2]{\{ #1 \;\vert\; #2 \}}
\newcommand{\bigset}[2]{\bigl\{ #1 \;\big\vert\; #2 \bigr\}}
\newcommand{\Bigset}[2]{\Bigl\{ #1 \;\Big\vert\; #2 \Bigr\}}

\newcommand{\struct}[1]{( #1 )}
\newcommand{\bigstruct}[1]{\bigl( #1 \bigr)}

\newtheorem{thm}{Theorem}
\newtheorem{cor}[thm]{Corollary}
\newtheorem{lem}[thm]{Lemma}
\newtheorem{conj}[thm]{Conjecture}

\newcounter{txtitem}
\renewcommand{\thetxtitem}{(\roman{txtitem})}
\newcommand{\txtitem}{\refstepcounter{txtitem}\thetxtitem}

\begin{document}
\title{Multipartite entanglement measures}
\author{{Sz}il{\'a}rd {Sz}alay}
\email{szalay.szilard@wigner.mta.hu}
\affiliation{
Strongly Correlated Systems ``Lend{\"u}let'' Research Group,
Institute for Solid State Physics and Optics,
MTA Wigner Research Centre for Physics,
H-1121 Budapest, Konkoly-Thege Mikl{\'o}s {\'u}t 29-33, Hungary}
\date{\today}

\begin{abstract}

The main concern of this paper is
how to define proper measures of \emph{multipartite} entanglement for mixed quantum states.
Since the structure of partial separability and multipartite entanglement
is getting complicated if the number of subsystems exceeds two,
one cannot expect the existence of an ultimate scalar entanglement measure, which grasps
even a small part of the rich hierarchical structure of multipartite entanglement,
and some higher-order structure characterizing that is needed.
In this paper we make some steps in this direction.
 
\textit{First,} we reveal the lattice-theoretic structure of the partial separability classification,
introduced earlier 
[\href{http://dx.doi.org/10.1103/PhysRevA.86.032341}{{Sz.}~Szalay and Z.~K{\"o}k{\'e}nyesi, Phys.~Rev.~A \textbf{86}, 032341 (2012)}].
It turns out that, \emph{mathematically},
the structure of the entanglement classes is the up-set lattice of
the structure of the different kinds of partial separability,
which is the down-set lattice of the lattice of the partitions of the subsystems.
It also turns out that, \emph{physically},
this structure is related to the LOCC convertibility:
If a state from a class can be mapped into another one,
then that class can be found higher in the hierarchy.

\textit{Second,} we introduce the notion of \emph{multipartite monotonicity}, expressing
that a given set of entanglement monotones,
while measuring the different kinds of entanglement,
shows also the same hierarchical structure as the entanglement classes.
Then we construct such hierarchies of entanglement measures,
and we propose a physically well-motivated one,
being the direct multipartite generalization 
of the entanglement of formation based on the entanglement entropy,
motivated by the notion of statistical distinguishability.
The multipartite monotonicity shown by this set of measures
motivates us to consider the measures 
to be the different manifestations of some ``unified'' notion of entanglement. 


\end{abstract}

\maketitle{}

\tableofcontents{}

\section{Introduction}
\label{sec:Intro}

In the description of entanglement \cite{Horodecki4,SzalayDissertation}
a hard problem, yet unsolved,
is how to step from the bipartite scenario to the multipartite one,
in particular, how to define proper measures of multipartite entanglement.
The state of a bipartite quantum system can be either separable or entangled,
while for more-than-two-partite systems
the partial separability properties have a complicated structure \cite{SzalayKokenyesiPartSep},
and a system of measures fitting to this structure, while being physically motivated, is not known.

The quantum entanglement in \emph{bipartite pure states} can be described completely
by the use of the \emph{singular value decomposition} (SVD), also called \emph{Schmidt decomposition} \cite{Schmidt-1907,BennettetalPureStates}.
This leads to a \emph{local unitary canonical form,}
which allows for the separation of the nonlocal parameters of the state (relevant for the description of correlations)
from the local (irrelevant) ones.
The Schmidt coefficients contain then all nonlocal properties of the state;
they show a simple structure
(what means that the pure bipartite entanglement \emph{itself} shows a simple structure);
and, in principle, every measure of entanglement can be expressed by them.
A well-known example is 
the entanglement entropy \cite{BennettetalPureStates}.
One can step from pure state entanglement measures to mixed ones by the use of \emph{convex roof extension} \cite{UhlmannConvRoofs}. 
A well-known example is 
the entanglement of formation \cite{BennettetalMixedStates}, which is the extension of the entanglement entropy.

For the case of \emph{multipartite pure states,}
local unitary canonical form is not known, in general \cite{AcinetalGenSchmidt3QB,Carteret-2000}.
A \emph{higher-order singular value decomposition} (HOSVD, or \emph{Vidal decomposition}) \cite{Vidal-2003b}
can be formulated, which is a sequence of bipartite singular value decompositions.
Then a sequence of bipartite entanglement measures 
characterizes the multipartite entanglement in some sense.
There are also other decomposition methods \cite{Hackbusch-2012}, such as 
the \emph{parallel factors method} (PARAFAC), also called \emph{canonical decomposition} (CANDECOMP) \cite{Hitchcock-1927,Bro-1997},
or the \emph{Tucker decomposition} \cite{Tucker-1966,Lathauwer-2000,Jiang-2015}.
Although these approaches are very important in numerical techniques 
of the quantum mechanics of strongly correlated systems \cite{SzalayIntJQuantChemReview,LegezaBachBook},
they do not give us as deep an insight 
into the structure and quantitative description of multipartite entanglement
as the SVD did in the bipartite case.

A very different approach is
to build up the partial separability, or multipartite entanglement structure 
from the grounds \cite{SzalayKokenyesiPartSep},
and define different entanglement measures for the different kinds of partial separability.
The basic observation making this possible is that
the whole construction can be formulated by the use of the notion of
pure state entanglement with respect to a bipartite split,
which is relatively well understood.
In the present paper, we carry out this program.

The \textit{first part} of the paper is devoted to 
the \emph{classification} of multipartite entanglement.
After recalling the basic notions in the theory of entanglement of bipartite systems
in Section \ref{sec:EntBasics},
we build up those for the multipartite case
in Section \ref{sec:EntMulti}
in a more clarified treatment than was presented originally \cite{SzalayKokenyesiPartSep},
which makes it possible to achieve numerous new developments.
Our results can be formulated naturally in the language of 
lattice theory \cite{DaveyPriestley,Stanley}.
We work out the hierarchical structure of different kinds of partial separability,
which turns out to be 
the down-set lattice of the lattice of the partitions of the subsystems
(see Section \ref{sec:EntMulti.second}),
and also the structure of the entanglement classes,
which turns out to be also hierarchical,
being the up-set lattice of the lattice above
(see Section \ref{sec:EntMulti.classes}).
We clarify the meaning of this structure;
it is related to the LOCC (local operations and classical communication) convertibility:
If a state from a class can be mapped into another one,
then that class can be found higher in the hierarchy.

The \textit{second part} of the paper is devoted to 
the \emph{quantification} of multipartite entanglement.
After recalling the basic notions in the theory of measures of bipartite entanglement
in Section \ref{sec:MeasBasics},
we consider the $q$-sums and $q$-means together with their generalizations 
as useful tools for the construction of entanglement measures from entanglement measures
in Section \ref{sec:Means},
then we construct measures for the multipartite case
in Section \ref{sec:MeasMulti}.
The principle is that
all kinds of ``partial entanglement content'' of a given state
are quantified simultaneously
by the elements of a set of multipartite entanglement measures.
Besides the usual \emph{entanglement monotonicity} and \emph{discriminance} properties,
we introduce the \emph{multipartite monotonicity,}
which endows the set of multipartite entanglement measures
with the same hierarchical structure as the partial separability shows.
We succeed in constructing a hierarchy of multipartite entanglement measures
satisfying these requirements 
in Section \ref{sec:MeasMulti.second},
which are the direct generalizations 
of the entanglement entropy for pure states and the entanglement of formation for mixed states in the bipartite case.
These measures have the same information-geometrical meaning,
related to correlation measures based on statistical distinguishability,
as the entanglement entropy and the entanglement of formation.
The multipartite monotonicity shown by this set of measures
motivates us to consider these measures 
to be the different manifestations of some ``unified'' notion of entanglement.

The summary 
with some important discussions and a list of open questions
is left to Section \ref{sec:Summ}.
The table of contents on the first page
helps the reader to find the way in this complex but carefully composed paper:
The sectioning follows the structure of the theory faithfully. 
The proofs of propositions in the main text are given in appendixes.
In the theory of entanglement of mixed states, the central notion is the \emph{convexity:}
We deal mostly with convex or concave functions defined over convex sets.
Apart from the \textit{original results} in the body of the text 
(Sections \ref{sec:EntMulti}, \ref{sec:MeasMulti}, and \ref{sec:Summ}, together with some of the appendixes),
this paper is intended to be a self-contained discourse and toolbox on 
convexity \cite{BengtssonZyczkowski,BoydVandenbergheConvOpt}
and entanglement \cite{Horodecki4,BengtssonZyczkowski,GuhneTothEntDet,NielsenChuang,PreskillNotes,PetzQInfo,SzalayDissertation,Wilde,Wolf}.
In the spirit of this,
we give the \textit{necessary grounding} for these topics using a unified notation system 
(Sections \ref{sec:EntBasics}, \ref{sec:MeasBasics}, and \ref{sec:Means}, together with some of the appendixes).
We also recall some known proofs of theorems about entanglement measures (appendixes),
and also some useful calculations about convexity
to enlighten how this structure shows up \cite{BengtssonZyczkowski,BoydVandenbergheConvOpt}.

\section{Quantum states and entanglement: basics}
\label{sec:EntBasics}

Here we briefly recall the basic notions
arising in the description of the states of singlepartite
(Section \ref{sec:EntBasics.States})
and bipartite 
(Section \ref{sec:EntBasics.Ent})
quantum systems
\cite{NeumannFoundQM,NielsenChuang,PreskillNotes,PetzQInfo,SzalayDissertation,Wilde,Wolf,BengtssonZyczkowski}.
We fix some basic notational conventions for 
state vectors, 
pure and mixed states,
separable and entangled states.

\subsection{Quantum states}
\label{sec:EntBasics.States}

Entanglement theory deals with the \emph{states} of quantum systems.
A \emph{state vector} is an element of a Hilbert space, $\cket{\psi}\in\mathcal{H}$,
which is normalized with respect to the standard $2$-norm of the Hilbert space,
$\norm{\psi}=\sqrt{\bracket{\psi|\psi}}=1$.
In the paper, we consider the $1<\dim\mathcal{H}<\infty$ case only.
The \emph{pure state} of a quantum system is represented by
a one-dimensional subspace (\emph{ray}) in the Hilbert space
(which is actually an element of the projective Hilbert space),
which can be given by a state vector as the self-adjoint linear operator
$\pi=\cket{\psi}\bra{\psi}$,
being the projector projecting to the one-dimensional subspace 
spanned by the state vector $\cket{\psi}$.
(Note that the projectors are characterized by $\pi^2=\pi=\pi^\dagger$.
For projectors, having unit trace is equivalent to being of rank $1$.)
The set of \emph{pure states} over the Hilbert space $\mathcal{H}$ arises as
\begin{subequations}
\begin{equation}
\label{eq:pure}
\mathcal{P}(\mathcal{H}):=\Bigset{\pi\in\Lin_\text{SA}\mathcal{H}}{\pi^2=\pi,\norm{\pi}_\text{tr}=\tr\pi=1}.
\end{equation}
(If there is no ambiguity about the underlying Hilbert space,
we use the notation $\mathcal{P}:=\mathcal{P}(\mathcal{H})$.)
A \emph{mixed state} is represented by the \emph{convex combination} (or \emph{mixture}) of pure states,
and it represents the state of an ensemble of quantum systems $\set{(p_i,\pi_i)}{i=1,\dots,m}$,
described by the pure state $\pi_i$ with probability $p_i$.
The \emph{convex body} of \emph{mixed states} over the Hilbert space $\mathcal{H}$ arises as
\begin{equation}
\label{eq:mixed.conv}
\begin{split}
\mathcal{D}(\mathcal{H}):=\Conv&\mathcal{P}(\mathcal{H})
\equiv\Bigset{\varrho\in\Lin_\text{SA}\mathcal{H}}{\exists\pi_i\in\mathcal{P},\\
&p_i\geq0, \sum_i p_i=1 : \varrho=\sum_ip_i\pi_i}.
\end{split}
\end{equation}
(If there is no ambiguity about the underlying Hilbert space,
we use the notation $\mathcal{D}:=\mathcal{D}(\mathcal{H})$.)
This turns out to be equivalent to the positive semidefinite operators normalized with respect to the trace-norm,
\begin{equation}
\label{eq:mixed.op}
\mathcal{D}(\mathcal{H})=\Bigset{\varrho\in\Lin_\text{SA}\mathcal{H}}{\varrho\geq0,\norm{\varrho}_\text{tr}=\tr\varrho=1}.
\end{equation}
Geometrically, the pure states are the \emph{extremal points} 
of the convex body of the mixed states \cite{BengtssonZyczkowski}, 
\begin{equation}
\label{eq:pure.extr}
\mathcal{P}=\Extr\mathcal{D},
\end{equation}
\end{subequations}
a pure state cannot be mixed nontrivially.

Convexity is a central notion in quantum (and also in classical) probability theory \cite{BengtssonZyczkowski}.
State spaces are, in general, convex sets, which means that 
they are closed under \emph{convex combination}, called also \emph{mixing}.
That is, for convex combination coefficients $0\leq p_i\in\field{R}$, with $\sum_i p_i=1$, also called mixing weights,
if $\varrho_i$s are states, 
then their convex combination $\sum_ip_i\varrho_i$ is also a state.
Mixing is interpreted as forgetting some classical information
about the state by which the system is described,
so this is indeed a necessary property.
The main difference between classical and quantum probability theory is that
in the quantum case, because of the superposition principle
(linear structure in the Hilbert space),
the pure state decomposition of a nonpure state is not unique,
contrary to the classical case.

\subsection{Entanglement}
\label{sec:EntBasics.Ent}

Entanglement theory, on the other hand, 
deals with the states of \emph{composite} quantum systems.
For example, for two subsystems, 
with state vectors being the normalized elements of the
associated Hilbert spaces $\mathcal{H}_1$ and $\mathcal{H}_2$,
the state vectors are the normalized elements of the tensor product Hilbert space,
$\mathcal{H}_{12}:=\mathcal{H}_1\otimes\mathcal{H}_2$.
We have again the sets of \emph{pure states} 
$\mathcal{P}_1:=\mathcal{P}(\mathcal{H}_1)$, 
$\mathcal{P}_2:=\mathcal{P}(\mathcal{H}_2)$, and 
$\mathcal{P}_{12}:=\mathcal{P}(\mathcal{H}_{12})$,
being the projectors onto one-dimensional subspaces in 
$\mathcal{H}_1$, 
$\mathcal{H}_2$, and 
$\mathcal{H}_{12}$,
and the sets of \emph{mixed states} 
$\mathcal{D}_1:=\mathcal{D}(\mathcal{H}_1)=\Conv\mathcal{P}_1$,
$\mathcal{D}_2:=\mathcal{D}(\mathcal{H}_2)=\Conv\mathcal{P}_2$, and
$\mathcal{D}_{12}:=\mathcal{D}(\mathcal{H}_{12})=\Conv\mathcal{P}_{12}$,
being the mixtures of pure states, 
and
$\mathcal{P}_1=\Extr\mathcal{D}_1$,
$\mathcal{P}_2=\Extr\mathcal{D}_2$, and
$\mathcal{P}_{12}=\Extr\mathcal{D}_{12}$,
for subsystem $1$, $2$ and the whole system $12$.
One can obtain the \emph{reduced} (or \emph{marginal}) \emph{states} by the use of the \emph{partial trace} operation,
for example $\tr_2:\mathcal{D}_{12}\to\mathcal{D}_1$, which is linear, 
and $\tr_2(X\otimes Y)= X (\tr Y)$.

If the state vector $\cket{\psi}\in\mathcal{H}_{12}$
can be written as an elementary tensor
$\cket{\psi}=\cket{\psi_1}\otimes\cket{\psi_2}$
with suitable state vectors $\cket{\psi_1}\in\mathcal{H}_1$ and $\cket{\psi_2}\in\mathcal{H}_2$,
then it is \emph{separable},
or else it is \emph{entangled},
e.g., 
$\cket{\psi}=
\frac1{\sqrt{2}}\bigl(\cket{\psi_1}\otimes\cket{\psi_2}+\cket{\psi'_1}\otimes\cket{\psi'_2}\bigr)$,
with $\bracket{\psi_1|\psi_1'}=\bracket{\psi_2|\psi_2'}=0$.
The set of \emph{separable pure states} is then
\begin{subequations}
\begin{equation}
\label{eq:setPsep}
\mathcal{P}_\text{sep} := \Bigset{\pi\in\mathcal{P}_{12}}{\exists \pi_1\in\mathcal{P}_1, \exists \pi_2\in\mathcal{P}_2: \pi = \pi_1\otimes\pi_2 },
\end{equation}
that is, the rank-$1$ projectors with separable images,
while the set of \emph{entangled pure states} is its complement 
$\cmpl{\mathcal{P}_\text{sep}}=\mathcal{P}\setminus\mathcal{P}_\text{sep}$,
and
the set of \emph{separable mixed states} is
\begin{equation}
\label{eq:setDsep}
\mathcal{D}_\text{sep} := \Conv\mathcal{P}_\text{sep},
\end{equation}
while the set of \emph{entangled mixed states} is its complement
$\cmpl{\mathcal{D}_\text{sep}}=\mathcal{D}\setminus\mathcal{D}_\text{sep}$.
This definition is motivated by that 
the separable mixed states can be created from uncorrelated (product) states
by the use of \emph{local} (quantum) \emph{operations and classical communication} (LOCC) 
\cite{WernerSep,ChitambaretalWoodyLOCC},
while entangled states cannot.
Two points have to be emphasized here.
On the one hand, the set of separable mixed states are closed under LOCC;
on the other hand, starting with an entangled state,
one can obtain separable states by means of LOCC.
Geometrically, the separable pure states are the \emph{extremal points}
of the convex body of the separable mixed states, 
\begin{equation}
\label{eq:setPsepextr}
\mathcal{P}_\text{sep} = \Extr\mathcal{D}_\text{sep}.
\end{equation}
The situation is summarized as
\begin{equation}
\begin{array}[c]{ccc}
\mathcal{D}_\text{sep}&\subset&\mathcal{D}_{12}\\
\rotatebox{90}{$\subset$}&&\rotatebox{90}{$\subset$}\\
\mathcal{P}_\text{sep}&\subset&\mathcal{P}_{12},
\end{array}
\end{equation}
\end{subequations}
which represents an important point of view in the sequel.

Thanks to the \emph{Schmidt decomposition} for bipartite state vectors \cite{Schmidt-1907,NielsenChuang},
it is easy to decide whether a pure state is separable or not:
For all $\pi\in\mathcal{P}_{12}$,
\begin{equation}
\label{eq:PsepDecide}
\pi\in\mathcal{P}_\text{sep} 
\quad\Longleftrightarrow\quad
\tr_2\pi\in\mathcal{P}_1
\quad\Longleftrightarrow\quad
\tr_1\pi\in\mathcal{P}_2.
\end{equation}
The mixed separability problem is, however, 
a hard optimization task \cite{Horodecki4,SzalayDissertation,SzalaySepCrit,GuhneTothEntDet}.

From the classical point of view, 
one faces several counterintuitive consequences 
following from the existence of entangled states.
These are based more or less on the fact that, as can be seen from \eqref{eq:PsepDecide},
entangled pure states have mixed marginals,
which is completely unimaginable for the classically thinking mind \cite{EPRpaper,SchrodingerEnt,Schrodinger2},
since in the classical case the marginals of a pure joint probability distribution are pure ones.

\section{Quantum states and entanglement for multipartite systems}
\label{sec:EntMulti}

In this section, we rebuild the \emph{partial separability classification}
of multipartite mixed states
in a more clarified way than was done originally \cite{SzalayKokenyesiPartSep,SzalayDissertation}.
This classification is complete in the sense of partial separability;
that is, it utilizes all the possible combinations of different kinds of partially separable pure states.
We also reveal the lattice theoretic structure behind the class structure.
For a quick summary on the very basic elements of lattice theory we use in the sequel,
see Appendix \ref{app:lattices.gen}, based on \cite{DaveyPriestley}.

The basic observation upon which the construction is built
is that
the whole construction can be formulated by the use of the notion of
pure state entanglement with respect to a bipartite split.
During the construction,
we separate the abstract hierarchy of (the labeling of) the partial separability properties
from the concrete hierarchy of the state sets of pure and mixed states,
which results in a very transparent building.
This building is of four floors.
The ground floor is the hierarchy of subsystems 
(Section \ref{sec:EntMulti.zeroth});
then the first and second floors are 
the hierarchic structures of the state sets of different partial separability properties
(Sections \ref{sec:EntMulti.first} and \ref{sec:EntMulti.second});
and the third floor is
the hierarchic structure of the classes of states showing different entanglement properties
(Section \ref{sec:EntMulti.classes}).

\subsection{Level 0: subsystems}
\label{sec:EntMulti.zeroth}

First of all, let us introduce some convenient notations.
For $n$-partite systems ($n>0$), 
the set of the \emph{labels} of the \emph{elementary subsystems} is $L=\{1,2,\dots,n\}$.
That is, for all $a\in L$, we have a Hilbert space $\mathcal{H}_a$, 
 with $1<\dim\mathcal{H}_a    <\infty$,
associated with the elementary subsystem of label $a$.
A \emph{subsystem} (not elementary in general) is then labeled by a subset $K\subseteq L$,
and has the Hilbert space $\mathcal{H}_K = \bigotimes_{a\in K}\mathcal{H}_a$ associated with it.
For $K=\emptyset$, we have the one-dimensional Hilbert space
$\mathcal{H}_\emptyset=\Span\{\cket{\psi}\}\cong\field{C}$.
For labeling the \emph{complementary subsystem,}
we have the notation $\pcmpl{K}=L\setminus K$.
We have the shorthand notation $\mathcal{H} \equiv \mathcal{H}_L$
for the Hilbert space of the whole system.
For a subsystem $K$, we have the set of \emph{pure states} \eqref{eq:pure},
\begin{subequations}
\begin{equation}
\mathcal{P}_K:=\mathcal{P}(\mathcal{H}_K),
\end{equation}
and the set of \emph{mixed states} \eqref{eq:mixed.conv},
\begin{equation}
\mathcal{D}_K:=\Conv\mathcal{P}_K,
\end{equation}
and, by construction,
\begin{equation}
\mathcal{P}_K=\Extr\mathcal{D}_K. 
\end{equation}
\end{subequations}
For $K=\emptyset$, we have $\mathcal{P}_\emptyset=\mathcal{D}_\emptyset=\{\cket{\psi}\bra{\psi}\}$.
We have the shorthand notation 
    $\mathcal{P}\equiv \mathcal{P}_L$
and $\mathcal{D}\equiv \mathcal{D}_L$ 
for the pure and mixed states of the whole system, respectively.
Let $K,K'\in L$, such that $K\subseteq K'$;
then let the linear map
 $\tr_K:\Lin\mathcal{H}_{K'}\to\Lin\mathcal{H}_{K'\setminus K}$, 
the \emph{partial trace over} $K$,
be defined as 
\begin{equation}
\label{eq:ptr}
\tr_K \bigotimes_{a'\in K'} X_{a'} := 
\Bigl( \prod_{a\in K}\tr X_a \Bigr)
\Bigl( \bigotimes_{a'\in K'\setminus K}X_{a'} \Bigr)
\end{equation}
for $X_a\in\Lin\mathcal{H}_a$,
adopting the convention that the empty product is $1\in\field{C}$,
and the empty tensorial product is the normalized $\cket{\psi}\bra{\psi}\in\mathcal{D}_\emptyset$.
(A slight abuse of the notation is that we use the same $\tr_K$ for all $K'$.)

In a formal sense,
the \emph{label of a subsystem} is an element of the power-set 
\begin{equation}
P_\text{0} := 2^L
\end{equation}
of the labels of the elementary subsystems $L$,
so we have the power-set lattice of subsystems \cite{DaveyPriestley}, 
\begin{equation}
\label{eq:poset0}
(P_\text{0},\subseteq,\cup,\cap,\pcmpl{\phantom{o}},\emptyset,L).
\end{equation}
The size of that is
$\abs{P_\text{0}}=2^{\abs{L}}=2^n$.

\subsection{Level I: partial separability hierarchy of the first kind}
\label{sec:EntMulti.first}

We would like to form mixtures from a given kind of partially separable pure states.
To this end, 
let $\alpha=\{K_1,K_2,\dots,K_\abs{\alpha}\} \equiv K_1|K_2|\dots|K_\abs{\alpha}$ 
denote a \emph{splitting} of the system,
that is, a \emph{partition} of the labels $L$
into \emph{parts}, being disjoint nonempty sets $K_i\subseteq L$, which together amount to $L$.
We have the set of all the possible partitions
\begin{equation}
\label{eq:PI}
\begin{split}
P_\text{I} := 
\Bigset{\alpha=K_1|K_2|\dots|K_{\abs{\alpha}}}{
\forall K\in\alpha: K\in P_\text{0}\setminus\{\emptyset\}, \\
\forall K,K'\in\alpha: K\neq K'\Rightarrow K\cap K'=\emptyset,
\bigcup_{K\in\alpha} K = L}.
\end{split}
\end{equation}
We call the partitions \emph{labels of the first kind},
and we use them for the labeling of such states.
The number of them for all $n$ is given by the $\abs{P_\text{I}}=B_n$
Bell numbers \cite{oeisA000110},
given by the recursive formula
$B_{n+1}=\sum_{k=0}^n\binom{n}{k}B_k$,
with $B_0=B_1=1$.
 
There is a natural (partial) order on the set of the partitions.
For two partitions $\beta,\alpha\in P_\text{I}$,
$\beta$ is a \emph{refinement} of $\alpha$
(``$\beta$ is \emph{finer} than $\alpha$'' or ``$\alpha$ is \emph{coarser} than $\beta$''),
denoted with $\beta\preceq\alpha$,
if $\alpha$ can be obtained from $\beta$ 
by joining some (maybe none) of the parts of $\beta$;
that is,
\begin{equation}
\label{eq:FirstLabRelDef}
\beta\preceq\alpha\quad\defn\quad \forall K'\in\beta, \exists K\in\alpha : K'\subseteq K.
\end{equation}
This defines a partial order on the set of partitions \cite{DaveyPriestley}, 
and $(P_\text{I},\preceq)$ is a \emph{poset} (partially ordered set).
(For a summary on the basic constructions in order theory,
see Appendix \ref{app:lattices.gen}.)
For example, for the tripartite case $1|2|3\preceq 1|23\preceq 123$.
(In the following, we omit the braces and also the comma in the cases when this does not cause confusion.)
Moreover, 
there are a top and a bottom element,
which are the full $n$-partite split and the trivial partition without split, respectively,
$\bot = 1|2|\dots|n\preceq\alpha\preceq\top=12\dots n$.

For the poset of partitions,
one can define
the \emph{greatest lower bound}, or \emph{meet}, $\alpha\wedge\alpha'$, and 
the  \emph{least upper bound}, or \emph{join}, $\alpha\vee\alpha'$, as
\begin{subequations}
\label{eq:partmj}
\begin{align}
\label{eq:part.meet}
\alpha\wedge\alpha' &:= \bigset{K\cap K'\neq\emptyset}{K\in\alpha,K'\in\alpha'},\\
\label{eq:part.join}
\alpha\vee\alpha' &:= \bigwedge\upset\{\alpha,\alpha'\},
\end{align}
\end{subequations}
so the set of partitions forms a lattice,
\begin{equation}
\label{eq:posetI}
\bigstruct{ P_\text{I}, \preceq,\vee,\wedge,1|2|\dots|n,12\dots n }.
\end{equation}
(The definition \eqref{eq:part.join} comes from a general construction \eqref{eq:joinconstr}.)

It is important that
the \emph{bipartitions} $K|\pcmpl{K}\in P_\text{I}$ 
can be used for the generation of all partitions,
\begin{equation}
\label{eq:partitionDual}
\alpha = \bigwedge_{K\in\alpha} K|\pcmpl{K}.
\end{equation} 
(For the proof, see Appendix \ref{app:lattices.dualI}.)
This turns out to be crucial later,
when the multipartite entanglement measures
are built upon bipartite ones.

For a partition $\alpha\in P_\text{I}$, 
we have the set of \emph{$\alpha$-separable pure states},
\begin{subequations}
\begin{equation}
\label{eq:setIP}
\mathcal{P}_\alpha:=\Bigset{\pi\in\Lin\mathcal{H}}
{\forall K\in\alpha, \exists \pi_K\in \mathcal{P}_K :
\pi = \bigotimes_{K\in\alpha} \pi_K},
\end{equation}
and the set of \emph{$\alpha$-separable mixed states},
\begin{equation}
\label{eq:setID}
\mathcal{D}_\alpha:=\Conv\mathcal{P}_\alpha,
\end{equation}
(that is, $\varrho$ is $\alpha$-separable if and only if 
it can be mixed by the use of $\alpha$-separable pure states) 
\cite{DurCiracTarrach3QBMixSep,DurCirac3QBMixSep,SeevinckUffinkMixSep,DurCiracMultipart,NagataKoashiImotoMultipart}.
It also holds by construction that
\begin{equation}
\label{eq:setIPextr}
\mathcal{P}_\alpha=\Extr\mathcal{D}_\alpha,
\end{equation}
\end{subequations}
there are no other extremal $\alpha$-separable states than the pure ones.
For the $1$-partite trivial split $\alpha = \{K_1\} = \{L\}$, we have that
 the $\{L\}$-separable pure and mixed states
    $\mathcal{P}_{\{L\}}=\mathcal{P}_L \equiv \mathcal{P}$
and $\mathcal{D}_{\{L\}}=\mathcal{D}_L \equiv \mathcal{D}$
are obviously all the pure and mixed states of the system.
Note that for all $\alpha\in P_\text{I}$,
the state sets $\mathcal{D}_\alpha$ are closed under LOCC;
that is, for all LOCC map $\Lambda:\mathcal{D}\to\mathcal{D}$,
\begin{equation}
\label{eq:LOCCcloseI}
\varrho\in\mathcal{D}_\alpha
\quad\Longrightarrow\quad
\Lambda(\varrho)\in\mathcal{D}_\alpha.
\end{equation}
(For the proof, see Appendix \ref{app:lattices.LOCCcloseI}.)

Note that these definitions 
only demand the separability with respect to a given split,
independently of whether the separability with respect to a finer split also holds.
That is,
the $\mathcal{P}_\alpha$ and $\mathcal{D}_\alpha$ sets are containing
(and also closed),
and the sets
\begin{subequations}
\begin{align}
\label{eq:statesIP}
P_{\text{I},\mathcal{P}}&:=\bigset{\mathcal{P}_\alpha}{\alpha\in P_\text{I}},\\
\label{eq:statesID}
P_{\text{I},\mathcal{D}}&:=\bigset{\mathcal{D}_\alpha}{\alpha\in P_\text{I}}
\end{align}
\end{subequations}
are posets with respect to the inclusion,
$(P_{\text{I},\mathcal{P}}, \subseteq)$,
$(P_{\text{I},\mathcal{D}}, \subseteq)$.
Moreover, the set-theoretical inclusion perfectly resembles the ordering of the respective partitions,
\begin{subequations}
\label{eq:orderisomI}
\begin{align}
\label{eq:orderisomIP}
\beta\preceq\alpha \quad\Longleftrightarrow\quad
 \mathcal{P}_\beta \subseteq \mathcal{P}_\alpha,
\intertext{and}
\label{eq:orderisomID}
\beta\preceq\alpha \quad\Longleftrightarrow\quad
 \mathcal{D}_\beta \subseteq \mathcal{D}_\alpha
\end{align}
\end{subequations}
(that is, separability with respect to a finer split 
implies that with respect to a coarser one),
so the posets $(P_\text{I},\preceq)$, 
$(P_{\text{I},\mathcal{P}},\subseteq)$, and
$(P_{\text{I},\mathcal{D}},\subseteq)$ are isomorphic.
(That has already been proven in \cite{SzalayKokenyesiPartSep,SzalayDissertation}
in a different construction.
We give a more basic proof, 
which uses only the notion of bipartite separability \eqref{eq:PsepDecide},
in Appendix \ref{app:lattices.isomI}.)

Do the other structures, meet $\wedge$ and join $\vee$ \eqref{eq:partmj}, resemble
the natural, set-theoretical intersection $\cap$ and union $\cup$ 
for the state sets \eqref{eq:statesIP} and \eqref{eq:statesID}?
We know from \eqref{eq:wvGen} that
$\alpha\wedge\alpha'\preceq\alpha,\alpha'\preceq\alpha\vee\alpha'$;
this leads to
$\mathcal{P}_{\alpha\wedge\alpha'}\subseteq\mathcal{P}_\alpha,\mathcal{P}_{\alpha'}\subseteq\mathcal{P}_{\alpha\vee\alpha'}$
and
$\mathcal{D}_{\alpha\wedge\alpha'}\subseteq\mathcal{D}_\alpha,\mathcal{D}_{\alpha'}\subseteq\mathcal{D}_{\alpha\vee\alpha'}$
due to \eqref{eq:orderisomIP} 
and \eqref{eq:orderisomID}.
From these we have
\begin{subequations}
\begin{align}
\label{eq:meetjoinintersectunionIP}
\mathcal{P}_{\alpha\wedge\alpha'}&\subseteq\mathcal{P}_\alpha\cap\mathcal{P}_{\alpha'},&\quad
\mathcal{P}_\alpha\cup\mathcal{P}_{\alpha'}&\subseteq\mathcal{P}_{\alpha\vee\alpha'},\\
\label{eq:meetjoinintersectunionID}
\mathcal{D}_{\alpha\wedge\alpha'}&\subseteq\mathcal{D}_\alpha\cap\mathcal{D}_{\alpha'},&\quad
\mathcal{D}_\alpha\cup\mathcal{D}_{\alpha'}&\subseteq\mathcal{D}_{\alpha\vee\alpha'}.
\end{align}
\end{subequations}
These are what we have by the use of only the \eqref{eq:orderisomIP} and \eqref{eq:orderisomID}
isomorphisms of the orderings. 
However, there is more to be known for pure states if one takes into consideration
the \eqref{eq:setIP} definition of the $\mathcal{P}_\alpha$ sets of $\alpha$-separable pure states.
In this case it can be proven that
\begin{equation}
\label{eq:meetintersectIPeq}
\mathcal{P}_\alpha\cap\mathcal{P}_{\alpha'}=\mathcal{P}_{\alpha\wedge\alpha'};
\end{equation}
that is, a \emph{pure} state is separable under the splits $\alpha$ and $\alpha'$
\emph{if and only if} it is separable under their meet $\alpha\wedge\alpha'$ \eqref{eq:part.meet}.
(For the proof, see Appendix \ref{app:lattices.isomIP}.)
This means that $P_{\text{I},\mathcal{P}}$ is closed under intersection,
and
\begin{equation}
\label{eq:posetIP}
\bigstruct{ P_{\text{I},\mathcal{P}},\subseteq,\cap,\mathcal{P}_{1|2|\dots|n},\mathcal{P}_{12\dots n} }
\end{equation}
is a meet-semilattice,
and, due to \eqref{eq:orderisomIP} and \eqref{eq:meetintersectIPeq},
this structure is isomorphic to that of $P_\text{I}$ given in \eqref{eq:posetI},
\begin{equation}
\begin{split}
&\bigstruct{ P_{\text{I},\mathcal{P}},\subseteq,\cap,\mathcal{P}_{1|2|\dots|n},\mathcal{P}_{12\dots n} }\\
&\quad\cong
\bigstruct{ P_\text{I},\preceq,\wedge,1|2|\dots|n,12\dots n }.
\end{split}
\end{equation}
(Note that, by \eqref{eq:joinconstr},
one can also define the join for the meet $\cap$;
however, this would not lead to the set-theoretical union, but for 
$\cap\downset\{\mathcal{P}_\alpha,\mathcal{P}_{\alpha'}\}=\mathcal{P}_{\alpha\vee\alpha'}$.
Of course, $P_{\text{I},\mathcal{P}}$ is not closed under the set-theoretical union.)
A corollary of \eqref{eq:meetintersectIPeq} and \eqref{eq:partitionDual} is
that
\begin{equation}
\label{eq:IPintersect}
\mathcal{P}_\alpha = \bigcap_{K\in\alpha} \mathcal{P}_{K|\cmpl{K}};
\end{equation}
that is, a \emph{pure} state is separable under a split $\alpha$,
\emph{if and only if} it is separable under all bipartitions $K|\cmpl{K}$,
where $K\in\alpha$.
A corollary of \eqref{eq:IPintersect} and \eqref{eq:PsepDecide} is that
it is easy to decide whether a pure state is $\alpha$-separable or not:
For all $\pi\in\mathcal{P}$,
\begin{equation}
\label{eq:IPDecide}
\pi\in\mathcal{P}_\alpha
\quad\Longleftrightarrow\quad
\forall K\in\alpha:\; \tr_{\cmpl{K}}\pi\in\mathcal{P}_K.
\end{equation}
(For the proof, see Appendix \ref{app:lattices.IPdecision}.)
On the other hand, because of the convex hull construction \eqref{eq:setID},
there are no such results for mixed states,
we have only the poset
\begin{equation}
\label{eq:posetID}
\bigstruct{ P_{\text{I},\mathcal{D}} \subseteq,\mathcal{D}_{1|2|\dots|n},\mathcal{D}_{12\dots n} },
\end{equation}
and, due to \eqref{eq:orderisomID},
this structure is isomorphic to that of $P_\text{I}$ given in \eqref{eq:posetI},
\begin{equation}
\begin{split}
&\bigstruct{ P_{\text{I},\mathcal{D}},\subseteq,\mathcal{D}_{1|2|\dots|n},\mathcal{D}_{12\dots n} }\\
&\quad\cong
\bigstruct{ P_\text{I},\preceq,1|2|\dots|n,12\dots n },
\end{split}
\end{equation}
as was mentioned before.
The mixed separability problem is, again, 
a hard optimization task \cite{Horodecki4,SzalayDissertation,SzalaySepCrit,GuhneTothEntDet}.

Note that a complementary notion can also be defined.
A pure state $\pi\in\mathcal{P}$ is
\emph{$\alpha$-entangled,} if it is not $\alpha$-separable;
that is, $\pi\in\cmpl{\mathcal{P}_\alpha}=\mathcal{P}\setminus\mathcal{P}_\alpha$.
For the preparation of these states, nonlocal operations are needed among the $K\in\alpha$ subsystems.
Note that $\cmpl{\mathcal{P}_\alpha}$ is not closed.
A mixed state $\varrho\in\mathcal{D}$ is
\emph{$\alpha$-entangled,} if it is not $\alpha$-separable;
that is, $\varrho\in\cmpl{\mathcal{D}_\alpha}=\mathcal{D}\setminus\mathcal{D}_\alpha$.
For the preparation of these states, it is not enough to use $\alpha$-separable states only;
there is also a need for $\alpha$-entangled states.
Note that $\cmpl{\mathcal{D}_\alpha}$ is neither convex nor closed.
For these complementary state sets we have the sets 
\begin{subequations}
\label{eq:statesIc}
\begin{align}
\label{eq:statesIPc}
P_{\text{I},\cmpl{\mathcal{P}}}&:=\bigset{\cmpl{\mathcal{P}_\alpha}}{\alpha\in P_\text{I}},\\
\label{eq:statesIDc}
P_{\text{I},\cmpl{\mathcal{D}}}&:=\bigset{\cmpl{\mathcal{D}_\alpha}}{\alpha\in P_\text{I}},
\end{align}
\end{subequations}
which are posets with respect to the inclusion,
$(P_{\text{I},\cmpl{\mathcal{P}}}, \subseteq)$,
$(P_{\text{I},\cmpl{\mathcal{D}}}, \subseteq)$.
For these complementary state sets we have the reverse \eqref{eq:orderisomI} order,
\begin{subequations}
\label{eq:orderisomIc}
\begin{align}
\label{eq:orderisomIPc}
\beta\preceq\alpha \quad\Longleftrightarrow\quad 
\cmpl{\mathcal{P}_\beta} \supseteq \cmpl{\mathcal{P}_\alpha},
\intertext{and}
\label{eq:orderisomIDc}
\beta\preceq\alpha \quad\Longleftrightarrow\quad 
\cmpl{\mathcal{D}_\beta} \supseteq \cmpl{\mathcal{D}_\alpha}
\end{align}
\end{subequations}
(that is, entanglement with respect to a coarser split 
implies that with respect to a finer one),
so the posets $(P_\text{I},\preceq)$,
$(P_{\text{I},\cmpl{\mathcal{P}}},\supseteq)$, and
$(P_{\text{I},\cmpl{\mathcal{D}}},\supseteq)$ are isomorphic.

\subsection{Examples}
\label{sec:EntMulti.firstExamples}

\begin{figure}
\includegraphics{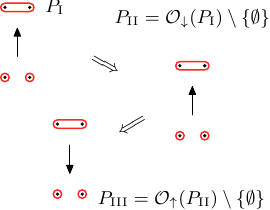}
\caption{Lattices of the labels of the first and second kinds and class labels,
$P_\text{I}$, $P_\text{II}$, and $P_\text{III}$,
are illustrated for $n=2$.
The partitions $\alpha\in P_\text{I}$ are denoted by small pictograms;
the labels of the second kind $\vs{\alpha}\in P_\text{II}$ are down-sets of partitions
(only the maximal elements are drawn).
The class labels $\vvs{\alpha}\in P_\text{III}$ are up-sets of labels of the second kind
(only the minimal elements are drawn).
The order relation is denoted with an arrow: 
$\beta\rightarrow\alpha$ means $\beta\preceq\alpha$,
$\vs{\beta}\rightarrow\vs{\alpha}$ means $\vs{\beta}\preceq\vs{\alpha}$, and
$\vvs{\beta}\rightarrow\vvs{\alpha}$ means $\vvs{\beta}\preceq\vvs{\alpha}$.
By means of \eqref{eq:orderisomI}
and \eqref{eq:orderisomII},
the lattices $P_\text{I}$ and $P_\text{II}$ resemble the inclusion of the sets of 
$\alpha$-separable pure ($\mathcal{P}_\alpha$), 
$\alpha$-separable mixed ($\mathcal{D}_\alpha$), and
$\vs{\alpha}$-separable pure ($\mathcal{P}_{\vs{\alpha}}$),
$\vs{\alpha}$-separable mixed states ($\mathcal{D}_{\vs{\alpha}}$);
(see Sections \ref{sec:EntMulti.first} and \ref{sec:EntMulti.second}).
The lattice $P_\text{III}$ is the class hierarchy (see Section \ref{sec:EntMulti.classes}),
and by means of \eqref{eq:LOCCconvhierarchy}, it is related to the LOCC convertibility
of the classes ($\mathcal{C}_{\vvs{\alpha}}$).
}
\label{fig:labellattices2}
\end{figure}

\begin{figure} 
\includegraphics{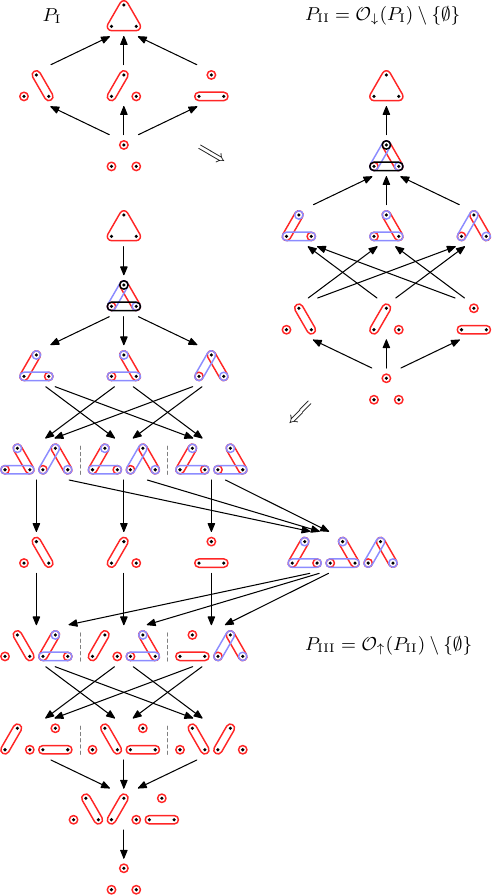}
\caption{Lattices of the labels of the first and second kinds and class labels,
$P_\text{I}$, $P_\text{II}$, and $P_\text{III}$,
are illustrated for $n=3$.
The partitions $\alpha\in P_\text{I}$ are denoted by small pictograms;
the labels of the second kind $\vs{\alpha}\in P_\text{II}$ are down-sets of partitions,
in which case the different elements are drawn with different colors
(only the maximal elements are drawn).
The class labels $\vvs{\alpha}\in P_\text{III}$ are up-sets of labels of the second kind,
these are written side by side
(only the minimal elements are drawn).
The order relation is denoted with an arrow: 
$\beta\rightarrow\alpha$ means $\beta\preceq\alpha$, 
$\vs{\beta}\rightarrow\vs{\alpha}$ means $\vs{\beta}\preceq\vs{\alpha}$, and
$\vvs{\beta}\rightarrow\vvs{\alpha}$ means $\vvs{\beta}\preceq\vvs{\alpha}$.
By means of \eqref{eq:orderisomI}
and \eqref{eq:orderisomII},
the lattices $P_\text{I}$ and $P_\text{II}$ resemble the inclusion of the sets of 
$\alpha$-separable pure ($\mathcal{P}_\alpha$), 
$\alpha$-separable mixed ($\mathcal{D}_\alpha$), and
$\vs{\alpha}$-separable pure ($\mathcal{P}_{\vs{\alpha}}$),
$\vs{\alpha}$-separable mixed states ($\mathcal{D}_{\vs{\alpha}}$);
(see Sections \ref{sec:EntMulti.first} and \ref{sec:EntMulti.second}).
The lattice $P_\text{III}$ is the class hierarchy (see Section \ref{sec:EntMulti.classes}),
and by means of \eqref{eq:LOCCconvhierarchy}, it is related to the LOCC convertibility
of the classes ($\mathcal{C}_{\vvs{\alpha}}$).
}
\label{fig:labellattices3}
\end{figure}

Writing out some examples explicitly might not be useless here.
The lattice $P_\text{I}$ for the cases $n=2$ and $3$ can be seen in the upper-left parts
of Figures \ref{fig:labellattices2} and \ref{fig:labellattices3}.
As we have learned in \eqref{eq:orderisomIP} and \eqref{eq:orderisomID},
we need to draw only this lattice 
for the inclusion hierarchy of the sets of $\alpha$-separable pure and mixed states $\mathcal{P}_\alpha$ \eqref{eq:setIP} and $\mathcal{D}_\alpha$ \eqref{eq:setID}.

For the \emph{bipartite} case we have $\mathcal{H}_{12}=\mathcal{H}_1\otimes\mathcal{H}_2$,
and we get back the content of Section \ref{sec:EntBasics.Ent}.
The sets of $\alpha$-separable pure states are
\begin{align*}
\mathcal{P}_{12}  &\equiv\mathcal{P}(\mathcal{H}_{12}),\\
\mathcal{P}_{1|2} &=\bigset{\pi\in\mathcal{P}_{12}}{\pi=\pi_1\otimes\pi_2}=\mathcal{P}_\text{sep}.
\end{align*}
Note that $\mathcal{P}_{1|2}\subseteq\mathcal{P}_{12}$.
The sets of $\alpha$-separable mixed states are
\begin{align*}
\mathcal{D}_{12}  &= \Conv\mathcal{P}_{12}
\equiv\mathcal{D}(\mathcal{H}_{12}),\\
\mathcal{D}_{1|2} &= \Conv\mathcal{P}_{1|2}=\mathcal{D}_\text{sep}.
\end{align*}
Note that, again, $\mathcal{D}_{1|2}\subseteq\mathcal{D}_{12}$.

For the \emph{tripartite} case we have $\mathcal{H}_{123}=\mathcal{H}_1\otimes\mathcal{H}_2\otimes\mathcal{H}_3$.
The sets of $\alpha$-separable pure states are
\begin{align*}
\mathcal{P}_{123}   &\equiv\mathcal{P}(\mathcal{H}_{123}),\\
\mathcal{P}_{a|bc}  &=\bigset{\pi\in\mathcal{P}_{123}}{\pi=\pi_a\otimes\pi_{bc}},\\
\mathcal{P}_{1|2|3} &=\bigset{\pi\in\mathcal{P}_{123}}{\pi=\pi_1\otimes\pi_2\otimes\pi_3},
\end{align*}
with all bipartitions $a|bc$ of $\{1,2,3\}$.
Note that $\mathcal{P}_{1|2|3}\subseteq\mathcal{P}_{a|bc}\subseteq\mathcal{P}_{123}$.
Note, on the other hand, the manifestation of \eqref{eq:meetintersectIPeq}:
If a pure state is separable under the splits $2|13$ and $3|12$, then it is separable under
$2|13\wedge3|12=1|2|3$, that is, fully separable.
The sets of $\alpha$-separable mixed states are
\begin{align*}
\mathcal{D}_{123}   &= \Conv\mathcal{P}_{123}
\equiv\mathcal{D}(\mathcal{H}_{123}),\\
\mathcal{D}_{a|bc}  &= \Conv\mathcal{P}_{a|bc},\\
\mathcal{D}_{1|2|3} &= \Conv\mathcal{P}_{1|2|3}.
\end{align*}
Note that, again, $\mathcal{D}_{1|2|3}\subseteq\mathcal{D}_{a|bc}\subseteq\mathcal{D}_{123}$.
Note, on the other hand, that there is no \eqref{eq:meetintersectIPeq}-like result for mixed states:
If a mixed state is separable under the splits $2|13$ and $3|12$, 
then it is not necessarily fully separable, 
we have only \eqref{eq:meetjoinintersectunionID}
(see, e.g., \cite{AcinetalMixThreeQB,SeevinckUffinkMixSep}).

\noindent

\subsection{Level II: partial separability hierarchy of the second kind}
\label{sec:EntMulti.second}

An important observation in the theory of entanglement of multipartite mixed states
\cite{AcinetalMixThreeQB,SeevinckUffinkMixSep} is
that there are mixed states
which
cannot be mixed by the use of any given nontrivial $\alpha$-separable pure states,
while they can be mixed by the use of pure states of different nontrivial $\alpha$-separability.
For example,
for the tripartite case in Section \ref{sec:EntMulti.firstExamples},
there are states $\varrho\notin \mathcal{D}_{a|bc}$,
which can be mixed by the use of bipartite entanglement in subsystems $12$, $13$, and $23$;
that is, $\varrho\in \Conv(\mathcal{P}_{1|23}\cup\mathcal{P}_{2|13}\cup\mathcal{P}_{3|12})$.
Such states should not be considered fully tripartite-entangled,
since they can be mixed without the use of genuine tripartite entanglement,
and these kinds of situations have to be handled
\cite{AcinetalMixThreeQB,SeevinckUffinkMixSep}.

So we would also like to form mixtures from different kinds of partially separable pure states.
To this end, let $\vs{\alpha}$ be 
a nonempty \emph{down-set} \eqref{eq:downset} in $P_\text{I}$,
that is, 
a nonempty $\vs{\alpha}=\{\alpha_1,\alpha_2,\dots,\alpha_{\abs{\vs{\alpha}}}\}\subseteq P_\text{I}$ set,
which contains every partition which is finer than its maximal elements
(see Appendix \ref{app:lattices.gen}).
We have the set of all the possible nonempty down-sets 
\begin{equation}
\label{eq:PII}
\begin{split}
&P_\text{II} := \mathcal{O}_\downarrow(P_\text{I})\setminus\{\emptyset\} \\
&\equiv \Bigset{\vs{\alpha}\in 2^{P_\text{I}}\setminus\{\emptyset\}}{
\forall \alpha\in\vs{\alpha}:\; \beta\preceq\alpha \;\Rightarrow\;
\beta\in\vs{\alpha}}.
\end{split}
\end{equation}
We call the nonempty down-sets of partitions \emph{labels of the second kind},
and we use them for the labeling of such states.
(For a nonempty down-set $\vs{\alpha}$,
the set of its maximal elements, $\max\vs{\alpha}$, 
was called \emph{proper label}
and denoted in the same way previously \cite{SzalayKokenyesiPartSep,SzalayDissertation}.
Since the set of maximal elements of a nonempty down-set $\vs{\alpha}$ 
determines $\vs{\alpha}=\downarrow\max\vs{\alpha}$ uniquely,
and vice versa,
both $\max\vs{\alpha}$ and $\vs{\alpha}$ are equally suitable for the labeling of the 
sets of states with the given partial separability properties. 
The former one is perhaps more natural and expressive in some sense,
while the latter one leads to a simpler and more transparent mathematical construction.)

The $P_\text{II}$ set of nonempty down-sets of the lattice $P_\text{I}$ 
forms a lattice
with respect to the inclusion, intersection, and union \cite{DaveyPriestley},
so we have
\begin{equation}
\label{eq:posetII}
\begin{split}
&\bigstruct{ P_\text{II},\preceq,\wedge,\vee,\bot,\top }\\
&\;= \bigstruct{ \mathcal{O}_\downarrow(P_\text{I})\setminus\{\emptyset\},\subseteq,\cap,\cup,\{1|2|\dots|n\},\downset\{12\dots n\}\equiv P_\text{I} }.
\end{split}
\end{equation}
For example, for the tripartite case
$\downset\{1|2|3\} = \{1|2|3\} \preceq \downset\{1|23\}\preceq \downset\{1|23,2|13\}\preceq\downset\{1|23,2|13,3|12\}\preceq \downset\{123\}=P_\text{I}$.

For a down-set $\vs{\alpha}\in P_\text{II}$, 
we have the set of \emph{$\vs{\alpha}$-separable pure states}
\begin{subequations}
\begin{equation}
\label{eq:setIIP}
\mathcal{P}_{\vs{\alpha}}:=\bigcup_{\alpha\in\vs{\alpha}} \mathcal{P}_\alpha
=\bigcup_{\alpha\in\max\vs{\alpha}} \mathcal{P}_\alpha,
\end{equation}
(that is, $\pi$ is $\vs{\alpha}$-separable if and only if
it is $\alpha$-separable for at least one $\alpha\in \vs{\alpha}$;
on the other hand, because of \eqref{eq:orderisomIP},
it is enough to use $\max\vs{\alpha}$ for the calculation of the union)
and the set of \emph{$\vs{\alpha}$-separable mixed states}
\begin{equation}
\label{eq:setIID}
\mathcal{D}_{\vs{\alpha}}:=\Conv\mathcal{P}_{\vs{\alpha}}
\end{equation}
(that is, $\varrho$ is $\vs{\alpha}$-separable if and only if 
it can be mixed by the use of \emph{any} $\alpha$-separable pure states
for which $\alpha\in\vs{\alpha}$).
It also holds by construction that
\begin{equation}
\label{eq:setIIPextr}
\mathcal{P}_{\vs{\alpha}}=\Extr\mathcal{D}_{\vs{\alpha}},
\end{equation}
\end{subequations}
so there are no other extremal $\vs{\alpha}$-separable states than the pure ones.
For the $\vs{\alpha}$ containing only the $1$-partite trivial split $\vs{\alpha}=\{\alpha\} = \{\{L\}\}$, we have that
 the $\{\{L\}\}$-separable pure and mixed states
    $\mathcal{P}_{\{\{L\}\}}=\mathcal{P}_L \equiv \mathcal{P}$
and $\mathcal{D}_{\{\{L\}\}}=\mathcal{D}_L \equiv \mathcal{D}$
are obviously all the pure and mixed states of the system.
Note that for all $\vs{\alpha}\in P_\text{II}$,
the state sets $\mathcal{D}_{\vs{\alpha}}$ are closed under LOCC;
that is, for all LOCC map $\Lambda:\mathcal{D}\to\mathcal{D}$,
\begin{equation}
\label{eq:LOCCcloseII}
\varrho\in\mathcal{D}_{\vs{\alpha}}
\quad\Longrightarrow\quad
\Lambda(\varrho)\in\mathcal{D}_{\vs{\alpha}}.
\end{equation}
(For the proof, see Appendix \ref{app:lattices.LOCCcloseII}.)
Note that if $\abs{\max\vs{\alpha}}>1$, then
$\mathcal{D}_{\vs{\alpha}} \supset
 \cup_{\alpha\in\vs{\alpha}}\mathcal{D}_\alpha$;
that is, an $\vs{\alpha}$-separable mixed state does not need to be $\alpha$-separable
for any particular split $\alpha\in\vs{\alpha}$.

Note that these definitions 
only demand the separability with respect to any of the given splits,
independently of whether the separability with respect to finer splits also holds.
That is,
the $\mathcal{P}_{\vs{\alpha}}$ and $\mathcal{D}_{\vs{\alpha}}$ sets are containing
(and also closed),
and the sets
\begin{subequations}
\begin{align}
\label{eq:statesIIP}
P_{\text{II},\mathcal{P}}&:=\bigset{\mathcal{P}_{\vs{\alpha}}}{{\vs{\alpha}}\in P_\text{II}},\\
\label{eq:statesIID}
P_{\text{II},\mathcal{D}}&:=\bigset{\mathcal{D}_{\vs{\alpha}}}{{\vs{\alpha}}\in P_\text{II}}
\end{align}
\end{subequations}
are posets with respect to the inclusion,
$(P_{\text{II},\mathcal{P}}, \subseteq)$,
$(P_{\text{II},\mathcal{D}}, \subseteq)$.
Moreover, the set-theoretical inclusion perfectly resembles the ordering (inclusion)
\eqref{eq:posetII}
of the respective labels of the second kind,
\begin{subequations}
\label{eq:orderisomII}
\begin{align}
\label{eq:orderisomIIP}
\vs{\beta}\preceq\vs{\alpha} \quad\Longleftrightarrow\quad \mathcal{P}_{\vs{\beta}} \subseteq \mathcal{P}_{\vs{\alpha}},
\intertext{and}
\label{eq:orderisomIID}
\vs{\beta}\preceq\vs{\alpha} \quad\Longleftrightarrow\quad \mathcal{D}_{\vs{\beta}} \subseteq \mathcal{D}_{\vs{\alpha}},
\end{align}
\end{subequations}
(that is, a separability lower in the hierarchy implies a higher one),
so the posets $(P_\text{II},\preceq)$, 
$(P_{\text{II},\mathcal{P}},\subseteq)$, and
$(P_{\text{II},\mathcal{D}},\subseteq)$ are isomorphic.
(We recall the proof in Appendix \ref{app:lattices.isomII} from \cite{SzalayKokenyesiPartSep,SzalayDissertation} in a slightly modified form, adjusted to the present construction.)

Do the other structures meet $\wedge$ and join $\vee$ \eqref{eq:posetII} resemble
the natural, set-theoretical intersection $\cap$ and union $\cup$ 
for the state sets \eqref{eq:statesIIP} and \eqref{eq:statesIID}?
We know from \eqref{eq:wvGen} that
$\vs{\alpha}\wedge\vs{\alpha}'\preceq\vs{\alpha},\vs{\alpha}'\preceq\vs{\alpha}\vee\vs{\alpha}'$,
this leads to
$\mathcal{P}_{\vs{\alpha}\wedge\vs{\alpha}'}\subseteq\mathcal{P}_{\vs{\alpha}},\mathcal{P}_{\vs{\alpha}'}\subseteq\mathcal{P}_{\vs{\alpha}\vee\vs{\alpha}'}$,
and
$\mathcal{D}_{\vs{\alpha}\wedge\vs{\alpha}'}\subseteq\mathcal{D}_{\vs{\alpha}},\mathcal{D}_{\vs{\alpha}'}\subseteq\mathcal{D}_{\vs{\alpha}\vee\vs{\alpha}'}$
due to \eqref{eq:orderisomIIP} 
and \eqref{eq:orderisomIID}.
From these we have
\begin{subequations}
\label{eq:meetjoinintersectunionII}
\begin{align}
\label{eq:meetjoinintersectunionIIP}
\mathcal{P}_{\vs{\alpha}\wedge\vs{\alpha}'}&\subseteq\mathcal{P}_{\vs{\alpha}}\cap\mathcal{P}_{\vs{\alpha}'},&\quad
\mathcal{P}_{\vs{\alpha}}\cup\mathcal{P}_{\vs{\alpha}'}&\subseteq\mathcal{P}_{\vs{\alpha}\vee\vs{\alpha}'},\\
\label{eq:meetjoinintersectunionIID}
\mathcal{D}_{\vs{\alpha}\wedge\vs{\alpha}'}&\subseteq\mathcal{D}_{\vs{\alpha}}\cap\mathcal{D}_{\vs{\alpha}'},&\quad
\mathcal{D}_{\vs{\alpha}}\cup\mathcal{D}_{\vs{\alpha}'}&\subseteq\mathcal{D}_{\vs{\alpha}\vee\vs{\alpha}'}.
\end{align}
\end{subequations}
These are what we have by the use of only the \eqref{eq:orderisomIIP} and \eqref{eq:orderisomIID}
isomorphisms of the orderings.
However, there is more to be known for pure states if one takes into consideration
the \eqref{eq:setIIP} definition of the $\mathcal{P}_{\vs{\alpha}}$ sets of $\vs{\alpha}$-separable pure states.
In this case, it can be proven that
\begin{equation}
\label{eq:meetjoinintersectunionIIPeq}
\mathcal{P}_{\vs{\alpha}}\cap\mathcal{P}_{\vs{\alpha}'}=\mathcal{P}_{\vs{\alpha}\wedge\vs{\alpha}'},\quad
\mathcal{P}_{\vs{\alpha}}\cup\mathcal{P}_{\vs{\alpha}'}=\mathcal{P}_{\vs{\alpha}\vee\vs{\alpha}'}.
\end{equation}
(For the proof, see Appendix \ref{app:lattices.isomIIP}.)
This means that $P_{\text{II},\mathcal{P}}$ is closed under intersection and union,
and
\begin{equation}
\label{eq:posetIIP}
\bigstruct{ P_{\text{II},\mathcal{P}},\subseteq,\cap,\cup,\mathcal{P}_{\downset\{1|2|\dots|n\}},\mathcal{P}_{\downset\{12\dots n\}} }
\end{equation}
is a lattice,
and due to \eqref{eq:orderisomIIP} and \eqref{eq:meetjoinintersectunionIIPeq},
this structure is isomorphic to that of $P_\text{II}$ given in \eqref{eq:posetII},
\begin{equation}
\begin{split}
&\bigstruct{ P_{\text{II},\mathcal{P}},\subseteq,\cap,\cup,\mathcal{P}_{\downset\{1|2|\dots|n\}},\mathcal{P}_{\downset\{12\dots n\}} }\\
&\quad\cong
\bigstruct{ P_\text{II},\preceq,\wedge,\vee,\downset\{1|2|\dots|n\},\downset\{12\dots n\} }.
\end{split}
\end{equation}
It is again easy to decide
whether a pure state is $\vs{\alpha}$-separable or not:
By definition \eqref{eq:setIIP}, we have to decide if it is $\alpha$-separable \eqref{eq:IPDecide}
for at least one $\alpha\in\vs{\alpha}$.
On the other hand, because of the convex hull construction \eqref{eq:setIID},
there is no such result for mixed states;
we have only the poset
\begin{equation}
\label{eq:posetIID}
\bigstruct{ P_{\text{II},\mathcal{D}},\subseteq,\mathcal{D}_{\downset\{1|2|\dots|n\}},\mathcal{D}_{\downset\{12\dots n\}} },
\end{equation}
and, due to \eqref{eq:orderisomIID},
this structure is isomorphic to that of $P_\text{II}$ given in \eqref{eq:posetII},
\begin{equation}
\begin{split}
&\bigstruct{ P_{\text{II},\mathcal{D}},\subseteq,\mathcal{D}_{\downset\{1|2|\dots|n\}},\mathcal{D}_{\downset\{12\dots n\}} }\\
&\quad\cong
\bigstruct{ P_\text{II},\preceq,\downset\{1|2|\dots|n\},\downset\{12\dots n\} },
\end{split}
\end{equation}
as was mentioned before.
The mixed separability problem is, again, 
a hard optimization task \cite{Horodecki4,SzalayDissertation,SzalaySepCrit,GuhneTothEntDet}.

Note that a complementary notion can also be defined.
A pure state $\pi\in\mathcal{P}$ is
\emph{$\vs{\alpha}$-entangled,} if it is not $\vs{\alpha}$-separable;
that is, $\pi\in\cmpl{\mathcal{P}_{\vs{\alpha}}}=\mathcal{P}\setminus\mathcal{P}_{\vs{\alpha}}$;
that is, it is not separable under \emph{any} $\alpha\in\vs{\alpha}$ splits.
Note that $\cmpl{\mathcal{P}_{\vs{\alpha}}}$ is not closed.
A mixed state $\varrho\in\mathcal{D}$ is
\emph{$\vs{\alpha}$-entangled,} if it is not $\vs{\alpha}$-separable,
that is, $\varrho\in\cmpl{\mathcal{D}_{\vs{\alpha}}}=\mathcal{D}\setminus\mathcal{D}_{\vs{\alpha}}$.
For the preparation (by mixing) of these states, it is not enough to use $\vs{\alpha}$-separable states only;
there is also a need for $\vs{\alpha}$-entangled states.
Note that $\cmpl{\mathcal{D}_{\vs{\alpha}}}$ is neither convex nor closed.
For these complementary state sets we have the sets 
\begin{subequations}
\label{eq:statesIIc}
\begin{align}
\label{eq:statesIIPc}
P_{\text{II},\cmpl{\mathcal{P}}}&:=\bigset{\cmpl{\mathcal{P}_{\vs{\alpha}}}}{\vs{\alpha}\in P_\text{II}},\\
\label{eq:statesIIDc}
P_{\text{II},\cmpl{\mathcal{D}}}&:=\bigset{\cmpl{\mathcal{D}_{\vs{\alpha}}}}{\vs{\alpha}\in P_\text{II}},
\end{align}
\end{subequations}
which are posets with respect to the inclusion,
$(P_{\text{II},\cmpl{\mathcal{P}}}, \subseteq)$,
$(P_{\text{II},\cmpl{\mathcal{D}}}, \subseteq)$.
Because of the reverse order \eqref{eq:orderisomIc} of the $\alpha$-entangled state sets,
the inclusion hierarchy of the 
$\cmpl{\mathcal{P}_{\vs{\alpha}}}$ and $\cmpl{\mathcal{D}_{\vs{\alpha}}}$
$\vs{\alpha}$-entangled sets
is $\mathcal{O}_\uparrow(P_\text{I})\setminus\{P_\text{I}\}$,
given by the \emph{up-set} lattice $\mathcal{O}_\uparrow(P_\text{I})$ of $P_\text{I}$,
and for these complementary state sets we have then the reverse \eqref{eq:orderisomII} order
\begin{subequations}
\label{eq:orderisomIIc}
\begin{align}
\label{eq:orderisomIIPc}
\vs{\beta}\preceq\vs{\alpha} \quad\Longleftrightarrow\quad 
\cmpl{\mathcal{P}_{\vs{\beta}}} \supseteq \cmpl{\mathcal{P}_{\vs{\alpha}}},
\intertext{and}
\label{eq:orderisomIIDc}
\vs{\beta}\preceq\vs{\alpha} \quad\Longleftrightarrow\quad 
\cmpl{\mathcal{D}_{\vs{\beta}}} \supseteq \cmpl{\mathcal{D}_{\vs{\alpha}}}
\end{align}
\end{subequations}
(that is, entanglement higher in the hierarchy implies a lower one),
so the posets $(P_\text{II},\preceq)$,
$(P_{\text{II},\cmpl{\mathcal{P}}},\supseteq)$, and
$(P_{\text{II},\cmpl{\mathcal{D}}},\supseteq)$ are isomorphic.

Note that in this framework, complete in the sense of partial separability,
one can also describe the notion of $k$-separability 
\cite{AcinetalMixThreeQB,GuhneTothMultipartite,SeevinckUffinkMixSep}
and $k$-producibility
\cite{SeevinckUffinkMultipartite,GuhneTothMultipartite,TothGuhneMultipartite}.
A mixed state is \emph{$k$-separable}, if it can be mixed by the use of pure states
being separable into \emph{at least} $k$ parts.
That is, the set of $k$-separable states is given by 
$\mathcal{D}_\text{$k$-sep} := \mathcal{D}_{\vs{\beta}_k}$,
where the $\vs{\beta}_k \in P_\text{II}$ label of the second kind is such that
\begin{subequations}
\begin{equation}
\label{eq:labksep}
\begin{split}
\vs{\beta}_k:\quad
\forall\beta   \in\vs{\beta}_k:\; &\abs{\beta} \geq k\;\text{and}\\
\forall\beta\notin\vs{\beta}_k:\; &\abs{\beta}  <   k.
\end{split}
\end{equation}
(This is related to the natural gradation on the lattice of partitions $P_\text{I}$.)
These labels form a chain (a completely ordered set),
$\{1|2|\dots|n\}=\vs{\beta}_n
\preceq\dots
\preceq\vs{\beta}_{k+1}
\preceq\vs{\beta}_k
\preceq\dots\preceq\vs{\beta}_1=P_\text{I}$, leading to
$\mathcal{D}_{\{1|2|\dots|n\}}=\mathcal{D}_\text{$n$-sep}
\subseteq\dots
\subseteq\mathcal{D}_\text{$(k+1)$-sep}
\subseteq\mathcal{D}_\text{$k$-sep}
\subseteq\dots\subseteq\mathcal{D}_\text{$1$-sep}=\mathcal{D}$
by \eqref{eq:orderisomIID}.
A mixed state is \emph{$k$-producible}, if it can be mixed by the use of pure states
being separable with respect to splits containing parts \emph{at most} of size $k$.
That is, the set of $k$-producible states is given by
$\mathcal{D}_\text{$k$-prod} := \mathcal{D}_{\vs{\gamma}_k}$,
where the $\vs{\gamma}_k \in P_\text{II}$
label of the second kind is such that 
\begin{equation}
\label{eq:labkprod}
\begin{split}
\vs{\gamma}_k:\quad
\forall\gamma   \in\vs{\gamma}_k, \forall K\in\gamma:\; &\abs{K} \leq k\;\text{and}\\
\forall\gamma\notin\vs{\gamma}_k, \exists K\in\gamma:\; &\abs{K}  >   k.
\end{split}
\end{equation}
\end{subequations}
These labels form a chain,
$\{1|2|\dots|n\}=\vs{\gamma}_1
\preceq\dots
\preceq\vs{\gamma}_{k-1}
\preceq\vs{\gamma}_k
\preceq\dots\preceq\vs{\gamma}_n=P_\text{I}$, leading to
$\mathcal{D}_{\{1|2|\dots|n\}}=\mathcal{D}_\text{$1$-prod}
\subseteq\dots
\subseteq\mathcal{D}_\text{$(k-1)$-prod}
\subseteq\mathcal{D}_\text{$k$-prod}
\subseteq\dots\subseteq\mathcal{D}_\text{$n$-prod}=\mathcal{D}$
by \eqref{eq:orderisomIID}.

\subsection{Examples}
\label{sec:EntMulti.secondExamples}

Writing out some examples explicitly might not be useless here.
The lattices $P_\text{II}$ for the cases $n=2$ and $3$ can be seen in the upper-right parts
of Figures \ref{fig:labellattices2} and \ref{fig:labellattices3}.
As we have learned in \eqref{eq:orderisomIIP} and \eqref{eq:orderisomIID},
we need to draw only this lattice 
for the inclusion hierarchy of the sets of $\vs{\alpha}$-separable pure and mixed states $\mathcal{P}_{\vs{\alpha}}$ \eqref{eq:setIIP} and $\mathcal{D}_{\vs{\alpha}}$ \eqref{eq:setIID}.

For the \emph{bipartite} case, 
we do not have additional structure over that of the first kind 
(see Section \ref{sec:EntMulti.firstExamples}),
and we get back the content of Section \ref{sec:EntBasics.Ent}.
The sets of $\vs{\alpha}$-separable pure states are
\begin{align*}
\mathcal{P}_{\downset\{12\}}  &=\mathcal{P}_{12}\equiv\mathcal{P}(\mathcal{H}_{12}),\\
\mathcal{P}_{\downset\{1|2\}} &=\mathcal{P}_{1|2}=\mathcal{P}_\text{sep}.
\end{align*}
Note that $\mathcal{P}_{\downset\{1|2\}}\subseteq\mathcal{P}_{\downset\{12\}}$.
The sets of $\vs{\alpha}$-separable mixed states are
\begin{align*}
\mathcal{D}_{\downset\{12\}}  &=\mathcal{D}_{12}\equiv\mathcal{D}(\mathcal{H}_{12}),\\
\mathcal{D}_{\downset\{1|2\}} &=\mathcal{D}_{1|2}=\mathcal{D}_\text{sep}.
\end{align*}
Note that, again, $\mathcal{D}_{\downset\{1|2\}}\subseteq\mathcal{D}_{\downset\{12\}}$.

For the \emph{tripartite} case we \emph{do} have additional structure over that of the first kind
(see Section \ref{sec:EntMulti.firstExamples}).
The sets of $\vs{\alpha}$-separable pure states are
\begin{align*}
\mathcal{P}_{\downset\{123\}}  &= \mathcal{P}_{123} \equiv\mathcal{P}(\mathcal{H}_{123}),\\
\mathcal{P}_{\downset\{1|23,2|13,3|12\}}  &= \mathcal{P}_{1|23}\cup\mathcal{P}_{2|13}\cup\mathcal{P}_{3|12},\\
\mathcal{P}_{\downset\{b|ac,c|ab\}}  &= \mathcal{P}_{b|ac}\cup\mathcal{P}_{c|ab},\\
\mathcal{P}_{\downset\{a|bc\}}  &= \mathcal{P}_{a|bc},\\
\mathcal{P}_{\downset\{1|2|3\}} &= \mathcal{P}_{1|2|3},
\end{align*}
with all bipartitions $a|bc$ of $\{1,2,3\}$.
Note that 
$\mathcal{P}_{\downset\{1|2|3\}}\subseteq 
 \mathcal{P}_{\downset\{a|bc\}}\subseteq
 \mathcal{P}_{\downset\{a|bc,b|ac\}}\subseteq
 \mathcal{P}_{\downset\{a|bc,b|ac,c|ab\}}\subseteq
 \mathcal{P}_{\downset\{123\}}$.
Note, on the other hand, how \eqref{eq:meetjoinintersectunionIIPeq} works.
The sets of $\vs{\alpha}$-separable mixed states are
\begin{align*}
\mathcal{D}_{\downset\{123\}}  &= \Conv \mathcal{P}_{\downset\{123\}}\equiv\mathcal{D}(\mathcal{H}_{123}),\\
\mathcal{D}_{\downset\{1|23,2|13,3|12\}}  &= \Conv \mathcal{P}_{\downset\{1|23,2|13,3|12\}},\\
\mathcal{D}_{\downset\{b|ac,c|ab\}}  &= \Conv \mathcal{P}_{\downset\{b|ac,c|ab\}},\\
\mathcal{D}_{\downset\{a|bc\}}  &= \Conv\mathcal{P}_{\downset\{a|bc\}},\\
\mathcal{D}_{\downset\{1|2|3\}} &= \Conv\mathcal{P}_{\downset\{1|2|3\}}.
\end{align*}
Note that, again,
$\mathcal{D}_{\downset\{1|2|3\}}\subseteq 
 \mathcal{D}_{\downset\{a|bc\}}\subseteq
 \mathcal{D}_{\downset\{a|bc,b|ac\}}\subseteq
 \mathcal{D}_{\downset\{a|bc,b|ac,c|ab\}}\subseteq
 \mathcal{D}_{\downset\{123\}}$.
Note, on the other hand, that there is no \eqref{eq:meetjoinintersectunionIIPeq}-like result for mixed states,
we have only \eqref{eq:meetjoinintersectunionIID}.

\subsection{Level III: partial separability classes}
\label{sec:EntMulti.classes}

The state sets $\mathcal{D}_{\vs{\alpha}}$
of given $\vs{\alpha}$-separability are containing, \eqref{eq:orderisomIID};
that is, if a state is $\vs{\alpha}$-separable, $\varrho\in\mathcal{D}_{\vs{\alpha}}$,
it can also be $\vs{\beta}$-separable 
for a $\vs{\beta}$ lower in the hierarchy $(P_\text{II},\preceq)$.
Now we construct the partial separability classes, 
which are the sets of states 
having well-defined partial separability properties,
that is, being $\vs{\alpha}$-separable for given $\vs{\alpha}$s but not separable
under any $\vs{\beta}\preceq\vs{\alpha}$.

The \emph{partial separability classes} are defined 
as the intersections of the $\mathcal{D}_{\vs{\alpha}}$ sets
of states of different partial separability.
First we select a sublattice of $P_\text{II}$,
\begin{equation}
\bigstruct{ P_\text{II*},\preceq,\wedge,\vee }
\subseteq 
\bigstruct{ P_\text{II},\preceq,\wedge,\vee }
\end{equation}
by the use of which we can tune how fine/coarse the arising classification is
and what kinds of entanglement are taken into account.
The elements of this (sub)lattice give rise to a (sub)hierarchy,
based on which the classification is carried out.
If the whole lattice is taken, $P_\text{II*} = P_\text{II}$,
then we get the complete classification in the sense of partial separability,
which utilizes all the possible combinations of different kinds of partially separable pure states \cite{SzalayKokenyesiPartSep,SzalayDissertation}.
If only the principal elements of $P_\text{II}$ are taken,
that is,  $P_\text{II*} = \set{\vs{\alpha}\in P_\text{II}}{ \exists\alpha\in P_\text{I}: \vs{\alpha}= \downset\{\alpha\}}$,
then we get an incomplete classification
introduced in \cite{DurCiracTarrach3QBMixSep,DurCirac3QBMixSep}.
If $P_\text{II*}=\set{\vs{\beta}_k}{k=1,2,\dots,n}$ is 
the set of labels of the second kind labeling the different $k$-separability
properties \eqref{eq:labksep},
then we get an intermediate classification,
based on $k$-separability
\cite{AcinetalMixThreeQB,GuhneTothMultipartite,SeevinckUffinkMixSep}.
If $P_\text{II*}=\set{\vs{\gamma}_k}{k=1,2,\dots,n}$ is 
the set of labels of the second kind labeling the different $k$-producibility
properties \eqref{eq:labkprod},
then we get another intermediate classification,
based on $k$-producibility
\cite{SeevinckUffinkMultipartite,GuhneTothMultipartite,TothGuhneMultipartite}.

Now, we need to obtain all the classes, the possible nonempty intersections of the 
$\mathcal{D}_{\vs{\alpha}}$ sets (for $\vs{\alpha}\in P_\text{II*}$).
In general, the intersections can be 
labeled by a nonempty $\vvs{\alpha}\subseteq P_\text{II*}$ as 
\begin{equation}
\label{eq:classDef}
\mathcal{C}_{\vvs{\alpha}} :=
\bigcap_{\vs{\alpha}\notin\vvs{\alpha}} \cmpl{\mathcal{D}_{\vs{\alpha}}} \cap 
\bigcap_{\vs{\alpha}   \in\vvs{\alpha}}       \mathcal{D}_{\vs{\alpha}}  .
\end{equation}
However, because of the inclusions \eqref{eq:orderisomIID},
some of the intersections are \emph{empty by construction}, 
\begin{equation}
\label{eq:classemptybyconstruction}
\exists \vs{\alpha}\in \vvs{\alpha},\;
\exists \vs{\beta}\notin \vvs{\alpha}\;:\;
\vs{\alpha} \preceq \vs{\beta}
\quad\Longrightarrow\quad
\mathcal{C}_{\vvs{\alpha}}=\emptyset.
\end{equation}
(This comes from \eqref{eq:orderisomIID} and elementary set-algebra:
If $A\subseteq B$, then $\cmpl{B}\cap A = A\setminus B=\emptyset$.)
It will turn out later that if a class 
is not empty by construction, then its label $\vvs{\alpha}$
is a nonempty element of the \emph{up-set lattice} of $P_\text{II*}$
(see Appendix \ref{app:lattices.gen}),
which is now denoted with
\begin{equation}
\label{eq:PIII}
\begin{split}
&P_\text{III} := \mathcal{O}_\uparrow(P_\text{II*})\setminus\{\emptyset\} \\
&= \Bigset{\vvs{\alpha}\in 2^{P_\text{II*}}\setminus\{\emptyset\}}{
\forall \vs{\alpha}\in\vvs{\alpha}:\; \vs{\alpha}\preceq\vs{\beta}
\;\Rightarrow\;
\vs{\beta}\in\vvs{\alpha}}.
\end{split}
\end{equation}
Again, 
the $P_\text{III}$ set of nonempty up-sets of the lattice $P_\text{II*}$ 
forms a lattice
with respect to the inclusion, intersection, and union \cite{DaveyPriestley},
so we have
\begin{equation}
\label{eq:posetIII}
\bigstruct{ P_\text{III},\preceq,\wedge,\vee }
= \bigstruct{ \mathcal{O}_\uparrow(P_\text{II*})\setminus\{\emptyset\},\subseteq,\cap,\cup }.
\end{equation}

With the above definitions in hand, we can prove that
$P_\text{III}$ is sufficient for the labeling of the classes in the above sense;
that is,
\begin{equation}
\label{eq:classlabels}
\mathcal{C}_{\vvs{\alpha}}\neq\emptyset
\quad\Longrightarrow\quad
\vvs{\alpha}\in P_\text{III}.
\end{equation}
(For the proof, see Appendix \ref{app:lattices.classes}.)
We also have the set
\begin{equation}
\label{eq:statesIIIC}
P_{\text{III},\mathcal{C}}:=\bigset{\mathcal{C}_{\vvs{\alpha}}}{{\vvs{\alpha}}\in P_\text{III}}.
\end{equation}
From \eqref{eq:classlabels}, 
by the inclusion rules \eqref{eq:orderisomIID} and \eqref{eq:orderisomIIDc},
 it immediately follows that
the classes \eqref{eq:classDef} can be written as
\begin{equation}
\label{eq:classDefm}
\mathcal{C}_{\vvs{\alpha}}=
\bigcap_{\vs{\alpha}\in\max\pcmpl{\vvs{\alpha}}} \cmpl{\mathcal{D}_{\vs{\alpha}}}  \cap 
\bigcap_{\vs{\alpha}\in\min       \vvs{\alpha}}        \mathcal{D}_{\vs{\alpha}}  ,
\end{equation}
with the notation $\pcmpl{\vvs{\alpha}}=P_\text{II*}\setminus\vvs{\alpha}$.
(That is, 
because of \eqref{eq:orderisomIIP},
it is enough to use $\max\pcmpl{\vvs{\alpha}}$ and $\min\vvs{\alpha}$ 
for the calculation of the intersections.)
A conjecture is that the reverse implication also holds in \eqref{eq:classlabels},
even for the most detailed case when $P_\text{II*}=P_\text{II}$
\cite{SzalayKokenyesiPartSep,SzalayDissertation}.
\begin{conj}
\label{conj:classdef}
 The classes \eqref{eq:classDef} given by all $\vvs{\alpha}\in P_\text{III}$
are nonempty; hence,
\begin{equation}
\label{eq:classlabelsconj}
\mathcal{C}_{\vvs{\alpha}}\neq\emptyset
\quad\Longleftrightarrow\quad
\vvs{\alpha}\in P_\text{III}.
\end{equation}
\end{conj}
An advantage of the formulation by the labeling constructions
is that, roughly speaking, by useing that,
``we have separated the \emph{algebraic} and the \emph{geometric part}'' of the problem 
of the nonemptiness of the classes.
At this point, it seems that 
we have tackled all the \emph{algebraic} issues of the problem,
and this conjecture cannot be proven without the investigation of the \emph{geometry} of $\mathcal{D}$,
more precisely, the geometry of the different $\mathcal{P}_{\alpha}$ sets of extremal points.

Having the \eqref{eq:classDef} definition of the classes $\mathcal{C}_{\vvs{\alpha}}$ 
for $\vvs{\alpha}\in P_\text{III}$ in hand,
we can also reconstruct the original state sets $\mathcal{D}_{\vs{\alpha}}$.
By the definition \eqref{eq:classDef} we have
\begin{subequations}
\begin{equation}
\label{eq:CinD}
\mathcal{C}_{\vvs{\alpha}} \subseteq \mathcal{D}_{\vs{\alpha}}
\quad\Longleftrightarrow\quad
\vs{\alpha}\in\vvs{\alpha},
\end{equation}
so we need to collect every class $\mathcal{C}_{\vvs{\alpha}}$
where $\vs{\alpha}\in\vvs{\alpha}$ to reconstruct $\mathcal{D}_{\vs{\alpha}}$,
\begin{equation}
\label{eq:DbyC}
\mathcal{D}_{\vs{\alpha}} = 
\bigcup_{\substack{\forall \vvs{\alpha}\in P_\text{III}:\\\vs{\alpha}\in\vvs{\alpha}}}
\mathcal{C}_{\vvs{\alpha}}.
\end{equation}
\end{subequations}
These classes are labeled 
by the elements of the up-set of the principal element $\upset\{\vs{\alpha}\}$
(being a principal element in $\mathcal{O}_\uparrow(P_\text{III})$);
that is,
\begin{subequations}
\begin{equation}
\label{eq:forDbyC2}
\bigset{\vvs{\alpha}\in P_\text{III}}{\vs{\alpha}\in\vvs{\alpha}}
= \upset\bigl\{ \upset\{ \vs{\alpha} \} \bigr\}.
\end{equation}
(For the proof, see Appendix \ref{app:lattices.reconstrDII}.)
Using this, we have
\begin{equation}
\label{eq:DbyC2}
\mathcal{D}_{\vs{\alpha}}
= \bigcup_{\vvs{\alpha}\preceq\upset\{\upset\{\vs{\alpha}\}\}} \mathcal{C}_{\vvs{\alpha}}. 
\end{equation}
\end{subequations}

After these technicalities, 
let us take a wider look at the construction.
We have that, interestingly, 
the partial separability classes \eqref{eq:classDefm} are also endowed 
with a hierarchical (lattice) structure \eqref{eq:posetIII}.
Although this structure arises in a very natural way in the construction,
the meaning of this hierarchy is not fully understood at this point.
Now we clarify that.

Since at this point we do not have a well-established notion,
based on which it could be meaningful to say that 
states in a given class $\mathcal{C}_{\vvs{\alpha}}$
are ``more entangled'' than states in class $\mathcal{C}_{\vvs{\beta}}$,
we are free to adopt (and we would really like to adopt)
the hierarchy $\struct{P_\text{III},\preceq}$ for this purpose,
if doing this makes any sense.
Being more entangled is a notion strongly related to LOCC convertibility,
so one should make a trial of this direction.
Here we use the notations for the \emph{strong} and \emph{weak LOCC convertibility}
among different classes (different definite partial separability properties),
\begin{subequations}
\label{eq:defLOCCconv}
\begin{align}
\label{eq:defLOCCconv.strong}
\begin{split}
&\mathcal{C}_{\vvs{\beta}} \overset{\text{LOCC}_\text{s}}{\longrightarrow} \mathcal{C}_{\vvs{\alpha}}
\quad \defn \\
&\quad\forall \varrho \in \mathcal{C}_{\vvs{\beta}},\; \exists \text{$\Lambda$ LOCC map
such that}\;\Lambda(\varrho) \in \mathcal{C}_{\vvs{\alpha}},
\end{split}\\
\label{eq:defLOCCconv.weak}
\begin{split}
&\mathcal{C}_{\vvs{\beta}} \overset{\text{LOCC}_\text{w}}{\longrightarrow} \mathcal{C}_{\vvs{\alpha}}
\quad \defn \\
&\quad\exists \varrho \in \mathcal{C}_{\vvs{\beta}},\; \exists \text{$\Lambda$ LOCC map
such that}\;\Lambda(\varrho) \in \mathcal{C}_{\vvs{\alpha}}.
\end{split}
\end{align}
\end{subequations}
(Note that we do not consider the question whether 
all states in $\mathcal{C}_{\vvs{\alpha}}$ can be reached by LOCC 
from the states in $\mathcal{C}_{\vvs{\beta}}$.)
The LOCC convertibility
$\mathcal{C}_{\vvs{\beta}} \overset{\text{LOCC}_\text{s/w}}{\longrightarrow} \mathcal{C}_{\vvs{\alpha}}$
is also denoted with
$\mathcal{C}_{\vvs{\beta}}\geq_\text{s/w}\mathcal{C}_{\vvs{\alpha}}$.
(In the literature, the same arrows and ordering signs are used \cite{BennettetalEquivalences}
for the notion of convertibility of states.
Here we use the notion of convertibility of classes.)
Using that the state sets $\mathcal{D}_{\vs{\alpha}}$ are closed under LOCC \eqref{eq:LOCCcloseII},
we have that $\forall\vvs{\alpha},\vvs{\beta}\in P_\text{III}$ 
\begin{equation}
\label{eq:LOCCconvhierarchy}
\mathcal{C}_{\vvs{\beta}} \overset{\text{LOCC}_\text{s}}{\longrightarrow} \mathcal{C}_{\vvs{\alpha}}
\quad\Longrightarrow\quad
\mathcal{C}_{\vvs{\beta}} \overset{\text{LOCC}_\text{w}}{\longrightarrow} \mathcal{C}_{\vvs{\alpha}}
\quad\Longrightarrow\quad
\vvs{\beta}\preceq\vvs{\alpha}.
\end{equation}
(The first implication is obvious;
for the proof of the second one, see Appendix \ref{app:lattices.LOCCconv}.)
From this, we have that $P_{\text{III},\mathcal{C}}$
is a poset with respect to the LOCC convertibility,
\begin{equation}
\label{eq:posetIIIC}
\struct{P_{\text{III},\mathcal{C}},\geq_\text{s}}.
\end{equation}
(For the proof, see Appendix \ref{app:lattices.classLOCCorder}.)
So we can move by LOCC along the hierarchy $\struct{P_\text{III}, \preceq}$;
however, it can happen, that $\mathcal{C}_{\vvs{\beta}}$
cannot be converted to $\mathcal{C}_{\vvs{\alpha}}$ for all $\vvs{\alpha}$ which is 
$\vvs{\beta}\preceq\vvs{\alpha}$.
A conjecture is that this is not the case;
that is, the converse also holds in \eqref{eq:LOCCconvhierarchy}.
\begin{conj}
\label{conj:LOCCconvhierarchyfull}
$\forall\vvs{\alpha},\vvs{\beta}\in P_\text{III}$ we have that
\begin{equation}
\label{eq:LOCCconvhierarchyfull}
\mathcal{C}_{\vvs{\beta}} \overset{\text{LOCC}_\text{s}}{\longrightarrow} \mathcal{C}_{\vvs{\alpha}}
\quad\Longleftrightarrow\quad
\vvs{\beta}\preceq\vvs{\alpha},
\end{equation}
that is, the two posets are isomorphic,
\begin{equation} 
\struct{P_{\text{III},\mathcal{C}},\geq_\text{s}}\cong\struct{P_\text{III}, \preceq}.
\end{equation}
\end{conj}
Note that if Conjecture \ref{conj:LOCCconvhierarchyfull} is true,
then the notions of the 
strong and the weak convertibility \eqref{eq:defLOCCconv} would coincide.
Note also that
Conjecture \ref{conj:LOCCconvhierarchyfull} implies
Conjecture \ref{conj:classdef}, since 
one could not convert $\mathcal{C}_{\vvs{\beta}}$ to $\mathcal{C}_{\vvs{\alpha}}$
if the latter were empty.
For the proof of \eqref{eq:LOCCconvhierarchy}
it has been enough to use the set-theoretical notions of the construction,
however, this does not seem to be the case 
for the proof of \eqref{eq:LOCCconvhierarchyfull},
one has to construct explicit protocol even for the weak LOCC convertibility of classes.
Anyway, \eqref{eq:LOCCconvhierarchy} may be enough for saying
that
states in a given class $\mathcal{C}_{\vvs{\beta}}$
are more entangled than states in class $\mathcal{C}_{\vvs{\alpha}}$,
if $\vvs{\beta}\preceq\vvs{\alpha}$.

As coarse-grained cases, we may consider only the $k$-separability
or the $k$-producibility properties.
We have the $P_\text{II*}=\set{\vs{\beta}_k}{k=1,2,\dots,n}$
and the $P_\text{II*}=\set{\vs{\gamma}_k}{k=1,2,\dots,n}$
lattices of the labels of the second kind, labeling the different
$k$-separability \eqref{eq:labksep}, respectively $k$-producibility \eqref{eq:labkprod}
 properties in the two cases.
Since these form chains,
the arising hierarchies \eqref{eq:PIII} of the classes are also chains in both cases.
For the labeling of the \emph{$k$-separability classes},
we have the nonempty up-sets $\vvs{\beta}_k=\upset\{\vs{\beta}_k\}
\in\mathcal{O}_\uparrow(P_\text{II*})\setminus\{\emptyset\}= P_\text{III}$,
with the hierarchy 
$\{P_\text{I}\}=\vvs{\beta}_1
\preceq\dots
\preceq\vvs{\beta}_k
\preceq\vvs{\beta}_{k+1}
\preceq\dots\preceq\vvs{\beta}_n=P_\text{II*}$,
leading to the $k$-separability classes by \eqref{eq:classDefm}, being
$\mathcal{C}_{\vvs{\beta}_k}\equiv\mathcal{C}_\text{$k$-sep ent}
=\cmpl{\mathcal{D}_\text{$(k+1)$-sep}}\cap\mathcal{D}_\text{$k$-sep}
\equiv \mathcal{D}_\text{$k$-sep}\setminus \mathcal{D}_\text{$(k+1)$-sep}$ for $k=1,2,\dots,n$,
called \emph{$k$-separable entangled}.
That is, a state is $k$-separable entangled if 
it can be mixed by the use of $k$-separable states, 
but cannot be mixed by the use of $k+1$-separable (``more separable'') states.
If Conjecture \ref{conj:LOCCconvhierarchyfull} holds,
then the strong LOCC hierarchy 
$\mathcal{C}_\text{$k$-sep ent}\geq_\text{s}\mathcal{C}_\text{$(k+1)$-sep ent}$ follows.
For the labeling of the \emph{$k$-producibility classes},
we have the nonempty up-sets $\vvs{\gamma}_k=\upset\{\vs{\gamma}_k\}
\in\mathcal{O}_\uparrow(P_\text{II*})\setminus\{\emptyset\}= P_\text{III}$,
with the hierarchy 
$\{P_\text{I}\}=\vvs{\gamma}_n
\preceq\dots
\preceq\vvs{\gamma}_k
\preceq\vvs{\gamma}_{k-1}
\preceq\dots\preceq\vvs{\gamma}_1=P_\text{II*}$,
leading to the $k$-producibility classes by \eqref{eq:classDefm}, being
$\mathcal{C}_{\vvs{\gamma}_k}\equiv\mathcal{C}_\text{$k$-prod ent}
=\cmpl{\mathcal{D}_\text{$(k-1)$-prod}}\cap\mathcal{D}_\text{$k$-prod}
\equiv \mathcal{D}_\text{$k$-prod}\setminus \mathcal{D}_\text{$(k-1)$-prod}$ for $k=1,2,\dots,n$,
called \emph{genuine $k$-partite entangled}.
That is, a state is genuine $k$-partite entangled if 
it can be mixed by the use of $k$-producible states
(entanglement among at most $k$ elementary subsystems),
but cannot be mixed by the use of $k-1$-producible (``less entangled'') states
(entanglement among, at most, $k-1$ elementary subsystems).
If Conjecture \ref{conj:LOCCconvhierarchyfull} holds,
then the strong LOCC hierarchy 
$\mathcal{C}_\text{$k$-prod ent}\geq_\text{s}\mathcal{C}_\text{$(k-1)$-prod ent}$ follows.
So, in these two cases,
when the $P_\text{III}$ hierarchies of the labels of the classes are chains,
we have the expressive meaning for the (same) hierarchies of the classes themselves.

\subsection{Examples}
\label{sec:EntMulti.classesExamples}

Writing out some examples explicitly might not be useless here.
The lattices $P_\text{III}$ for $P_\text{II*}=P_\text{II}$ 
for the cases $n=2$ and $3$ can be seen in the lower-left parts of Figures \ref{fig:labellattices2} and \ref{fig:labellattices3}.
The classes have this hierarchical structure,
however, this does not manifest itself in inclusion hierarchy (the classes are disjoint),
but the meaning of this is the LOCC convertibility \eqref{eq:LOCCconvhierarchy}.

For the \emph{bipartite} case, 
we get back the content of Section \ref{sec:EntBasics.Ent},
\begin{align*}
\mathcal{C}_{\upset\{\downset\{12\}\}} 
&= \cmpl{\mathcal{D}_{\downset\{1|2\}}} \cap
         \mathcal{D}_{\downset\{12\}} 
= \mathcal{D}_{12}\setminus\mathcal{D}_{1|2} 
= \mathcal{C}_\text{ent},\\
\mathcal{C}_{\upset\{\downset\{1|2\}\}} 
&= \mathcal{D}_{\downset\{1|2\}} \cap
   \mathcal{D}_{\downset\{12\}} = \mathcal{D}_{1|2} 
=  \mathcal{C}_\text{sep},
\end{align*}
being the \emph{entangled} and \emph{separable} state classes.
Note that every entangled bipartite state can be converted to a separable one
by means of LOCC,
so Conjecture \ref{conj:LOCCconvhierarchyfull} holds in the bipartite case,
$\mathcal{C}_\text{ent}\geq_\text{s}\mathcal{C}_\text{sep}$.

For the \emph{tripartite} case, we have $1+18+1=20$ classes,
shown in Table \ref{tab:classes3}.
The meaning of these is discussed in \cite{SzalayKokenyesiPartSep,SzalayDissertation}.

\begin{table*}
\setlength{\tabcolsep}{6pt}
\begin{tabular}{|l||c|ccc|ccc|c|c||lll|}
\hline
Class \hfill (name)& 
\begin{sideways}$\mathcal{D}_{\downset\{1|2|3\}}$\end{sideways}  & 
\begin{sideways}$\mathcal{D}_{\downset\{a|bc\}}$\end{sideways}  & 
\begin{sideways}$\mathcal{D}_{\downset\{b|ac\}}$\end{sideways}  & 
\begin{sideways}$\mathcal{D}_{\downset\{c|ab\}}$\end{sideways}  & 
\begin{sideways}$\mathcal{D}_{\downset\{b|ac,c|ab\}}$\end{sideways}  & 
\begin{sideways}$\mathcal{D}_{\downset\{a|bc,c|ab\}}$\end{sideways}  &
\begin{sideways}$\mathcal{D}_{\downset\{a|bc,b|ac\}}$\end{sideways}  &
\begin{sideways}$\mathcal{D}_{\downset\{1|23,2|13,3|12\}}\;$\end{sideways}  &
\begin{sideways}$\mathcal{D}_{\downset\{123\}}$\end{sideways}  &
in \cite{SzalayKokenyesiPartSep} &
in \cite{SeevinckUffinkMixSep} &
in \cite{DurCirac3QBMixSep} \\
\hline
\hline
$\mathcal{C}_{\upset\{ \downset\{123\} \}}$    \hfill(tripartite entangled)                   & $\nsubset$ & $\nsubset$ & $\nsubset$ & $\nsubset$ & $\nsubset$ & $\nsubset$ & $\nsubset$ & $\nsubset$ & $\subset$  & 1       & 1       & 1      \\ 
\hline
$\mathcal{C}_{\upset\{ \downset\{1|23,2|13,3|12\} \}}$                                         & $\nsubset$ & $\nsubset$ & $\nsubset$ & $\nsubset$ & $\nsubset$ & $\nsubset$ & $\nsubset$ & $\subset$  & $\subset$  & 2.1     & 2.1     & 1       \\
$\mathcal{C}_{\upset\{ \downset\{b|ac,c|ab\} \}}$                                              & $\nsubset$ & $\nsubset$ & $\nsubset$ & $\nsubset$ & $\subset$  & $\nsubset$ & $\nsubset$ & $\subset$  & $\subset$  & 2.2.a   & 2.1     & 1       \\
$\mathcal{C}_{\upset\{ \downset\{a|bc,b|ac\},\downset\{a|bc,c|ab\} \}}$                        & $\nsubset$ & $\nsubset$ & $\nsubset$ & $\nsubset$ & $\nsubset$ & $\subset$  & $\subset$  & $\subset$  & $\subset$  & 2.3.a   & 2.1     & 1       \\
$\mathcal{C}_{\upset\{ \downset\{1|23,2|13\},\downset\{1|23,3|12\},\downset\{2|13,3|12\} \}}$  & $\nsubset$ & $\nsubset$ & $\nsubset$ & $\nsubset$ & $\subset$  & $\subset$  & $\subset$  & $\subset$  & $\subset$  & 2.4     & 2.1     & 1       \\
$\mathcal{C}_{\upset\{ \downset\{a|bc\} \}}$                                                   & $\nsubset$ & $\subset$  & $\nsubset$ & $\nsubset$ & $\nsubset$ & $\subset$  & $\subset$  & $\subset$  & $\subset$  & 2.5.a   & 2.4,3,2 & 2.3,2,1 \\
$\mathcal{C}_{\upset\{ \downset\{a|bc\},\downset\{b|ac,c|ab\} \}}$       \hfill(roundabout)    & $\nsubset$ & $\subset$  & $\nsubset$ & $\nsubset$ & $\subset$  & $\subset$  & $\subset$  & $\subset$  & $\subset$  & 2.6.a   & 2.4,3,2 & 2.3,2,1 \\
$\mathcal{C}_{\upset\{ \downset\{b|ac\},\downset\{c|ab\} \}}$                                  & $\nsubset$ & $\nsubset$ & $\subset$  & $\subset$  & $\subset$  & $\subset$  & $\subset$  & $\subset$  & $\subset$  & 2.7.a   & 2.7,6,5 & 3.3,2,1 \\
$\mathcal{C}_{\upset\{ \downset\{1|23\},\downset\{2|13\},\downset\{3|12\} \}}$ \hspace{1cm} \hfill(semiseparable)   & $\nsubset$ & $\subset$  & $\subset$  & $\subset$  & $\subset$  & $\subset$  & $\subset$  & $\subset$  & $\subset$  & 2.8     & 2.8     & 4       \\
\hline
$\mathcal{C}_{\upset\{ \downset\{1|2|3\}\}}$      \hfill(fully separable)                      & $\subset$  & $\subset$  & $\subset$  & $\subset$  & $\subset$  & $\subset$  & $\subset$  & $\subset$  & $\subset$  & 3       & 3       & 5       \\
\hline
\end{tabular}
\caption{Partial separability classes of mixed tripartite states; cf.~the lower-left part of Figure \ref{fig:labellattices3}.
Additionally, we show the labels of classes in \cite{SzalayKokenyesiPartSep,SzalayDissertation},
and the classifications obtained by
Seevinck and Uffink \cite{SeevinckUffinkMixSep} and
D\"ur and Cirac \cite{DurCirac3QBMixSep}.}
\label{tab:classes3}
\end{table*}

\section{Entanglement measures: basics}
\label{sec:MeasBasics}

A very basic question of entanglement theory is how to quantify entanglement \cite{Horodecki4}.
There are many different measures of entanglement obtained by the use of two main approaches,
the operational and the axiomatic ones
\cite{EltschkaSiewertEntMeas,PlenioVirmaniEntMeas,HorodeckiEntMeas,VidalEntMon}.
Here we follow more-or-less the axiomatic way, because,
on the one hand, 
it clearly distinguishes between relevant and irrelevant properties of quantities,
and, on the other hand, 
it allows experimenting.

Starting with this section,
we mainly deal with real-valued functions over state spaces, which are convex sets.
On convex sets it is meaningful to define \emph{convex},
\begin{subequations}
\begin{equation}
\label{eq:conv}
f\Bigl(\sum_i p_i \varrho_i\Bigr) \leq \sum_i p_i f(\varrho_i),
\end{equation}
and \emph{concave},
\begin{equation}
\label{eq:conc}
g\Bigl(\sum_i p_i \varrho_i\Bigr) \geq \sum_i p_i g(\varrho_i)
\end{equation}
\end{subequations}
functions.
Since mixing is interpreted as forgetting some classical information
concerning the identity of a $\varrho_i$ member of an ensemble $\{(p_i,\varrho_i)\}$,
convexity and concavity reflect how the given function
is behaving in this process.
For a collection of tools on convexity,
see Sections 2 and 3 of \cite{BoydVandenbergheConvOpt}.

\subsection{Mixedness of states}
\label{sec:MeasBasics.Mix}

Before turning to measuring entanglement of multipartite states,
in this and the next subsections,
we recall some important notions in the characterization of states considered as a whole,
without respect to the existence of subsystems (tensor product structure in the Hilbert space).

The mixedness of a quantum state can be characterized
by real-valued functions
called \emph{entropies} \cite{MarshallMajorization,BengtssonZyczkowski}.
\begin{subequations}
The most widely used of them is the \emph{von Neumann entropy} \cite{Neumann-1927,OhyaPetzQEntr,Bravyi-2003},
\begin{equation}
\label{eq:NeumannEntr}
S(\varrho)=-\tr(\varrho\ln\varrho).
\end{equation}
Other notable entropies are the one-parameter families of
\emph{quantum Tsallis entropies} \cite{FuruichiTsallis,HavrdaCharvat-1967,AczelDaroczy-1975,Tsallis-1988},
\begin{equation}
\label{eq:TsallisEntr}
S^\text{Ts}_q(\varrho)=\frac{1}{1-q}\bigl(\tr\varrho^q-1\bigr),\quad q>0
\end{equation}
(with $S^\text{R}_1:=\lim_{q\to1} S^\text{R}_q = S$),
and
\emph{quantum R\'enyi entropies} \cite{Renyi-1961}
\begin{equation}
\label{eq:RenyiEntr}
S^\text{R}_q(\varrho)=\frac{1}{1-q}\ln\tr\varrho^q,\quad q>0
\end{equation}
(with $S^\text{Ts}_1:=\lim_{q\to1} S^\text{Ts}_q = S$).
The \emph{concurrence-squared} is a qubit-normalized version of the $q=2$ Tsallis entropy,
\begin{equation}
\label{eq:ConcSquared}
C^2(\varrho) = 2S^\text{Ts}_2(\varrho) = 2 (1-\tr\varrho^2);
\end{equation}
\end{subequations}
for qubits, it obeys $0\leq C^2(\varrho)\leq 1$.
(The same holds if $\log_2$ is used in the definitions of von Neumann and R\'enyi entropies.)

All of these are non-negative, vanishing exactly for pure states,
\begin{subequations}
\begin{align}
\label{eq:entrNNeg}
S(\varrho), S^\text{Ts}_q(\varrho), S^\text{R}_q(\varrho) &\geq0, \\
\label{eq:entrDiscr}
S(\varrho), S^\text{Ts}_q(\varrho), S^\text{R}_q(\varrho)    &=0 
\quad\Longleftrightarrow\quad \varrho \in\mathcal{P}\subset\mathcal{D},
\end{align}
and their maximal values are
\begin{equation}
\label{eq:entrBound}
S(\varrho), S^\text{R}_q(\varrho) \leq \ln \dim \mathcal{H},\quad
S^\text{Ts}_q(\varrho) \leq \frac{(\dim \mathcal{H})^{1-q}-1}{1-q}.
\end{equation}
\end{subequations}
It is also important to know that
not all R{\'e}nyi entropies are concave \eqref{eq:conc} \cite{BengtssonZyczkowski},
\begin{subequations}
\label{eq:entrConcavity}
\begin{align}
\label{eq:entrConcavity.Neumann}
S\Bigl(\sum_i p_i \varrho_i\Bigr) &\geq \sum_i p_i S(\varrho_i),\\
\label{eq:entrConcavity.Tsallis}
S^\text{Ts}_q\Bigl(\sum_i p_i \varrho_i\Bigr) &\geq \sum_i p_i S^\text{Ts}_q(\varrho_i)
\quad\text{for all $q>0$},\\
\label{eq:entrConcavity.Renyi}
S^\text{R}_q \Bigl(\sum_i p_i \varrho_i\Bigr) &\geq \sum_i p_i S^\text{R}_q (\varrho_i)
\quad\text{if $q\leq1$}.
\end{align}
\end{subequations}
(For some useful tools in matrix analysis,
see Appendix \ref{app:EntMon.cnvcnc} and \cite{CarlenIneqs,PetzQInfo,OhyaPetzQEntr}.)

A common property of these functions is that they are monotonically increasing
in \emph{bistochastic quantum channels} $\Phi$,
\begin{subequations}
\begin{align}
S\bigl(\Phi(\varrho)\bigr)&\geq S(\varrho),\\
S^\text{Ts}_q\bigl(\Phi(\varrho)\bigr)&\geq S^\text{Ts}_q(\varrho),\\
S^\text{R}_q\bigl(\Phi(\varrho)\bigr) &\geq S^\text{R}_q(\varrho).
\end{align}
\end{subequations}
(For the theory of quantum channels, see, for example,
\cite{Wilde,Wolf,NielsenChuang,PetzQInfo,BengtssonZyczkowski}.)

In quantum probability theory, contrary to the classical,
the entropy is not monotonically decreasing for the restriction to subsystems (partial trace),
e.g., 
using the notation $\varrho_K=\tr_{K'}\varrho_{KK'}$,
$S(\varrho_{KK'})\ngeq S(\varrho_K)$, for the disjoint subsystems $K$ and $K'$.
(For pure states, this is entanglement itself;
 see \eqref{eq:PsepDecide} and \eqref{eq:entrDiscr}.)
However, the \emph{subadditivity} holds in some cases \cite{RaggioTsallis,AudenaertTsallisSubadd},
\begin{subequations}
\label{eq:subadd}
\begin{align}
\label{eq:subadd.Neumann}
S(\varrho_{KK'})&\leq S(\varrho_K)+S(\varrho_{K'}),\\
\label{eq:subadd.Tsallis}
S^\text{Ts}_q(\varrho_{KK'})&\leq S^\text{Ts}_q(\varrho_K)+S^\text{Ts}_q(\varrho_{K'})\quad\text{for $q>1$.}
\end{align}
\end{subequations}
Unfortunately, the R{\'e}nyi entropies are not subadditive \cite{vanDam-2002}.

\subsection{Distinguishability of states}
\label{sec:MeasBasics.Dist}

There are several quantities measuring the distinguishability of two quantum states;
here we consider only
the \emph{Umegaki relative entropy} or \emph{quantum Kullback-Leibler divergence} \cite{Umegaki-1962,OhyaPetzQEntr}.
For the density matrices $\varrho,\omega\in\mathcal{D}$, it is given as
\begin{equation}
\label{eq:KullbackLeiblerDiv}
D^\text{KL}(\varrho\Vert\omega)=\tr\varrho(\ln\varrho-\ln\omega).
\end{equation}
This expresses the \emph{statistical distinguishability} 
of the state $\varrho$ from the state $\omega$ \cite{BengtssonZyczkowski,HiaiPetzQRelEntr}.
It is non-negative and vanishes if and only if the two states are equal,
\begin{subequations}
\begin{align}
\label{eq:RelEntrNNeg}
D^\text{KL}(\varrho\Vert\omega) &\geq0, \\
\label{eq:RelEntrDiscr}
D^\text{KL}(\varrho\Vert\omega)  &=0 \quad\Longleftrightarrow\quad \varrho=\omega. 
\end{align}
\end{subequations}
It is not a distance, but only a divergence, since it is not symmetric,
and only a weak version of the triangle inequality holds \cite{BengtssonZyczkowski}.
It is also jointly convex,
\begin{equation}
D^\text{KL}\Bigl(\sum_ip_i\varrho_i\Big\Vert\sum_ip_i\omega_i\Bigr)\leq
\sum_ip_iD^\text{KL}(\varrho_i\Vert\omega_i),
\end{equation}
from which the convexity \eqref{eq:conv} follows in both arguments separately.
An important property of the relative entropy is that it is monotonically decreasing
in \emph{quantum channels} $\Phi$,
\begin{equation}
D^\text{KL}\bigl(\Phi(\varrho)\Vert\Phi(\omega)\bigr) \leq D^\text{KL}(\varrho\Vert\omega).
\end{equation}
For nice summaries on the properties and meaning of the relative entropy,
see, for example, \cite{OhyaPetzQEntr,SagawaNotes,BengtssonZyczkowski}.
There are also R\'enyi and Tsallis versions 
\cite{PetzQuasiEntr,HiaietalRenyiDiv,MullerLennertetalQRenyiDiv,WildeWinterYangSandwichedRenyi,RajagopalSandwichedRTDiv}.

\subsection{LOCC monotonicity: entanglement measures}
\label{sec:MeasBasics.LOCCmon}

The most fundamental property of \emph{entanglement measures} 
\cite{EltschkaSiewertEntMeas,PlenioVirmaniEntMeas,HorodeckiEntMeas,VidalEntMon}
is the monotonicity under LOCC
(local operation and classical communication, 
\cite{BennettetalMixedStates,ChitambaretalWoodyLOCC}).
An $f:\mathcal{D}\to\field{R}$ is \emph{(nonincreasing) monotonic under LOCC}, if
\begin{subequations}
\label{eq:meas}
\begin{equation}
\label{eq:meas.mon}
f\bigl(\Lambda(\varrho)\bigr) \leq f(\varrho)
\end{equation}
for any LOCC transformation $\Lambda$,
which expresses that
\textit{(i)} like any reasonable notion of correlation, 
entanglement does not increase locally
and
\textit{(ii)} while classical correlation does, entanglement does not increase by classical communication (``classical interaction'') either.
An $f:\mathcal{D}\to\field{R}$ is \emph{nonincreasing on average under LOCC}, if
\begin{equation}
\label{eq:meas.average}
\sum_i p_i f(\varrho_i') \leq f(\varrho),
\end{equation}
for all $\varrho\mapsto\{(p_i,\varrho_i')\}$ ensembles resulted from LOCC transformation $\Lambda$,
where the LOCC is constituted as $\Lambda=\sum_i \Lambda_i$,
where the $\Lambda_i$s are the suboperations of the LOCC 
realizing the outcomes of selective measurements,
and $\varrho_i'=\frac1{p_i}\Lambda_i(\varrho)$,
with $p_i=\tr\Lambda_i(\varrho)$.
This latter condition is stronger than the former one
if the function is \emph{convex} \eqref{eq:conv},
\begin{equation}
\label{eq:meas.conv}
f\Bigl(\sum_i p_i \varrho_i\Bigr) \leq \sum_i p_i f(\varrho_i),
\end{equation}
\end{subequations}
for all ensembles $\{(p_i,\varrho_i)\}$,
which expresses that entanglement cannot increase for mixing.
This is a plausible property,
since mixing is interpreted as forgetting some classical information
concerning the identity of a $\varrho_i$ member of an ensemble $\{(p_i,\varrho_i)\}$,
which can be done locally \cite{VidalEntMon}.
An $f:\mathcal{D}\to\field{R}$ is called an \emph{entanglement monotone}
if~\eqref{eq:meas.average} and~\eqref{eq:meas.conv} hold \cite{VidalEntMon}.
There is common agreement
that LOCC-monotonicity~\eqref{eq:meas.mon} is the only necessary postulate
for a function to be an \emph{entanglement measure} \cite{Horodecki4};
however, the stronger condition~\eqref{eq:meas.average} 
is often satisfied too,
and it is often easier to prove.
(On the other hand, the description of forgetting classical information is debated by some authors;
then convexity is not demanded \cite{PlenioLogNegnConv,PlenioVirmaniEntMeas},
and the only requirement for an entanglement measure is \eqref{eq:meas.average}.)

If $f$ is defined only for pure states,
$f:\mathcal{P}\to\field{R}$,
then only~\eqref{eq:meas.average} makes sense; the restriction of that is
that a pure function is \emph{nonincreasing on average under pure LOCC},
or \emph{entanglement monotone}, if
\begin{equation}
\label{eq:averagePure}
\sum_i p_i f(\pi_i') \leq f(\pi).
\end{equation}
Here $\pi\mapsto\{(p_i,\pi_i')\}$ is the ensemble of pure states
arising from the \emph{pure} LOCC suboperations $\Lambda_i$,
and $\pi_i'=\frac1{p_i}\Lambda_i(\pi) \in \mathcal{P}$
with $p_i=\tr\Lambda_i(\pi)$.
That is, \emph{mathematically,} one can decompose the LOCC $\Lambda$ into
\emph{pure} suboperations having only one Kraus operator each,
leading to much simpler constructions.
Note that not all $\pi_i'$ results of these operations may be accessible \emph{physically},
only the outcomes of the LOCC, which are formed by partial mixtures of this ensemble \cite{HorodeckiEntMeas}.

Clearly, functions obeying any particular one of the requirements in \eqref{eq:meas} and \eqref{eq:averagePure}
form a convex cone, that is, 
their sums and multiples by non-negative real numbers also obey the particular requirement.

Since fully separable states can be reversibly converted
into each other by means of LOCC,
it follows that if a function obeys \eqref{eq:meas.mon},
then it takes the same (minimal) value for all fully separable states \cite{VidalEntMon}.

\subsection{Discriminance: indicator functions}
\label{sec:MeasBasics.Discr}

In the sequel, we extensively use another property 
of functions $f:\mathcal{D}\to\field{R}$ on state spaces,
which is the \emph{discriminance}
with respect to a \emph{convex} set $\mathcal{D}_*\subseteq\mathcal{D}$;
that is,
\begin{subequations}
\begin{equation}
\label{eq:measdiscr}
\varrho\in\mathcal{D}_*\quad\Longleftrightarrow\quad f(\varrho)=0.
\end{equation}
So the vanishing of the function gives a necessary and sufficient criterion for that subset.
In this paper we deal only with functions having this property,
which are often called \emph{indicator functions} with respect to a kind of state.
Discriminance with respect to $\mathcal{D}_\text{sep}$
is an important property for functions measuring bipartite entanglement 
(Section \ref{sec:EntBasics.Ent}).

If $f$ is defined only for pure states,
$f:\mathcal{P}\to\field{R}$,
then the discriminance for the \emph{closed} set $\mathcal{P}_*\subseteq\mathcal{P}$ is
\begin{equation}
\label{eq:measdiscrPure}
\pi\in\mathcal{P}_*\quad\Longleftrightarrow\quad f(\pi)=0.
\end{equation}
\end{subequations}
Discriminance with respect to $\mathcal{P}_\text{sep}$
is an important property for functions measuring pure bipartite entanglement 
(Section \ref{sec:EntBasics.Ent}).

\subsection{Local entropies: pure state measures}
\label{sec:MeasBasics.Pure}

A possible way of obtaining entanglement measures for mixed states
is to obtain measures for pure states first,
then to extend them to the whole set of mixed states.
In the present and the following two subsections, we recall this way of construction.
 
It is proven by Vidal \cite{VidalEntMon,HorodeckiEntMeas} that any properly chosen function
applied to one of the reduced density matrices of a pure state
leads to a measure of pure state entanglement in the sense of \eqref{eq:averagePure}.
\begin{thm}\label{thm:pureEntMon}
Let $F:\mathcal{D}(\mathcal{H}_K)\to\field{R}$ be\\
(i) a symmetric and extensible function of the eigenvalues, and\\
(ii) concave \eqref{eq:conc}, 
\begin{equation}
\label{eq:pureVidalconc}
F\Bigl(\sum_i p_i \varrho_i\Bigr) \geq \sum_i p_i F(\varrho_i);
\end{equation}
then $f:\mathcal{P}\to\field{R}$ defined as 
\begin{equation}
\label{eq:pureVidal}
f_K(\pi):=F(\tr_{\cmpl{K}}\pi)
\end{equation}
 is an
entanglement monotone \eqref{eq:averagePure}.
\end{thm}
(We recall the simpler proof of Horodecki \cite{HorodeckiEntMeas}
in Appendix \ref{app:EntMon.pure}.
It turns out that, roughly speaking, 
the entanglement monotonicity \eqref{eq:averagePure} is 
actually the concavity on the subsystem.)

This construction characterizes the entanglement of the subsystem $K$ 
with the rest of the system $\cmpl{K}$,
that is, bipartite entanglement with respect to the split $K|\cmpl{K}$.
For the role of $F$, entropies are usually used;
see Section \ref{sec:MeasBasics.examples}.

\subsection{Convex roof extensions: mixed state measures}
\label{sec:MeasBasics.CnvRoof}

The pure state entanglement measures can be extended to mixed states 
by the use of the so-called \emph{convex roof extension}
\cite{BennettetalMixedStates,UhlmannFidelityConcurrence,UhlmannConvRoofs,RothlisbergerLehmannLossNumericalConvRoof,LibCreme}.
It is motivated by the practical approach of the optimal mixing of the mixed state
from pure states, that is, using as little of pure state entanglement as possible.
For a \emph{continuous} function $f:\mathcal{P}\to\field{R}$,
its convex roof extension $\cnvroof{f}:\Conv\mathcal{P}\equiv\mathcal{D}\to\field{R}$ is defined as
\begin{equation}
\label{eq:cnvroofext}
\cnvroof{f}(\varrho)=\min_{\sum_i p_i \pi_i=\varrho}  \sum_i p_i f(\pi_i),
\end{equation}
where the minimization
takes place over all $\{(p_i,\pi_i)\}$ pure state decompositions of $\varrho$. 
It follows from Schr\"odinger's mixture theorem \cite{SchrodingerMixtureThm}, also called
Gisin-Hughston-Jozsa-Wootters lemma \cite{GisinMixtureThm,HughstonJozsaWoottersMixtureThm},
that the decompositions of a mixed state into an ensemble of $m$ pure states
are labeled by the elements of a Stiefel manifold, which is a \emph{compact} complex manifold. 
On the other hand,
the Carath{\'e}odory theorem ensures that we need only \emph{finite} $m$,
or, to be more precise, $m \leq (\rk\varrho)^2 \leq (\dim\mathcal{H})^2$, shown by Uhlmann \cite{UhlmannOptimalDecomp}.
These observations guarantee the existence of the minimum in~\eqref{eq:cnvroofext}.

Obviously, for pure states the convex roof extension is trivial \eqref{eq:pure.extr},
\begin{subequations}
\begin{equation}
\label{eq:cnvroofpure}
\forall\pi\in\mathcal{P}:\quad
\cnvroof{f}(\pi)=f(\pi).
\end{equation}
The convex roof extension of a function is convex~\eqref{eq:meas.conv},
\begin{equation}
\label{eq:cnvroofcnv}
\cnvroof{f}\Bigl(\sum_i p_i \varrho_i\Bigr) \leq \sum_i p_i \cnvroof{f}(\varrho_i);
\end{equation}
moreover, it is the largest convex function
taking the same values for pure states 
as the original function \cite{UhlmannOptimalDecomp}.
On the other hand, it is bounded by the bounds of the original function,
\begin{equation}
\label{eq:cnvroofbound}
\min_{\pi\in\mathcal{P}} f(\pi) \leq
\cnvroof{f}(\varrho) \leq 
\max_{\pi\in\mathcal{P}} f(\pi).
\end{equation}
\end{subequations}

It is proven by Vidal \cite{VidalEntMon,HorodeckiEntMeas} that
if a function $f:\mathcal{P}\to\field{R}$ is nonincreasing on average for pure states~\eqref{eq:averagePure},
then its convex roof extension is also nonincreasing on average for mixed states~\eqref{eq:meas.average}.
That is, we have the following theorem.
\begin{subequations}
\begin{thm}
\label{thm:averageConvRoof}
For a continuous $f:\mathcal{P}\to\field{R}$,
\begin{equation}
\label{eq:averageConvRoof}
\sum_i p_i f(\pi'_i) \leq f(\pi)
\quad\Longrightarrow\quad
\sum_i p_i \cnvroof{f}(\varrho'_i) \leq \cnvroof{f}(\varrho)
\end{equation}
for all $\pi\mapsto\{(p_i,\pi'_i)\}$ and $\varrho\mapsto\{(p_i,\varrho'_i)\}$ ensembles
resulting from LOCC.
\end{thm}
(We recall the simpler proof of Horodecki \cite{HorodeckiEntMeas} in Appendix \ref{app:EntMon.cnvRoofs}.)
Because of \eqref{eq:cnvroofcnv} and \eqref{eq:averageConvRoof}, 
$\cnvroof{f}(\varrho)$ is also an 
entanglement monotone \eqref{eq:meas.average} and \eqref{eq:meas.conv}.

It is remarkable that in Theorem \ref{thm:pureEntMon} a reverse implication holds
in the bipartite case:
All bipartite mixed entanglement monotones 
(satisfying \eqref{eq:meas.average} and \eqref{eq:meas.conv}) 
restricted for pure states
can be expressed by an $F$
satisfying (i) and (ii) of Theorem \ref{thm:pureEntMon} 
applied to the reduced density matrix.

The convex roof extension preserves the discriminance property \eqref{eq:measdiscrPure}
if we additionally assume that $f\geq0$,
\begin{equation}
\label{eq:cnvroofDisc}
\begin{split}
\Bigl(\pi\in\mathcal{P}_*\;&\Leftrightarrow\; f(\pi)=0\Bigr)\\
\quad&\Longrightarrow\quad
\Bigl(\varrho\in\Conv \mathcal{P}_* = \mathcal{D}_*\;\Leftrightarrow\; \cnvroof{f}(\varrho)=0 \Bigr),
\end{split}
\end{equation}
which can be used for the detection of mixed state entanglement.
(For the proof, see Appendix \ref{app:EntMon.cnvRoofsDisc}.)
Note that this property is based more or less only on that
$\mathcal{P}_*=\Extr\mathcal{D}_*$ and 
$\mathcal{D}_*=\Conv\mathcal{P}_*$.

The convex roof extension also preserves the invariance properties of a function.
For a $G\in\LieGrp{GL}(\mathcal{H})$,
\begin{equation}
\label{eq:cnvroofinv}
\begin{split}
\forall &\pi\in\mathcal{P}:\; f(G\pi G^\dagger)=f(\pi)\\
&\Longleftrightarrow\quad
\forall \varrho\in\mathcal{D}:\; \cnvroof{f}(G\varrho G^\dagger)=\cnvroof{f}(\varrho)
\end{split}
\end{equation}
\end{subequations}
(For the proof, see Appendix \ref{app:EntMon.cnvRoofsInv}.)
Another important property of the convex roof construction is the monotonicity.
For functions $f,g:\mathcal{P}\to\field{R}$,
\begin{subequations}
\begin{equation}
\label{eq:cnvroofMon}
\forall \pi\in\mathcal{P}:\; f(\pi)\leq g(\pi)
\quad\Longleftrightarrow\quad
\forall \varrho\in\mathcal{D}:\; \cnvroof{f}(\varrho)\leq \cnvroof{g}(\varrho).
\end{equation}
(For the proof, see Appendix \ref{app:EntMon.cnvRoofsMon}.)
It is also easy to check the following properties
\begin{align}
\label{eq:cnvroof.c}
\cnvroof{(cf)} &= c\cnvroof{f} \quad \text{for $c\geq0$,}\\
\label{eq:cnvroof.sum}
\cnvroof{(f+g)} &\geq \cnvroof{f}+\cnvroof{g},\\
\label{eq:cnvroof.min}
\cnvroof{\bigl(\min\{f,g\}\bigr)} &\leq \min\{\cnvroof{f},\cnvroof{g}\}.
\end{align}
\end{subequations}
(For the proof, see Appendix \ref{app:EntMon.cnvRoofsEtc}.)

\subsection{Examples}
\label{sec:MeasBasics.examples}

For recalling some well-known examples,
let us consider the \emph{bipartite} case,
with the notations of Section \ref{sec:EntBasics.Ent}.
Particular choices for functions fulfilling the requirements in Theorem \ref{thm:pureEntMon} 
are some entropies given in Section \ref{sec:MeasBasics.Mix}.
Since the entangled pure states are the ones which have mixed marginals \eqref{eq:PsepDecide},
it is, at least, expressive to say that
\textit{``the more mixed the marginals, the more entangled is the state.''}
In particular, using the $F=S:\mathcal{D}_1\to\field{R}$ 
von Neumann entropy \eqref{eq:NeumannEntr}
in construction \eqref{eq:pureVidal}
leads to the 
\emph{``entanglement entropy,''}
\begin{subequations}
\begin{equation}
\label{eq:EntEnt}
E(\pi):= S(\tr_2\pi),
\end{equation}
which is also called simply \emph{``entanglement''} \cite{BennettetalPureStates}.
(Note that, because of the Schmidt decomposition,
the spectra of the marginals of a bipartite pure state 
are the same, apart from the multiplicity of the zero eigenvalues.)
Apart from the von Neumann entropy,
the Tsallis entropies \eqref{eq:TsallisEntr} for all $0<q$
and the R{\'e}nyi entropies \eqref{eq:RenyiEntr} for all $0<q<1$ \cite{VidalEntMon}
are known to be concave \eqref{eq:entrConcavity}, 
and all of them are symmetric and extensible functions of the eigenvalues.
They lead to the 
\emph{``Tsallis} or \emph{R{\'e}nyi entropy of entanglement,''}
\begin{align}
E^\text{Ts}_q(\pi)&:=S^\text{Ts}_q(\tr_2\pi)& &\text{for $0<q$},\\
E^\text{R}_q(\pi) &:=S^\text{R}_q (\tr_2\pi)& &\text{for $0<q<1$}.
\end{align}
A particular choice is the \emph{``concurrence (of entanglement),''}
with the concurrence \eqref{eq:ConcSquared},
\begin{equation}
E^\text{C}(\pi):= C(\tr_2\pi).
\end{equation}
\end{subequations}
All of the above functions
measure the \emph{pure bipartite entanglement}
in the sense that
they satisfy \eqref{eq:averagePure} by Theorem \ref{thm:pureEntMon},
and they are indicators of pure separability,
that is, discriminant \eqref{eq:measdiscrPure} with respect to $\mathcal{P}_\text{sep}$
of \eqref{eq:setPsep}
by \eqref{eq:PsepDecide} and \eqref{eq:entrDiscr}.

Having these pure measures in hand,
thanks to Theorem \ref{thm:averageConvRoof},
we can extend them to mixed states by the use of convex roof extension \eqref{eq:cnvroofext}.
The resulting measures are called
\emph{``entanglement of formation''} \cite{BennettetalMixedStates},
\emph{``Tsallis} or \emph{R{\'e}nyi entanglement of formation''} \cite{VidalEntMon}, and
\emph{``concurrence of formation''} \cite{WoottersConc},
\begin{subequations}
\begin{align}
\label{eq:EntOF}
E^\text{oF}(\varrho) &:= \cnvroof{E}(\varrho),& &\\
\label{eq:TsEntOF}
E^\text{Ts oF}_q(\varrho) &:= \cnvroof{E^\text{Ts}_q}(\varrho),& &\text{for $0<q$},\\
\label{eq:REntOF}
E^\text{R oF}_q(\varrho)  &:= \cnvroof{E^\text{R}_q}(\varrho), & &\text{for $0<q<1$},\\
\label{eq:COF}
E^\text{C oF}(\varrho) &:= \cnvroof{E^\text{C}}(\varrho).& &
\end{align}
\end{subequations}
All of these functions
measure the \emph{mixed bipartite entanglement}
in the sense that
they satisfy \eqref{eq:meas.average} and \eqref{eq:meas.conv} 
by Theorem \ref{thm:averageConvRoof},
and they are indicators of mixed separability,
that is, discriminant \eqref{eq:measdiscr} with respect to $\mathcal{D}_\text{sep}$
by \eqref{eq:cnvroofDisc}.
A remarkable result of Wootters is
a closed formula for the minimization in the convex roof extension in the
entanglement of formation
(through that for the concurrence of formation)
for the case when $\dim\mathcal{H}_1=\dim\mathcal{H}_2=2$,
that is, for two qubits \cite{HillWoottersConc,WoottersConc}.

\section{Sums and Means: a detour}
\label{sec:Means}

In the sequel, we will need to construct entanglement measures 
as functions of more basic ones in a systematic way.
For these, we need 
some properties to hold,
such as monotonicity, homogeneity, concavity, and permutation invariance in many cases.
The $q$-sums and $q$-means, or the more general quasi-sums and quasi-arithmetic means
turn out to be suitable tools in this situation.
$q$-means equate things, which is sometimes undesirable for our investigations,
so $q$-sums turn out to be more suitable in these cases.
Apart from this, they share the properties most important for us,
such as monotonicity, convexity or concavity, and vanishing properties.
Moreover, the $q$-sums and $q$-means of homogeneous functions of a given degree
is of the same degree,
which is a property which seems to be of great importance
in the topic of entanglement of pure states.

\subsection{The meaning of sums and means}
\label{sec:Means.General}

Let us suppose that we have a non-negative quantity $X$,
which can characterize $m$ different entities as $X_1,\dots,X_m$,
and which can also characterize these entities ``together''
as a \emph{total} value $X_\text{tot}$.
Suppose, moreover, that we have a ``law'' telling us that 
the total value of this quantity and
the values for the individual entities
are connected by a summation for their $q$-th powers, as
\begin{equation*}
Y=X_\text{tot}^q = X_1^q + \ldots + X_m^q.
\end{equation*}
Then, on the one hand, the total value is
\begin{equation*}
X_\text{tot} = \bigl(X_1^q + \ldots + X_m^q\bigr)^{1/q},
\end{equation*}
which is called \emph{$q$-sum}.
(For $q\geq1$, this is the same as the $q$-norm, 
restricted for the positive hyperoctant of $\field{R}^m$;
however, we do not need to have vector space structure for the $m$-tuples $(X_1,\dots,X_m)$.
This is why we do not use the name $q$-norm.)
On the other hand,
a natural question is what is the ``mean'' value of this quantity in this situation,
that is, 
what is the \emph{uniform} value for all $X_j$ which leads to the same $Y$ under the same law,
\begin{equation*}
Y = X_1^q + \ldots + X_m^q = X_\text{mean}^q+ \ldots + X_\text{mean}^q = mX_\text{mean}^q.
\end{equation*}
This leads to
\begin{equation*}
X_\text{mean} = \Bigl[\frac1m\bigl(X_1^q+\ldots+X_m^q\bigr)\Bigr]^{1/q},
\end{equation*}
which is called \emph{$q$-mean} (or \emph{H\"older mean}).

For $q=1$, we get back the \emph{sum} and the \emph{arithmetic mean}
for $X_\text{tot}$ and $X_\text{mean}$.
Well-known examples are 
the total and the mean resistance of $m$ resistors connected in series (or total and mean conductance in parallel)
or the total and the mean capacity of $m$ capacitors connected in parallel.
For $q=-1$, we get back the \emph{harmonic sum} and the \emph{harmonic mean}
for $X_\text{tot}$ and $X_\text{mean}$.
Well-known examples are
the total and the mean resistance of $m$ resistors connected in parallel (or total and mean conductance in series)
or the total and the mean capacity of $m$ capacitors connected in series.
For $q=2$, we get back the \emph{quadratic sum} and the \emph{quadratic mean}
for $X_\text{tot}$ and $X_\text{mean}$.
If we consider an $m$-dimensional hypercuboid
 of edges of length $X_j$,
then the quadratic sum of the length of the edges is the length of the diagonal, while
then the quadratic mean of those
 is the uniform length of edges of an $m$-dimensional hypercube
having diagonal of the same length as the original hypercuboid.

A conceptually (but mathematically not too much) different situation is
when the ``law'' is about products,
\begin{equation*} 
Y=X_1\cdot\ldots\cdot X_m.
\end{equation*}
This leads to the
\begin{equation*}
X_\text{mean} = \bigl(X_1\cdot\ldots\cdot X_m\bigr)^{1/m}
\end{equation*}
\emph{geometric mean}. 
We will see that this can be obtained as the $q$-mean for $q=0$.

If we consider an $m$-dimensional hypercuboid
 of edges of length $X_j$ again,
then the geometric mean of the length of the edges
 is the length of the edge of a hypercube of the same volume.
In this case, the meaning of $Y$ is the volume.

A more general, but still relevant, situation is when the ``law''
involves summation of more distorted values, as
\begin{equation*}
Y=h(X_\text{tot}) = h(X_1) + \ldots + h(X_m)
\end{equation*}
for some invertible $h$.
Then, on the one hand, the total value is
\begin{equation*}
X_\text{tot}  = h^{-1}\bigl(h(X_1)+\ldots+h(X_m)\bigr),
\end{equation*}
which might be called, say, \emph{quasi-sum}.
On the other hand,
for the uniform value
this leads to the \emph{quasi-arithmetic mean} (or \emph{Kolmogorov mean}),
\begin{equation*}
X_\text{mean} = h^{-1}\Bigl(\frac1m\bigl(h(X_1)+\ldots+h(X_m)\bigr)\Bigr).
\end{equation*}
The $h(x)=x^q$ gives back the $q$-mean for $q\neq0$,
while $h(x)=\ln(x)$ gives back the geometric mean.

For the definitions and properties of $q$-sums, $q$-means, 
quasi-sums and quasi-arithmetic means,
see Appendixes \ref{app:Means.Power} and \ref{app:Means.Quasi}.

\subsection{Sums and means of entanglement measures}
\label{sec:Means.Meas}

Some $q$-sums and $q$-means 
preserve entanglement monotonicity and discriminance for pure states.
We have, in general, the following lemma about entanglement monotonicity.
\begin{subequations}
\begin{lem}
\label{lem:GMeasure}
Let $f_j:\mathcal{P}\to\field{R}$ be non-negative functions for $j=1,\dots,m$,
which are pure entanglement monotones \eqref{eq:averagePure}.
If $G:\field{R}^m\to\field{R}$ is monotonically increasing in all arguments,
and concave,
then
$G(f_1,\dots,f_m):\mathcal{P}\to\field{R}$
is an entanglement monotone \eqref{eq:averagePure};
that is,
\begin{equation}
\sum_ip_iG(f_1,\dots,f_m)(\pi'_i)\leq G(f_1,\dots,f_m)(\pi)
\end{equation}
for all $\pi\mapsto\{(p_i,\pi'_i)\}$ ensembles resulting from LOCC.
\end{lem}
This is a simple consequence of the monotonicity and the concavity
(see Appendix \ref{app:MeanMon.lemGMeasure}).
Note that similar results can be proven for mixed states
for the properties 
\eqref{eq:meas.mon} and
\eqref{eq:meas.average};
however, the convexity \eqref{eq:meas.conv} would, of course, fail.
Because of the monotonicity and \eqref{eq:qSumCon.cave} and \eqref{eq:qMeanCon.cave}, we have the following.
\begin{cor}
\label{lem:MeanMeasure}
The $q$-sum \eqref{eq:qSums} and $q$-mean \eqref{eq:qMeans} 
of $f_j:\mathcal{P}\to\field{R}$ entanglement monotones \eqref{eq:averagePure}
are entanglement monotones \eqref{eq:averagePure} for $q\leq1$;
that is,
\begin{align}
\begin{split}
\sum_ip_iN_q(f_1,\dots,f_m)(\pi'_i)\leq N_q(f_1,\dots,f_m)(\pi)\\
\text{for $0\neq q\leq1$},
\end{split}\\
\begin{split}
\sum_ip_iM_q(f_1,\dots,f_m)(\pi'_i)\leq M_q(f_1,\dots,f_m)(\pi)\\
\text{for $q\leq1$},
\end{split}
\end{align}
for all $\pi\mapsto\{(p_i,\pi'_i)\}$ ensemble resulting from LOCC.
\end{cor}
\end{subequations}

We have, in general, the following lemma about entanglement discriminance.
\begin{lem}
\label{lem:GDiscr}
Let $f_j:\mathcal{P}\to\field{R}$ be non-negative functions for $j=1,\dots,m$,
which are discriminant 
with respect to a $\mathcal{P}_*\subseteq\mathcal{P}$ set \eqref{eq:measdiscrPure}.
If $G:\field{R}^m\to\field{R}$ obeys the vanishing properties
\begin{equation}
G(\ve{x}) = 0\quad\Longleftarrow\quad \exists j:\; x_j=0,
\end{equation}
then $G(f_1,\dots,f_m):\mathcal{P}\to\field{R}$
is also discriminant with respect to the same set \eqref{eq:measdiscrPure}.
\end{lem}
(This is obvious.)
Because of \eqref{eq:qSumVanish} and \eqref{eq:qMeanVanish}, we have the following.
\begin{cor}
\label{lem:MeanDiscr} 
The $q$-sum \eqref{eq:qSums} and $q$-mean \eqref{eq:qMeans} of indicators 
with respect to a $\mathcal{P}_*\subseteq\mathcal{P}$ set \eqref{eq:measdiscrPure}
are also indicators with respect to the same set \eqref{eq:measdiscrPure}.
\end{cor}
A more interesting situation arises when the different $f_j$ functions
are discriminant with respect to different sets,
as we will see in the sequel.

\section{Entanglement measures for multipartite systems}
\label{sec:MeasMulti}

In Section \ref{sec:EntMulti},
we have introduced the different meaningful kinds of partial separability,
and built up a hierarchy of those.
Now we construct a hierarchy of entanglement measures 
which resembles this hierarchic structure.

In Section \ref{sec:MeasBasics}, we had two main requirements for quantities 
measuring \emph{bipartite} entanglement:
the \emph{LOCC monotonicity} in the sense of \eqref{eq:meas.mon}
and the \emph{discriminance} \eqref{eq:measdiscr} with respect to the given partial separability.
For the hierarchy of quantities measuring \emph{multipartite} entanglement
we introduce a third requirement:
the \emph{multipartite monotonicity,} 
which reflects a natural relation among these quantities, 
and connects them to the hierarchy of entanglement.
By this property we can grasp the hierarchy of multipartite entanglement by the measures.
These three requirements seem to be mandatory.
Also, a fourth one should be satisfied by these measures:
\emph{being meaningful} in some sense.
This last one is quite hard to fulfill, but not impossible.

The construction of measures here reflects 
the construction of the partial separability hierarchy in Section \ref{sec:EntMulti}.
It is based on the measures of pure bipartite entanglement (Section \ref{sec:MeasMulti.zeroth}),
upon which the first and second kind hierarchies of measures of pure multipartite entanglement
are built (Sections \ref{sec:MeasMulti.first} and \ref{sec:MeasMulti.second}).
We turn to mixed states only in the final step (Section \ref{sec:MeasMulti.CnvRoof}),
by the use of convex roof extension,
as has been done in the bipartite construction in Section \ref{sec:MeasBasics.examples}.
Then follows the detection of the classes (Section \ref{sec:MeasMulti.classes}).

\subsection{Level 0: bipartite entanglement}
\label{sec:MeasMulti.zeroth}

Following Section~\ref{sec:MeasBasics.Pure}, 
let $F:\mathcal{D}_K\to\field{R}$ satisfy (i) and (ii) of Theorem \ref{thm:pureEntMon},
and
\begin{subequations}
\begin{align}
\label{eq:FNNeg}
F(\varrho)&\geq0,\\
\label{eq:FDiscr}
F(\varrho)&=0\quad\Longleftrightarrow\quad \varrho\in\mathcal{P}_K\subset\mathcal{D}_K.
\end{align}
\end{subequations}
With this, in the sense of Theorem \ref{thm:pureEntMon}, the function
\begin{equation}
\label{eq:fK}
f_K:=F\circ\tr_{\cmpl{K}}\;:\; \mathcal{P}\longrightarrow\field{R}
\end{equation}
is an entanglement monotone \eqref{eq:averagePure} 
indicator function \eqref{eq:measdiscrPure}
with respect to $\mathcal{P}_{K|\cmpl{K}}$, that is, 
for the pure bipartite entanglement with respect to the bipartite split $K|\pcmpl{K}$.
The latter is
\begin{equation}
\label{eq:defIndK}
f_K(\pi)=0 \quad\Longleftrightarrow\quad
\pi\in \mathcal{P}_{K|\pcmpl{K}},
\end{equation}
which is the consequence of \eqref{eq:FDiscr} and \eqref{eq:PsepDecide}.

\subsection{Level I: multipartite entanglement measures of the first kind}
\label{sec:MeasMulti.first}

In Section \ref{sec:EntMulti.first},
we have the $\struct{P_\text{I},\preceq}$ partial separability hierarchy of the first kind \eqref{eq:posetI}.
Now
we consider the $f_\alpha:\mathcal{P}\to\field{R}$ functions,
different for all $\alpha\in P_\text{I}$ labels of the first kind (partitions),
with the set of them
\begin{equation}
\label{eq:PIf}
P_{\text{I},f} := \bigset{f_\alpha:\mathcal{P}\to\field{R}}{\alpha\in P_\text{I}},
\end{equation}
and we formulate their important properties
expected for the measuring of the pure $\alpha$-entanglement.
Entanglement monotonicity \eqref{eq:averagePure}
is, of course, mandatory for all pure state measures.
The others are as follows.

For the $\alpha$ label of the first kind (partition), the
function $f_\alpha:\mathcal{P}\to\field{R}$ is called
a \emph{pure $\alpha$-indicator function} 
(or \emph{indicator function of the first kind with respect to $\mathcal{P}_\alpha$}),
if it is \emph{discriminant} \eqref{eq:measdiscrPure} with respect to $\mathcal{P}_\alpha$,
that is, if it vanishes exactly for $\alpha$-separable pure states \eqref{eq:setIP},
\begin{subequations}
\begin{equation}
\label{eq:defIndI}
f_\alpha(\pi)=0 
\quad\Longleftrightarrow\quad 
\pi\in \mathcal{P}_\alpha.
\end{equation}
Using \eqref{eq:IPintersect},
one can formulate the vanishing of the $\alpha$-indicator function $f_\alpha$
by the vanishing \eqref{eq:defIndK} of the functions $f_K$ of \eqref{eq:fK} as
\begin{equation}
\label{eq:IndIIndK}
f_\alpha = 0 \quad\Longleftrightarrow\quad \forall K\in\alpha: f_K=0.
\end{equation}
\end{subequations}
From the inclusion hierarchy \eqref{eq:orderisomIP},
we immediately have that the indicator functions \eqref{eq:defIndI} obey
\begin{subequations}
\begin{equation}
\label{FirstLabVanish}
\beta\preceq\alpha\quad\Longleftrightarrow\quad
\bigl(f_\beta=0\;\Rightarrow\;f_\alpha=0\bigr).
\end{equation}
That is, separability with respect to a finer split 
implies separability with respect to a coarser one, 
as it has to do.
We call this property \emph{weak multipartite monotonicity of the first kind}, 
and it provides the $P_{\text{I},f}$ set of functions
with the same hierarchical structure as that of $P_\text{I}$ in \eqref{eq:posetI}
and $P_{\text{I},\mathcal{P}}$ in \eqref{eq:statesIP}.
That is, if the implication 
$(f_\beta=0\;\Rightarrow\;f_\alpha=0)$ 
is denoted with $f_\beta\subseteq f_\alpha$,
then we have the isomorphism of the lattices
\begin{equation}
\struct{P_{\text{I},f},\subseteq} \cong \struct{P_\text{I},\preceq}.
\end{equation}
\end{subequations}
In addition to this, one can formulate
a stronger property for the set of functions $P_{\text{I},f}$,
having motivation in the theory of quantization of entanglement.
For the $P_\text{I}$ labels of the first kind (partition),
the set of functions $P_{\text{I},\mathcal{P}}$ is called
\emph{multipartite-monotonic of the first kind,} if
\begin{subequations}
\begin{equation}
\label{eq:FirstLabMon}
\beta\preceq\alpha\quad\Longleftrightarrow\quad f_\beta\geq f_\alpha.
\end{equation}
(The map $\alpha\mapsto f_\alpha$ is monotonically decreasing 
with respect to the labels of the first kind, 
and the pointwise relation of real-valued functions over the same domain.)
That is,
entanglement with respect to a coarser partition
cannot be higher than entanglement with respect to a finer one.
By this property we attempt to grasp the hierarchy of multipartite entanglement by the measures.
Since, e.g., the tripartite entanglement is considered to be a more powerful resource
than the bipartite entanglement \cite{BorstenGHZW},
one feels that a state can contain a smaller amount of that than of the bipartite entanglement.
(This may or may not seem to be plausible enough;
anyway, multipartite monotonicity holds automatically in some constructions.)
So, in this case we also have
\begin{equation}
\struct{P_{\text{I},f},\geq} \cong \struct{P_\text{I},\preceq}.
\end{equation}
\end{subequations}
The multipartite monotonicity \eqref{eq:FirstLabMon} is
indeed stronger than its weak version (vanishing implications) \eqref{FirstLabVanish},
since the latter one follows from the former one.

With the above definitions in hand, we construct 
multipartite monotonic \eqref{eq:FirstLabMon} hierarchies of entanglement measures 
for pure states for the hierarchy of the first kind,
consisting of entanglement monotonic \eqref{eq:averagePure} 
$\alpha$-indicator functions \eqref{eq:defIndI}.
Let us start with the construction of $\alpha$-indicators,
then check the monotonicity properties.

There are several ways of constructing $\alpha$-indicator functions \eqref{eq:defIndI},
based on the $K|\pcmpl{K}$-indicators \eqref{eq:fK} as in \eqref{eq:IndIIndK}.
Perhaps the simplest one is the \emph{sum,}
\begin{subequations}
\begin{equation}
\label{eq:IndISum}
f_\alpha:=\sum_{K\in\alpha}f_K.
\end{equation}
It clearly obeys $\alpha$-discriminance \eqref{eq:defIndI} through \eqref{eq:IndIIndK},
and entanglement monotonicity \eqref{eq:averagePure}.
(For convenience, one can also use the definition
$f_\alpha:=\frac12\sum_{K\in\alpha}f_K$, 
leading to $f_{K|\pcmpl{K}}=f_K$ for the bipartite splits.)
Another candidate is the arithmetic mean,
\begin{equation}
\label{eq:IndIArithm}
f_\alpha:= \frac1{\abs{\alpha}}\sum_{K\in\alpha}f_K 
= M_1\bigl(f_{K_1},\dots, f_{K_{\abs{\alpha}}}\bigr),
\end{equation}
which is just a sum, multiplied by a factor $\frac1{\abs{\alpha}}$,
which does not ruin the entanglement monotonicity and $\alpha$-discriminance.
One can notice that we can use $q$-sums \eqref{eq:qSums} and $q$-means \eqref{eq:qMeans}
with general parameters $q$,
\begin{align}
\label{eq:IndIqSum}
f_\alpha &:= N_q\bigl(f_{K_1},\dots, f_{K_{\abs{\alpha}}}\bigr),\quad 0<q\leq1,\\
\label{eq:IndIqMean}
f_\alpha &:= M_q\bigl(f_{K_1},\dots, f_{K_{\abs{\alpha}}}\bigr),\quad 0<q\leq1.
\end{align}
\end{subequations}
Indeed, $q$-sums and $q$-means are concave for $q\leq1$, 
see \eqref{eq:qSumCon.cave} and \eqref{eq:qMeanCon.cave},
which is needed for the entanglement monotonicity \eqref{eq:averagePure}
(see Corollary \ref{lem:MeanMeasure}),
while the proper vanishing properties
\eqref{eq:qSumVanish.and} and \eqref{eq:qMeanVanish.and} are satisfied for $0<q$, 
which is needed for the $\alpha$-discriminance \eqref{eq:defIndI} through \eqref{eq:IndIIndK}.

Now we would like to argue that, from the constructions above,
the simplest choice is the best motivated: the sum \eqref{eq:IndISum}.
First of all, as we have learned in Section \ref{sec:Means.General},
using $q$-sums or $q$-means would infer an underlying ``law,'' telling us that 
the sum of the $q$-th power of the functions $f_K$ is meaningful.
This seems to be true for $q=1$ only,
if we start with functions \eqref{eq:fK} based on entropies 
as in the bipartite case in Section \ref{sec:MeasBasics.examples}.
However, in this case, a sum may have more meaning than the arithmetic ($q=1$) mean.
Let us see why.
Taking $F=S$ with the von Neumann entropy \eqref{eq:NeumannEntr}, we simply get
\begin{equation}
\label{eq:IndISumNeumann}
f_\alpha(\pi) := S\bigl(\pi_{K_1}\bigr) + S\bigl(\pi_{K_2}\bigr)
+ \dots + S\bigl(\pi_{K_{\abs{\alpha}}}\bigr),
\end{equation}
the sum of the entropies of \emph{disjoint} subsystems given by the split $\alpha=K_1|K_2|\dots|K_{\abs{\alpha}}$.
(From here, we adopt the convenient notation that 
$\varrho_K=\tr_{\cmpl{K}}\varrho$,
$\pi_K=\tr_{\cmpl{K}}\pi$ (the latter is generally not pure)
in the writing out of a function,
if $\varrho$ or $\pi$ is argument of the function.)
This possesses an expressive meaning.
Let us consider an information-geometrical correlation measure,
called also \emph{relative entropy of correlation},
with respect to a partition $\alpha=K_1|K_2|\dots|K_{\abs{\alpha}}$,
\begin{subequations}
\begin{equation}
\label{eq:geomCorrImin}
\min_{\substack{\forall K\in\alpha:\\ \omega_K\in\mathcal{D}_K}} 
D^\text{KL}\bigl(\varrho\big\Vert
\omega_{K_1}\otimes\omega_{K_2}\otimes\dots\otimes\omega_{K_{\abs{\alpha}}}\bigr)
\end{equation}
being the minimal distinguishability of a state from the set of uncorrelated states,
with respect to the relative entropy \eqref{eq:KullbackLeiblerDiv}.
This characterizes \emph{all the correlations} (classical and quantum \cite{Modi-2010})
contained in the state $\varrho$ with respect to the split $\alpha=K_1|K_2|\dots|K_{\abs{\alpha}}$.
(In quantum information theory,
such geometry-based approach is widely used for the measuring of entanglement.
Perhaps the most relevant geometric entanglement measure
 is the relative entropy of entanglement \cite{Vedral-1997},
and there are several others \cite{Barnum-2001,Wei-2003,Chen-2010}. 
For an overview and references, see section 15.6 of \cite{BengtssonZyczkowski}.
Note that, contrary to these entanglement measures,
we have here a correlation measure.)
Moreover, it can be proven \cite{Modi-2010} that
\begin{equation}
\label{eq:geomCorrIArgmin}
\argmin_{\substack{\forall K\in\alpha:\\ \omega_K\in\mathcal{D}_K}} 
\Bigl\{ D^\text{KL}\Bigl(\varrho\;\Big\Vert\bigotimes_{K\in\alpha}\omega_K\Bigr) \Bigr\} 
= \set{\omega_K=\varrho_K}{\forall K\in\alpha};
\end{equation}
\end{subequations}
that is, the state least indistinguishable from $\varrho$
and uncorrelated with respect to $\alpha$ 
is formed by the marginals of $\varrho$.
(The proof is recalled in Appendix \ref{app:EntMon.geom}.)
So we can write out the relative entropy of correlation \eqref{eq:geomCorrImin} as
\begin{equation}
\label{eq:geomcorrI}
\begin{split}
\min_{\substack{\forall K\in\alpha:\\ \omega_K\in\mathcal{D}_K}} 
\Bigl\{ D^\text{KL}\Bigl(\varrho\;\Big\Vert\bigotimes_{K\in\alpha}\omega_K\Bigr) \Bigr\} 
&= D^\text{KL}\Bigl(\varrho\;\Big\Vert\bigotimes_{K\in\alpha}\varrho_K\Bigr) \\
=\sum_{K\in\alpha} S(\varrho_K) - S(\varrho) 
&=:I_\alpha(\varrho),
\end{split}
\end{equation}
which is a possible multipartite generalization of the \emph{mutual information} \cite{Herbut-2004},
which we call here \emph{$\alpha$-mutual information}.
(It is also called ``among-the-clusters correlation information'' \cite{Herbut-2004}.
For the finest split $1|2|\dots|n$, this is also called ``correlation information'' \cite{Herbut-2004},
or ``multipartite mutual information'' \cite{Yang-2009}, also considered by
Lindblad \cite{Lindblad-1973} and used \cite{Horodecki-1994} to describe correlations within multipartite quantum systems.)
Now, applying this to a pure state $\pi\in\mathcal{P}\subset\mathcal{D}$, 
since $S(\pi)=0$, we have that \eqref{eq:IndISum} is actually
\begin{equation}
\label{eq:IndISumNeumannMutInf}
\begin{split}
f_\alpha(\pi) &= \sum_{K\in\alpha}f_K(\pi)
\equiv \sum_{K\in\alpha} S(\pi_K)
= \sum_{K\in\alpha} S(\pi_K) - \underbrace{S(\pi)}_0\\
&= D^\text{KL}\Bigl(\pi\;\Big\Vert\bigotimes_{K\in\alpha}\pi_K\Bigr)
= I_\alpha(\pi).
\end{split}
\end{equation}
That is, for pure states,
the sum of the von Neumann entropies of \emph{disjoint} subsystems 
given by the split $\alpha=K_1|K_2|\dots|K_{\abs{\alpha}}$
is a meaningful quantity, and it characterizes the whole amount of correlation
contained in the state $\pi$ with respect to that split,
being the $\alpha$-mutual information above.

This reasoning enlightens also the meaning of entanglement itself
(which holds also in the bipartite case \eqref{eq:EntEnt}).
In classical probability theory,
pure states are always uncorrelated,
so if in the quantum case a pure state shows correlation,
then this correlation is considered to be of quantum origin,
and this correlation is \emph{defined to be the entanglement.}
From this point of view, it is plausible to think that
\textit{the quantum versions of classical correlation measures
applied to pure quantum states are pure entanglement measures,}
both in the bipartite and in the multipartite scenario.
However, the details should be clarified in this principle;
entanglement monotonicity should be checked for the concrete measures.
(For further discussion, see Section \ref{sec:Summ}.)

By this reasoning, let us define the
\emph{$\alpha$-entanglement entropy}, or simply \emph{$\alpha$-entanglement}
$E_\alpha:\mathcal{P}\to\field{R}$, as
\begin{equation}
\label{eq:EntEntI}
E_\alpha(\pi) := \frac12 I_\alpha(\pi) = \frac12\sum_{K\in\alpha}S(\pi_K),
\end{equation}
by the use of the von Neumann entropy \eqref{eq:NeumannEntr}.
This is the direct Level I multipartite generalization of the entanglement entropy \eqref{eq:EntEnt}.
(Note that while the $\alpha$-mutual information $I_\alpha$ 
is defined over the whole state space $\mathcal{D}$,
$E_\alpha$ is defined only for the pure states $\mathcal{P}$,
in accordance with \eqref{eq:EntEnt}.)
Note that
\begin{equation}
\label{eq:EntEntIbound}
0\leq E_\alpha(\pi) \leq \frac12 \ln \dim \mathcal{H} = \frac12 \sum_{a\in L} \ln \dim \mathcal{H}_a
\end{equation}
by \eqref{eq:entrBound}.

Until this point, we have taken into consideration
only the entanglement monotonicity \eqref{eq:averagePure}
and the $\alpha$-discriminance \eqref{eq:defIndI}.
The multipartite monotonicity \eqref{eq:FirstLabMon}
is an additional concern, which might be satisfied too.
For the measures \eqref{eq:EntEntI} based on the von Neumann entropy \eqref{eq:NeumannEntr},
the multipartite monotonicity \eqref{eq:FirstLabMon}
is a simple consequence of the subadditivity \eqref{eq:subadd.Neumann} of the von Neumann entropy.
The $q>1$ Tsallis entropies \eqref{eq:TsallisEntr} are also suitable \eqref{eq:subadd.Tsallis};
however, R{\'e}nyi entropies \eqref{eq:RenyiEntr} are not.
Note that, since means equate things,
using arithmetic mean \eqref{eq:IndIArithm} instead of sum \eqref{eq:IndISum}
ruins the multipartite monotonicity for these cases.

\subsection{Examples}
\label{sec:MeasMulti.firstExamples}

Writing out some examples explicitly might not be useless here
(cf.~Section \ref{sec:EntMulti.firstExamples}).
Here we consider the $\alpha$-entanglement entropy \eqref{eq:EntEntI},
arising from the construction \eqref{eq:IndISumNeumannMutInf} 
using the von Neumann entropy \eqref{eq:NeumannEntr}.
Since the resulting functions are 
multipartite monotonic \eqref{eq:FirstLabMon} indicator functions \eqref{eq:defIndI},
we can read off these relations from the
lattice $P_\text{I}$, which can be seen 
for the cases $n=2$ and $3$ 
in the upper-left parts of Figures \ref{fig:labellattices2} and \ref{fig:labellattices3}.
(Note that we have adopted the convenient notation that 
$\pi_K=\tr_{\cmpl{K}}\pi$ (generally not pure)
in the writing out of a function having $\pi$ as its argument.
On the other hand, from the definition of the partial trace \eqref{eq:ptr},
we have $\pi_L\equiv \pi$.)

For the \emph{bipartite} case,
we get back the content of Section \ref{sec:MeasBasics.examples},
\begin{align*}
E_{12}(\pi)  &= \frac12 S(\pi_{12}) = 0,\\
E_{1|2}(\pi) &= \frac12\bigl(S(\pi_1)+S(\pi_2)\bigr)=S(\pi_a).
\end{align*}
Note that the multipartite monotonicity \eqref{eq:FirstLabMon} holds,
$E_{1|2}(\pi)\geq E_{12} (\pi)$.
We have also the discriminance \eqref{eq:defIndI},
\begin{align*}
\pi&\in\mathcal{P}_{12}&
\quad&\Longleftrightarrow&\quad E_{12}(\pi)&=0,\\
\pi&\in\mathcal{P}_{1|2}&
\quad&\Longleftrightarrow&\quad E_{1|2}(\pi)&=0.
\end{align*}

For the \emph{tripartite} case,
\begin{align*}
E_{123}(\pi)   &= \frac12 S(\pi_{123}) = 0,\\
E_{a|bc}(\pi)  &= \frac12\bigl(S(\pi_a)+S(\pi_{bc})\bigr) = S(\pi_a),\\
E_{1|2|3}(\pi) &= \frac12\bigl(S(\pi_1)+S(\pi_2)+S(\pi_3)\bigr),
\end{align*}
with all bipartitions $a|bc$ of $\{1,2,3\}$.
Note that the multipartite monotonicity \eqref{eq:FirstLabMon} holds,
$E_{1|2|3}(\pi)\geq E_{a|bc}(\pi)\geq E_{123}(\pi)$.
We have also the discriminance \eqref{eq:defIndI},
\begin{align*}
\pi&\in\mathcal{P}_{123}&
\quad&\Longleftrightarrow&\quad E_{123}(\pi)&=0,\\
\pi&\in\mathcal{P}_{a|bc}&
\quad&\Longleftrightarrow&\quad E_{a|bc}(\pi)&=0,\\
\pi&\in\mathcal{P}_{1|2|3}&
\quad&\Longleftrightarrow&\quad E_{1|2|3}(\pi)&=0.
\end{align*}

\subsection{Level II: multipartite entanglement measures of the second kind}
\label{sec:MeasMulti.second}

In Section \ref{sec:EntMulti.second},
we have the $\struct{P_\text{II},\preceq}$ partial separability hierarchy of the second kind \eqref{eq:posetII}.
Now, similarly to Section \ref{sec:MeasMulti.first},
we consider the $f_{\vs{\alpha}}:\mathcal{P}\to\field{R}$ functions,
different for all $\vs{\alpha}\in P_\text{II}$ labels of the second kind,
with the set of them
\begin{equation}
\label{eq:PIIf}
P_{\text{II},f} := \bigset{f_{\vs{\alpha}}:\mathcal{P}\to\field{R}}{\vs{\alpha}\in P_\text{II}},
\end{equation}
and we formulate their important properties
expected for the measuring of the pure $\vs{\alpha}$-entanglement.
Entanglement monotonicity \eqref{eq:averagePure}
is, of course, mandatory for all pure state measures.
The others are as follows.

For the $\vs{\alpha}$ label of the second kind (nonempty down-set of partitions), the
function $f_{\vs{\alpha}}:\mathcal{P}\to\field{R}$ is called
\emph{pure $\vs{\alpha}$-indicator function} 
(or \emph{indicator function of the second kind with respect to $\mathcal{P}_{\vs{\alpha}}$}),
if it is \emph{discriminant} \eqref{eq:measdiscrPure} with respect to $\mathcal{P}_{\vs{\alpha}}$,
that is, if it vanishes exactly for $\vs{\alpha}$-separable pure states \eqref{eq:setIIP},
\begin{subequations}
\begin{equation}
\label{eq:defIndII}
f_{\vs{\alpha}}(\pi)=0
\quad\Longleftrightarrow\quad 
\pi\in \mathcal{P}_{\vs{\alpha}}.
\end{equation}
Using \eqref{eq:setIIP},
one can formulate the vanishing of the $\vs{\alpha}$-indicator function $f_{\vs{\alpha}}$
by the vanishing of the $\alpha$-indicator functions $f_\alpha$ of \eqref{eq:defIndI} as
\begin{equation}
\label{eq:IndIIIndI}
f_{\vs{\alpha}} = 0 \quad\Longleftrightarrow\quad \exists \alpha\in\vs{\alpha}: f_\alpha=0.
\end{equation}
\end{subequations}
From the inclusion hierarchy \eqref{eq:orderisomIIP},
we immediately have that the indicator functions \eqref{eq:defIndII} obey
\begin{subequations}
\begin{equation}
\label{SecondLabVanish}
\vs{\beta}\preceq\vs{\alpha}\quad\Longleftrightarrow\quad
\bigl(f_{\vs{\beta}}=0\;\Rightarrow\;f_{\vs{\alpha}}=0\bigr).
\end{equation}
That is, a separability lower in the hierarchy implies a higher one,
as it has to do.
We call this property \emph{weak multipartite monotonicity of the second kind},
and it provides the $P_{\text{II},f}$ set of functions
with the same hierarchical structure as that of $P_\text{II}$ in \eqref{eq:posetII}
and $P_{\text{II},\mathcal{P}}$ in \eqref{eq:statesIIP}.
That is, if the implication 
$(f_{\vs{\beta}}=0\;\Rightarrow\;f_{\vs{\alpha}}=0)$ 
is denoted with $f_{\vs{\beta}}\subseteq f_{\vs{\alpha}}$,
then we have the isomorphism of the lattices
\begin{equation}
\struct{P_{\text{II},f},\subseteq} \cong \struct{P_\text{II},\preceq}.
\end{equation}
\end{subequations}
In addition to this, one can formulate
a stronger property for the set of functions $P_{\text{I},f}$,
having some motivation in the theory of quantization of entanglement.
For the $P_\text{II}$ labels of the second kind,
the set of functions $P_{\text{II},\mathcal{P}}$ is called
\emph{multipartite-monotonic of the second kind,} if
\begin{subequations}
\begin{equation}
\label{eq:SecondLabMon}
\vs{\beta}\preceq\vs{\alpha}\quad\Longleftrightarrow\quad f_{\vs{\beta}}\geq f_{\vs{\alpha}}.
\end{equation}
(The map $\vs{\alpha}\mapsto f_{\vs{\alpha}}$ is monotonically decreasing 
with respect to the labels of the second kind,
and the pointwise relation of real-valued functions over the same domain.)
That is, 
entanglement higher in the hierarchy 
cannot be higher than entanglement lower in there.
By this property we attempt to grasp the hierarchy of multipartite entanglement by the measures.
(This may or may not seem to be plausible enough;
anyway, multipartite monotonicity holds automatically in some constructions.)
So, in this case we also have
\begin{equation}
\struct{P_{\text{II},f},\geq} \cong \struct{P_\text{II},\preceq}.
\end{equation}
\end{subequations}
The multipartite monotonicity \eqref{eq:SecondLabMon} is
indeed stronger than its weak version (vanishing implications) \eqref{SecondLabVanish},
since the latter one follows from the former one.

With the above definitions in hand, we construct
a multipartite monotonic \eqref{eq:SecondLabMon} hierarchy of entanglement measures
for pure states for the hierarchy of the second kind,
consisting of entanglement monotonic \eqref{eq:averagePure}
$\vs{\alpha}$-indicator functions \eqref{eq:defIndII}.
Let us start with the construction of $\vs{\alpha}$-indicators,
then check the monotonicity properties.

There are several ways of constructing $\vs{\alpha}$-indicator functions \eqref{eq:defIndII},
based on the $\alpha$-indicators \eqref{eq:defIndI}.
Perhaps the simplest one is the \emph{product},
\begin{subequations}
\begin{equation}
\label{eq:IndIIProd}
f_{\vs{\alpha}}:=\prod_{\alpha\in\vs{\alpha}}f_\alpha.
\end{equation}
Unfortunately, 
while it clearly obeys $\vs{\alpha}$-discriminance \eqref{eq:defIndII} through \eqref{eq:IndIIIndI},
it lacks for entanglement monotonicity \eqref{eq:averagePure}.
This is because the set of functions obeying \eqref{eq:averagePure}
is not closed under multiplication,
which is related to the fact that the product of two concave functions is not concave in general.
Moreover, a recent result of Eltschka \textit{et.~al.}~\cite{EltschkaetalEntMon,ViehmannetalRescalingEntMeasMix} suggests
that homogeneous functions obeying \eqref{eq:averagePure}
cannot be of arbitrarily high degree.
(See Theorem I.~in \cite{EltschkaetalEntMon}, concerning a special class of functions.)
This is an indication for using some $q$-sums \eqref{eq:qSums.q} or $q$-means \eqref{eq:qMeans.q},
since they do not change the degree.
The geometric mean \eqref{eq:qMeans.0},
\begin{equation}
\label{eq:IndIIGeom}
f_{\vs{\alpha}}:= \Bigl[\prod_{\alpha\in\vs{\alpha}}f_\alpha\Bigr]^{1/\abs{\vs{\alpha}}} 
=M_0\bigl(f_{\alpha_1},\dots, f_{\alpha_{\abs{\vs{\alpha}}}}\bigr)
\end{equation}
obeys $\vs{\alpha}$-discriminance \eqref{eq:defIndII} 
as the product \eqref{eq:IndIIProd} does,
and it turns out to be entanglement monotonic \eqref{eq:averagePure} \cite{SzalayKokenyesiPartSep,SzalayDissertation}.

One can notice that we can use $q$-sums \eqref{eq:qSums} and $q$-means \eqref{eq:qMeans}
with general parameters $q$,
\begin{align}
\label{eq:IndIIqSum}
f_{\vs{\alpha}} &:= N_q\bigl(f_{\alpha_1},\dots, f_{\alpha_{\abs{\vs{\alpha}}}}\bigr),\quad q<0,\\
\label{eq:IndIIqMean}
f_{\vs{\alpha}} &:= M_q\bigl(f_{\alpha_1},\dots, f_{\alpha_{\abs{\vs{\alpha}}}}\bigr),\quad q\leq0.
\end{align}
\end{subequations}
Indeed, $q$-sums and $q$-means are concave for $q\leq1$
(see \eqref{eq:qSumCon.cave} and  \eqref{eq:qMeanCon.cave}),
which is needed for the entanglement monotonicity \eqref{eq:averagePure}
(see Corollary \ref{lem:MeanMeasure}),
while the proper vanishing properties \eqref{eq:qSumVanish.or} and \eqref{eq:qMeanVanish.or} are satisfied for $q<0$ and $q\leq0$,
which is needed for the $\vs{\alpha}$-discriminance \eqref{eq:defIndII} through \eqref{eq:IndIIIndI}.

However, geometric means, or all $q\neq1$ $q$-means of indicator functions of the first kind
constructed from entropies in the way of Section \ref{sec:MeasMulti.first},
do not seem to make any sense in this situation.
As we have learned in Section~\ref{sec:Means.General},
using the $q$-mean of entropies would infer an underlying ``law'' telling us that 
the sum is the $q$th power of the functions $f_\alpha$ is meaningful,
which seems to be true only for $q=1$.
We have two ways of getting out from this deadlock.
The first one is to use some transformed quantities for the indicator functions of the first kind, 
the second one is to use the $-\infty$-mean \eqref{eq:qMeans.min}, that is, the minimum, 
which \emph{does} make sense.

To follow the \textit{first way,}
let us start with the $\alpha$-entanglement \eqref{eq:EntEntI},
$E_\alpha(\pi)=\frac12\sum_{K\in\alpha} S(\pi_K)$,
which is an \emph{``entropy-type''} quantity.
Only the sum of entropy-type quantities seems to be meaningful;
however, a sum does not fulfill the $\vs{\alpha}$-discriminancy \eqref{eq:defIndII} 
through \eqref{eq:IndIIIndI}.
Although, a product does fulfill the $\vs{\alpha}$-discriminancy \eqref{eq:defIndII}
through \eqref{eq:IndIIIndI},
the product of entropy-type quantities seems to be meaningless.
A product which is meaningful is the product of \emph{``probability-type''} quantities.
Indeed, in information theory 
(both classical \cite{CoverThomas} and quantum \cite{NielsenChuang,Wilde})
entropy-type quantities appear often as arguments of $\ee^{-x}$,
leading to probability-type quantities,
e.g., in coding, or hypothesis testing situations
\cite{BengtssonZyczkowski,NielsenChuang,PreskillNotes,Wilde,Wolf,PetzQInfo,CoverThomas}.
So, following this way, for the indicator functions of the first kind,
we use the $f_\alpha := g\circ E_\alpha$ transformed version of the
$\alpha$-entanglement \eqref{eq:EntEntI},
based on the $\alpha$-mutual information \eqref{eq:geomcorrI},
by the use of the continuous invertible function $g:\field{R}\to\field{R}$.
Then we take the geometric mean of these indicator functions,
preserving entanglement monotonicity and discriminance,
and then we do the transformation back in order to get an entropy-type quantity again.
This function $g$
\textit{(i)} should be the same for all $\alpha$ (for simplicity),
\textit{(ii)} should map from non-negative to non-negative values (for being meaningful),
\textit{(iii)} should map zero to zero (to preserve $\vs{\alpha}$-discriminancy \eqref{eq:IndIIIndI}),
\textit{(iv)} should be invertible,
\textit{(v)} should be monotonically increasing, and
\textit{(vi)} should be concave 
(this seems to be necessary but not sufficient 
for the entanglement monotonicity \eqref{eq:averagePure}).
A particular function obeying these requirements is
\begin{equation}
\label{eq:magicg}
g(x) := 1-\ee^{-x},
\end{equation}
being a perfect candidate for the conversion from entropy- to probability-type quantities.
With this, let
\begin{equation}
\label{eq:IndIImagicg}
\begin{split}
f_{\vs{\alpha}}&:= 
g^{-1}\Bigl(M_0\bigl(g(E_{\alpha_1}),\dots, g(E_{\alpha_{\abs{\vs{\alpha}}}})\bigr)\Bigr)\\
&= -\ln\Bigl(1-\Bigl[\prod_{\alpha\in\vs{\alpha}} \bigl(1-\ee^{-E_\alpha}\bigr) \Bigr]^{1/\abs{\vs{\alpha}}}\Bigr) \\
&= M_{\ln\circ g}\bigl(E_{\alpha_1},\dots, E_{\alpha_{\abs{\vs{\alpha}}}}\bigr),
\end{split}
\end{equation}
also formulated by the quasi-arithmetic mean \eqref{eq:MeanQuasiArithm}
for $h=\ln\circ g$.
It is far from obvious that this function is an entanglement monotone \eqref{eq:averagePure}.
(For the proof, see Appendix \ref{app:MeanMon.magic}.)
On the other hand, it is clearly an $\vs{\alpha}$-indicator \eqref{eq:defIndII} 
through \eqref{eq:IndIIIndI}.
Its only drawback is that it is not multipartite monotonic \eqref{eq:SecondLabMon}.
Using product instead of the geometric mean in the construction 
would give multipartite monotonicity; 
however, that would ruin entanglement monotonicity.
The first way seems to end here.

To follow the \textit{second way}, which is actually the simpler and also better motivated one, 
take \eqref{eq:IndIIqMean} or \eqref{eq:IndIIqSum} with $q\to-\infty$
with the indicator functions 
$f_\alpha(\pi)=I_\alpha(\pi)=\sum_{K\in\alpha} S(\pi_K)$
as in \eqref{eq:IndISumNeumannMutInf}, that is,
\begin{equation}
\label{eq:IndIImin}
f_{\vs{\alpha}}:= M_{-\infty}\bigl(I_{\alpha_1},\dots,I_{\alpha_{\abs{\vs{\alpha}}}}\bigr)
= \min\bigl(I_{\alpha_1},\dots, I_{\alpha_{\abs{\vs{\alpha}}}}\bigr).
\end{equation}
This also possesses an expressive meaning.
To clarify this, 
recall that the $\alpha$-mutual information $I_\alpha$ in \eqref{eq:geomcorrI}
characterizes all the correlations in the sense of statistical distinguishability,
that is, the distinguishability of the state from the closest (least distinguishable) 
uncorrelated state with respect to $\alpha$. 
Now the quantity $\min(I_{\alpha_1},\dots, I_{\alpha_{\abs{\vs{\alpha}}}})$
is the distinguishability of the state from the closest (least distinguishable)
uncorrelated state with respect to \emph{any} $\alpha\in\vs{\alpha}$,
and, using \eqref{eq:geomCorrIArgmin} as in \eqref{eq:geomcorrI},
we define the Level II version of the mutual information as
\begin{equation}
\label{eq:geomcorrII}
\begin{split}
\min_{\alpha\in\vs{\alpha}}\bigl\{I_\alpha(\varrho)\bigr\}
&=\min_{\alpha\in\vs{\alpha}} \min_{\substack{\forall K\in\alpha:\\ \omega_K\in\mathcal{D}_K}}
\Bigl\{ D^\text{KL}\Bigl(\varrho\Big\Vert\bigotimes_{K\in\alpha}\omega_K\Bigr) \Bigr\}\\ 
&= \min_{\alpha\in\vs{\alpha}}
\Bigl\{\sum_{K\in\alpha}S(\varrho_K)\Bigr\} - S(\varrho)
=:I_{\vs{\alpha}}(\varrho),
\end{split}
\end{equation}
which is also a geometric measure of correlation,
we call it \emph{$\vs{\alpha}$-mutual information}.
Now, applying this to a pure state $\pi\in\mathcal{P}\subset\mathcal{D}$,
since $S(\pi)=0$, we have that
\begin{equation}
\label{eq:IndIIminMutInf}
\begin{split}
f_{\vs{\alpha}}(\pi)=
\min_{\alpha\in\vs{\alpha}}\bigl\{I_\alpha(\pi)\bigr\}
&=\min_{\alpha\in\vs{\alpha}} \min_{\substack{\forall K\in\alpha:\\ \omega_K\in\mathcal{D}_K}}
\Bigl\{ D^\text{KL}\Bigl(\pi\Big\Vert\bigotimes_{K\in\alpha}\omega_K\Bigr) \Bigr\}\\ 
&= \min_{\alpha\in\vs{\alpha}}
\Bigl\{\sum_{K\in\alpha}S(\pi_K)\Bigr\}
=I_{\vs{\alpha}}(\pi).
\end{split}
\end{equation}
That is,
the minimal among the sums  of the von Neumann entropies of \emph{disjoint} subsystems
given by the different splits $\alpha\in\vs{\alpha}$
is a meaningful quantity,
characterizing the distinguishability of the state from the closest (least distinguishable)
uncorrelated state with respect to \emph{any} $\alpha\in\vs{\alpha}$.

By this reasoning,
let us define the \emph{$\vs{\alpha}$-entanglement entropy},
or simply \emph{$\vs{\alpha}$-entanglement}
$E_{\vs{\alpha}}:\mathcal{P}\to\field{R}$ as
\begin{equation}
\label{eq:EntEntII}
E_{\vs{\alpha}}(\pi) := \min_{\alpha\in\vs{\alpha}}\bigl\{E_\alpha(\pi)\bigr\},
\end{equation}
by the use of the $\alpha$-entanglement \eqref{eq:EntEntI}.
This is the Level II multipartite generalization of the entanglement entropy \eqref{eq:EntEnt}.
(Note that while the $\vs{\alpha}$-mutual information $I_{\vs{\alpha}}$ 
is defined over the whole state space $\mathcal{D}$,
$E_{\vs{\alpha}}$ is defined only for the pure states $\mathcal{P}$,
in accordance with \eqref{eq:EntEntI}.)
Note that
\begin{equation}
\label{eq:EntEntIIbound}
0\leq E_{\vs{\alpha}}(\pi) \leq \frac12 \ln \dim \mathcal{H} = \frac12 \sum_{a\in L} \ln \dim \mathcal{H}_a
\end{equation}
by \eqref{eq:EntEntIbound}.

This function is an entanglement monotone \eqref{eq:averagePure}
$\vs{\alpha}$-indicator \eqref{eq:IndIIIndI};
moreover, it can easily be checked that it is also multipartite monotonic \eqref{eq:SecondLabMon}.
Note that, because the $f_\alpha=E_\alpha$ Level I functions
are multipartite monotonic \eqref{eq:FirstLabMon},
in the minimization during the calculation of the $f_{\vs{\alpha}}=E_{\vs{\alpha}}$
Level II functions,
it is enough to consider only the functions labeled by $\max\vs{\alpha}$,
\begin{equation}
E_{\vs{\alpha}} = \min_{\alpha\in\vs{\alpha}}\{E_\alpha\} = 
\min_{\alpha\in\max\vs{\alpha}}\{E_\alpha\}.
\end{equation}

\subsection{Examples}
\label{sec:MeasMulti.secondExamples}

Writing out some examples explicitly might not be useless here
(cf.~Section \ref{sec:EntMulti.secondExamples}).
Here we consider the $\vs{\alpha}$-entanglement entropy \eqref{eq:EntEntII},
arising from the construction \eqref{eq:IndIIminMutInf}
based on the $\alpha$-entanglement entropy \eqref{eq:EntEntI},
arising from the construction \eqref{eq:IndISumNeumannMutInf}
using the von Neumann entropy \eqref{eq:NeumannEntr}.
Since the resulting functions are 
multipartite monotonic \eqref{eq:SecondLabMon} indicator functions \eqref{eq:defIndII},
we can read off these relations from the
lattice $P_\text{II}$, which can be seen 
for the cases $n=2$ and $3$ 
in the upper-right parts of Figures \ref{fig:labellattices2} and \ref{fig:labellattices3}.
(Note that we have adopted the convenient notation that 
$\pi_K=\tr_{\cmpl{K}}\pi$ (generally not pure)
in the writing out of a function having $\pi$ as its argument.
On the other hand, from the definition of the partial trace \eqref{eq:ptr},
we have $\pi_L\equiv \pi$.)

For the bipartite case, based on Section \ref{sec:MeasMulti.secondExamples},
we get back the content of Section \ref{sec:MeasBasics.examples},
\begin{align*}
E_{\downset\{12\}}(\pi)
&= \min\bigl\{E_{1|2}(\pi),E_{12}(\pi)\bigr\} 
= E_{12}(\pi) = 0,\\
E_{\downset\{1|2\}}(\pi) 
&= \min\bigl\{E_{1|2}(\pi)\bigr\} 
= E_{1|2}(\pi)  = S(\pi_a).
\end{align*}
Note that the multipartite monotonicity \eqref{eq:SecondLabMon} holds,
$E_{\downset\{1|2\}}(\pi)\geq E_{\downset\{12\}} (\pi)$.
We have also the discriminance \eqref{eq:defIndII},
\begin{align*}
\pi&\in\mathcal{P}_{\downset\{12\}}&
\quad&\Longleftrightarrow&\quad E_{\downset\{12\}}(\pi)  &= 0,\\
\pi&\in\mathcal{P}_{\downset\{1|2\}}&
\quad&\Longleftrightarrow&\quad E_{\downset\{1|2\}}(\pi) &= 0.
\end{align*}

For the \emph{tripartite} case, based on Section \ref{sec:MeasMulti.secondExamples},
\begin{align*} 
\begin{split} 
&E_{\downset\{123\}}(\pi)\\
&= \min\bigl\{E_{1|2|3}(\pi),E_{1|23}(\pi),E_{2|13}(\pi),E_{3|12}(\pi),E_{123}(\pi)\bigr\} \\
&= E_{123}(\pi)  = 0,
\end{split} \\ 
\begin{split} 
&E_{\downset\{1|23,2|13,3|12\}}(\pi) \\
&= \min\bigl\{E_{1|2|3}(\pi),E_{1|23}(\pi),E_{2|13}(\pi),E_{3|12}(\pi)\bigr\} \\ 
&= \min\bigl\{E_{1|23}(\pi),E_{2|13}(\pi),E_{3|12}(\pi)\bigr\} \\
&= \min\bigl\{S(\pi_1),S(\pi_2),S(\pi_2)\bigr\},
\end{split} \\ 
\begin{split} 
&E_{\downset\{b|ac,c|ab\}}(\pi)\\
&= \min\bigl\{E_{1|2|3}(\pi),E_{b|ac}(\pi),E_{c|ab}(\pi)\bigr\} \\ 
&= \min\bigl\{E_{b|ac}(\pi),E_{c|ab}(\pi)\bigr\} \\
&= \min\bigl\{S(\pi_b),S(\pi_c)\bigr\},
\end{split} \\ 
\begin{split} 
&E_{\downset\{a|bc\}}(\pi)\\
&= \min\bigl\{E_{1|2|3}(\pi),E_{a|bc}(\pi)\bigr\}\\
&= E_{a|bc}(\pi)  = S(\pi_a),
\end{split} \\ 
\begin{split} 
&E_{\downset\{1|2|3\}}(\pi)\\
&= \min\bigl\{E_{1|2|3}(\pi)\bigr\} \\
&= E_{1|2|3}(\pi)  = \frac12\bigl( S(\pi_1)+S(\pi_2)+S(\pi_3) \bigr).
\end{split} 
\end{align*}
Note that the multipartite monotonicity \eqref{eq:SecondLabMon} holds,
$E_{\downset\{1|2|3\}}(\pi)\geq 
E_{\downset\{a|bc\}}(\pi)\geq
E_{\downset\{a|bc,b|ac\}}(\pi)\geq
E_{\downset\{a|bc,b|ac,c|ab\}}(\pi)\geq
E_{\downset\{123\}}(\pi)$.
We have also the discriminance \eqref{eq:defIndII},
\begin{align*}
\pi&\in\mathcal{P}_{\downset\{123\}}&
\;&\Longleftrightarrow&\; E_{\downset\{123\}}(\pi)&=0,\\
\pi&\in\mathcal{P}_{\downset\{1|23,2|13,3|12\}}&
\;&\Longleftrightarrow&\; E_{\downset\{1|23,2|13,3|12\}}(\pi)&=0,\\
\pi&\in\mathcal{P}_{\downset\{b|ca,c|ab\}}&
\;&\Longleftrightarrow&\; E_{\downset\{b|ac,c|ab\}}(\pi)&=0,\\
\pi&\in\mathcal{P}_{\downset\{a|bc\}}&
\;&\Longleftrightarrow&\; E_{\downset\{a|bc\}}(\pi)&=0,\\
\pi&\in\mathcal{P}_{\downset\{1|2|3\}}&
\;&\Longleftrightarrow&\; E_{\downset\{1|2|3\}}(\pi)&=0.
\end{align*}

\subsection{Multipartite entanglement measures for mixed states}
\label{sec:MeasMulti.CnvRoof}

Now it is easy to step from the pure states to mixed ones,
thanks to the useful properties of the convex roof extension,
listed in Section \ref{sec:MeasBasics.CnvRoof}.
So for the functions $f_{\vs{\alpha}}\in P_{\text{II}}$,
we have its convex roof extension \eqref{eq:cnvroofext} over mixed states,
\begin{equation}
\label{eq:IndIIconvroof}
\cnvroof{f}_{\vs{\alpha}}(\varrho)=
\min_{\sum_i p_i\pi_i=\varrho}\sum_i p_i f_{\vs{\alpha}}(\pi_i),
\end{equation}
and let us define the set of these functions as
\begin{equation}
\label{eq:PIIfcnv}
P_{\text{II},\cnvroof{f}} := 
\bigset{\cnvroof{f}_{\vs{\alpha}}:\mathcal{D}\to\field{R}}{f_{\vs{\alpha}}\in P_{\text{II},f}}.
\end{equation}

If the function $f_{\vs{\alpha}}$ is an entanglement monotone, 
that is, nonincreasing on average for pure states \eqref{eq:averagePure}
(for example the $\vs{\alpha}$-entanglement entropy in \eqref{eq:EntEntII}),
then, thanks to Theorem \ref{thm:averageConvRoof},
its convex roof extension \eqref{eq:cnvroofext}
is also nonincreasing on average \eqref{eq:meas.average}
and also convex \eqref{eq:meas.conv},
so it is an entanglement monotone.

If the function $f_{\vs{\alpha}}$
is a pure $\vs{\alpha}$-indicator \eqref{eq:defIndII}
(for example the $\vs{\alpha}$-entanglement entropy in \eqref{eq:EntEntII}),
then, thanks to \eqref{eq:cnvroofDisc}, its convex roof extension \eqref{eq:IndIIconvroof}
is a mixed $\vs{\alpha}$-indicator,
\begin{subequations}
\begin{equation}
\label{eq:defIndIID}
\varrho\in \mathcal{D}_{\vs{\alpha}}
\quad\Longleftrightarrow\quad 
\cnvroof{f}_{\vs{\alpha}}(\varrho)=0.
\end{equation}
Again, by \eqref{eq:orderisomIID}, 
the \emph{weak multipartite monotonicity of the second kind} \eqref{SecondLabVanish}
for mixed states,
\begin{equation}
\label{SecondLabVanishMix}
\vs{\beta}\preceq\vs{\alpha}\quad\Longleftrightarrow\quad
\bigl(\cnvroof{f}_{\vs{\beta}}=0\;\Rightarrow\;\cnvroof{f}_{\vs{\alpha}}=0\bigr),
\end{equation}
\end{subequations}
follows from this automatically.
Again, this provides the $P_{\text{II},\cnvroof{f}}$ set of functions
with the same hierarchical structure as that of $P_\text{II}$ in \eqref{eq:posetII}
and $P_{\text{II},\mathcal{D}}$ in \eqref{eq:statesIID}.

If the set of functions $P_{\text{II},f}$ in \eqref{eq:PIIf}
is multipartite monotonic of the second kind \eqref{eq:SecondLabMon},
(for example the $\vs{\alpha}$-entanglement entropy in \eqref{eq:EntEntII}),
then, thanks to \eqref{eq:cnvroofMon}, the set \eqref{eq:PIIfcnv} 
of their convex roof extension \eqref{eq:IndIIconvroof}
is also \emph{multipartite monotonic of the second kind} for mixed states,
\begin{subequations}
\begin{equation}
\label{eq:SecondLabMonMix}
\vs{\beta}\preceq\vs{\alpha}\quad\Longleftrightarrow\quad 
\cnvroof{f}_{\vs{\beta}}\geq \cnvroof{f}_{\vs{\alpha}}.
\end{equation}
So, in this case we also have
\begin{equation}
\struct{P_{\text{II},\cnvroof{f}},\geq} \cong \struct{P_\text{II},\preceq}.
\end{equation}
\end{subequations}

By this reasoning,
let us define the \emph{$\vs{\alpha}$-entanglement of formation},
as the convex roof extension of the $\vs{\alpha}$-entanglement entropy \eqref{eq:EntEntII} as
\begin{equation}
\label{eq:EntIIOF}
E^\text{oF}_{\vs{\alpha}} := \cnvroof{E}_{\vs{\alpha}}.
\end{equation}
This is the multipartite generalization of the entanglement of formation \eqref{eq:EntOF}.
Note that in this case,
\begin{equation}
\cnvroof{E}_{\vs{\alpha}}
= \cnvroof{\Bigl( \min_{\alpha\in\vs{\alpha}}\{E_\alpha\} \Bigr)}
\leq \min_{\alpha\in\vs{\alpha}}\{ \cnvroof{E}_\alpha \},
\end{equation}
which is a consequence of \eqref{eq:cnvroof.min}.
Note that
\begin{equation}
\label{eq:EntIIOFbound}
0\leq \cnvroof{E}_{\vs{\alpha}}(\pi) \leq \frac12 \ln \dim \mathcal{H} = \frac12 \sum_{a\in L} \ln \dim \mathcal{H}_a
\end{equation}
by \eqref{eq:cnvroofbound} and \eqref{eq:EntEntIIbound}.

\subsection{Examples}
\label{sec:MeasMulti.CnvRoofExamples}

Writing out some examples explicitly might not be useless here
(cf.~Section \ref{sec:EntMulti.secondExamples}).
Here we consider the $\vs{\alpha}$-entanglement of formation \eqref{eq:EntIIOF},
which is the convex roof extension
of the $\vs{\alpha}$-entanglement entropy \eqref{eq:EntEntII},
arising from the construction \eqref{eq:IndIIminMutInf}
based on the $\alpha$-entanglement entropy \eqref{eq:EntEntI},
arising from the construction \eqref{eq:IndISumNeumannMutInf}
using the von Neumann entropy \eqref{eq:NeumannEntr}.
Since the resulting functions are 
multipartite monotonic \eqref{eq:SecondLabMon} indicator functions \eqref{eq:defIndIID},
we can read off these relations from the
lattice $P_\text{II}$, which can be seen 
for the cases $n=2$ and $3$ 
in the upper-right part of Figures \ref{fig:labellattices2} and \ref{fig:labellattices3}.

For the \emph{bipartite} case, based on Section \ref{sec:MeasMulti.secondExamples},
we get back the content of Section \ref{sec:MeasBasics.examples},
\begin{align*}
E^\text{oF}_{\downset\{12\}} (\varrho) 
&=\cnvroof{E}_{\downset\{12\}} (\varrho) 
= \cnvroof{E}_{12}(\varrho)  = 0, \\
E^\text{oF}_{\downset\{1|2\}} (\varrho)
&=\cnvroof{E}_{\downset\{1|2\}} (\varrho)
= \cnvroof{E}_{1|2}(\varrho) = E^\text{oF}(\varrho).
\end{align*}
Note that the multipartite monotonicity \eqref{eq:SecondLabMonMix} holds,
$E^\text{oF}_{\downset\{1|2\}}(\varrho)\geq E^\text{oF}_{\downset\{12\}} (\varrho)$.
We have also the discriminance \eqref{eq:defIndIID},
\begin{align*}
\varrho&\in\mathcal{D}_{\downset\{12\}}&
\quad&\Longleftrightarrow&\quad E^\text{oF}_{\downset\{12\}}(\varrho)  &= 0,\\
\varrho&\in\mathcal{D}_{\downset\{1|2\}}&
\quad&\Longleftrightarrow&\quad E^\text{oF}_{\downset\{1|2\}}(\varrho) &= 0.
\end{align*}

For the \emph{tripartite} case, based on Section \ref{sec:MeasMulti.secondExamples},
\begin{align*} 
\begin{split} 
&E^\text{oF}_{\downset\{123\}}(\varrho)
= \cnvroof{E}_{\downset\{123\}}(\varrho)
= \cnvroof{E}_{123}(\varrho)  = 0,
\end{split} \\ 
\begin{split} 
&E^\text{oF}_{\downset\{1|23,2|13,3|12\}}(\varrho) 
=\cnvroof{E}_{\downset\{1|23,2|13,3|12\}}(\varrho) \\
&\quad\leq \min\{\cnvroof{E}_{1|23}(\varrho),\cnvroof{E}_{2|13}(\varrho),\cnvroof{E}_{3|12}(\varrho)\}, 
\end{split} \\ 
\begin{split} 
&E^\text{oF}_{\downset\{b|ac,c|ab\}}(\varrho)
=\cnvroof{E}_{\downset\{b|ac,c|ab\}}(\varrho)\\
&\quad\leq \min\{\cnvroof{E}_{b|ac}(\varrho),\cnvroof{E}_{c|ab}(\varrho)\}, 
\end{split} \\ 
\begin{split} 
&E^\text{oF}_{\downset\{a|bc\}}(\varrho)
=\cnvroof{E}_{\downset\{a|bc\}}(\varrho)
= \cnvroof{E}_{a|bc}(\varrho), 
\end{split} \\ 
\begin{split} 
&E^\text{oF}_{\downset\{1|2|3\}}(\varrho)
=\cnvroof{E}_{\downset\{a|bc\}}(\varrho)
= \cnvroof{E}_{1|2|3}(\varrho).
\end{split} 
\end{align*}
Note that the multipartite monotonicity \eqref{eq:SecondLabMonMix} holds,
$E^\text{oF}_{\downset\{1|2|3\}}(\varrho)\geq 
E^\text{oF}_{\downset\{a|bc\}}(\varrho)\geq
E^\text{oF}_{\downset\{a|bc,b|ac\}}(\varrho)\geq
E^\text{oF}_{\downset\{a|bc,b|ac,c|ab\}}(\varrho)\geq
E^\text{oF}_{\downset\{123\}}(\varrho)$.
We have also the discriminance \eqref{eq:defIndIID},
\begin{align*}
\varrho&\in\mathcal{D}_{\downset\{123\}}&
\quad&\Longleftrightarrow&\quad E^\text{oF}_{\downset\{123\}}(\varrho)&=0,\\
\varrho&\in\mathcal{D}_{\downset\{1|23,2|13,3|12\}}&
\quad&\Longleftrightarrow&\quad E^\text{oF}_{\downset\{1|23,2|13,3|12\}}(\varrho)&=0,\\
\varrho&\in\mathcal{D}_{\downset\{b|ac,c|ab\}}&
\quad&\Longleftrightarrow&\quad E^\text{oF}_{\downset\{b|ac,c|ab\}}(\varrho)&=0,\\
\varrho&\in\mathcal{D}_{\downset\{a|bc\}}&
\quad&\Longleftrightarrow&\quad E^\text{oF}_{\downset\{a|bc\}}(\varrho)&=0,\\
\varrho&\in\mathcal{D}_{\downset\{1|2|3\}}&
\quad&\Longleftrightarrow&\quad E^\text{oF}_{\downset\{1|2|3\}}(\varrho)&=0.
\end{align*}

\subsection{Level III: detection of the classes}
\label{sec:MeasMulti.classes}

By the use of the mixed $\vs{\alpha}$-indicators \eqref{eq:defIndIID},
one can detect also the classes \eqref{eq:classDef},
\begin{equation}
\varrho \in \mathcal{C}_{\vvs{\alpha}} \quad\Longleftrightarrow\quad
\left\{
\begin{aligned}
&\Bigl( \forall\vs{\alpha}\notin\vvs{\alpha}:\; f_{\vs{\alpha}}\neq0 \Bigr)
\quad\text{and}\\
&\Bigl( \forall\vs{\alpha}\in   \vvs{\alpha}:\; f_{\vs{\alpha}} =  0 \Bigr),
\end{aligned}
\right.
\end{equation}
which is a simple consequence of \eqref{eq:defIndIID}.
Because of the weak multipartite monotonicity of the second kind for 
mixed states \eqref{SecondLabVanishMix} 
(vanishing implications, satisfied by a system of indicator functions \eqref{eq:defIndIID} automatically),
it is enough to consider only the functions
labeled by $\min\vvs{\alpha}$ and $\max\cmpl{\vvs{\alpha}}$,
\begin{equation}
\varrho \in \mathcal{C}_{\vvs{\alpha}} \quad\Longleftrightarrow\quad
\left\{
\begin{aligned}
&\Bigl( \forall\vs{\alpha}\in\max\cmpl{\vvs{\alpha}}:\; f_{\vs{\alpha}}\neq0 \Bigr)
\quad\text{and}\\
&\Bigl( \forall\vs{\alpha}\in\min      \vvs{\alpha} :\; f_{\vs{\alpha}} =  0 \Bigr),
\end{aligned}
\right.
\end{equation}
cf.~\eqref{eq:classDefm}.

\subsection{Examples}
\label{sec:MeasMulti.classesExamples}

Writing out some examples explicitly might not be useless here
(cf.~Section \ref{sec:EntMulti.classesExamples}).
For the detection of the classes,
here we consider the $\vs{\alpha}$-entanglement of formation \eqref{eq:EntIIOF},
which is the convex roof extension
of the $\vs{\alpha}$-entanglement entropy \eqref{eq:EntEntII},
arising from the construction \eqref{eq:IndIIminMutInf}
based on the measures of the first kind \eqref{eq:IndISumNeumannMutInf} 
using the von Neumann entropy \eqref{eq:NeumannEntr}.

For the bipartite case, 
we get back the content of Section \ref{sec:MeasBasics.examples},
\begin{align*}
\varrho\in\mathcal{C}_{\upset\{\downset\{12\}\}}= \mathcal{C}_\text{ent}
\quad&\Longleftrightarrow\quad
\left\{
\begin{aligned}
E^\text{oF}_{\downset\{1|2\}}(\varrho)&\neq0
\quad\text{and}\\
E^\text{oF}_{\downset\{12\}}(\varrho)&=0,
\end{aligned}
\right.\\
\varrho\in\mathcal{C}_{\upset\{\downset\{1|2\}\}}= \mathcal{C}_\text{sep}
\quad&\Longleftrightarrow\quad
\left\{
\begin{aligned}
E^\text{oF}_{\downset\{1|2\}}(\varrho)&=0
\quad\text{and}\\
E^\text{oF}_{\downset\{12\}}(\varrho)&=0,
\end{aligned}
\right.
\end{align*}
for the detection of the \emph{separable} and \emph{entangled} state classes.

For the tripartite case,
the detection of the classes is
shown in Table \ref{tab:classes3f}.

\begin{table*}
\setlength{\tabcolsep}{6pt}
\begin{tabular}{|l||c|ccc|ccc|c|c|}
\hline
Class & 
\begin{sideways}$E^\text{oF}_{\downset\{1|2|3\}}(\varrho)$\end{sideways}  & 
\begin{sideways}$E^\text{oF}_{\downset\{a|bc\}}(\varrho)$\end{sideways}  & 
\begin{sideways}$E^\text{oF}_{\downset\{b|ac\}}(\varrho)$\end{sideways}  & 
\begin{sideways}$E^\text{oF}_{\downset\{c|ab\}}(\varrho)$\end{sideways}  & 
\begin{sideways}$E^\text{oF}_{\downset\{b|ac,c|ab\}}(\varrho)$\end{sideways}  & 
\begin{sideways}$E^\text{oF}_{\downset\{a|bc,c|ab\}}(\varrho)$\end{sideways}  &
\begin{sideways}$E^\text{oF}_{\downset\{a|bc,b|ac\}}(\varrho)$\end{sideways}  &
\begin{sideways}$E^\text{oF}_{\downset\{1|23,2|13,3|12\}}(\varrho)\;$\end{sideways}  &
\begin{sideways}$E^\text{oF}_{\downset\{123\}}(\varrho)$\end{sideways} \\ 
\hline
\hline
$\mathcal{C}_{\upset\{ \downset\{123\} \}}$                                                    & $>0$ & $>0$ & $>0$ & $>0$ & $>0$ & $>0$ & $>0$ & $>0$ & $=0$ \\ 
\hline
$\mathcal{C}_{\upset\{ \downset\{1|23,2|13,3|12\} \}}$                                         & $>0$ & $>0$ & $>0$ & $>0$ & $>0$ & $>0$ & $>0$ & $=0$ & $=0$  \\
$\mathcal{C}_{\upset\{ \downset\{b|ac,c|ab\} \}}$                                              & $>0$ & $>0$ & $>0$ & $>0$ & $=0$ & $>0$ & $>0$ & $=0$ & $=0$  \\
$\mathcal{C}_{\upset\{ \downset\{a|bc,b|ac\},\downset\{a|bc,c|ab\} \}}$                        & $>0$ & $>0$ & $>0$ & $>0$ & $>0$ & $=0$ & $=0$ & $=0$ & $=0$  \\
$\mathcal{C}_{\upset\{ \downset\{1|23,2|13\},\downset\{1|23,3|12\},\downset\{2|13,3|12\} \}}$  & $>0$ & $>0$ & $>0$ & $>0$ & $=0$ & $=0$ & $=0$ & $=0$ & $=0$  \\
$\mathcal{C}_{\upset\{ \downset\{a|bc\} \}}$                                                   & $>0$ & $=0$ & $>0$ & $>0$ & $>0$ & $=0$ & $=0$ & $=0$ & $=0$  \\
$\mathcal{C}_{\upset\{ \downset\{a|bc\},\downset\{b|ac,c|ab\} \}}$                             & $>0$ & $=0$ & $>0$ & $>0$ & $=0$ & $=0$ & $=0$ & $=0$ & $=0$  \\
$\mathcal{C}_{\upset\{ \downset\{b|ac\},\downset\{c|ab\} \}}$                                  & $>0$ & $>0$ & $=0$ & $=0$ & $=0$ & $=0$ & $=0$ & $=0$ & $=0$  \\
$\mathcal{C}_{\upset\{ \downset\{1|23\},\downset\{2|13\},\downset\{3|12\} \}}$                 & $>0$ & $=0$ & $=0$ & $=0$ & $=0$ & $=0$ & $=0$ & $=0$ & $=0$  \\
\hline
$\mathcal{C}_{\upset\{ \downset\{1|2|3\}\}}$                                                   & $=0$ & $=0$ & $=0$ & $=0$ & $=0$ & $=0$ & $=0$ & $=0$ & $=0$  \\
\hline
\end{tabular}
\caption{Detection of the partial separability classes of mixed tripartite states
by indicator functions (cf., Table \ref{tab:classes3}).}
\label{tab:classes3f}
\end{table*}

\section{Summary, remarks and open questions}
\label{sec:Summ}

In this work, 
we have considered the entanglement
\emph{classification} and \emph{quantification} problem
for multipartite mixed states.

\subsection{On the classification of multipartite entanglement}
\label{sec:Summ.class}

In the \textit{first part} of the paper 
we have constructed the partial separability classification for multipartite quantum systems 
(Section \ref{sec:EntMulti}).
We have worked out 
the hierarchical structure of different kinds of partial separability $P_\text{II}$,
which has turned out to be 
the down-set lattice of the lattice of the partitions of the subsystems $P_\text{I}$
(Sections \ref{sec:EntMulti.first} and \ref{sec:EntMulti.second}),
and also the structure of the entanglement classes $P_\text{III}$,
which has turned out to be also hierarchical,
being the up-set lattice of the lattice above
(Section \ref{sec:EntMulti.classes}).
The hierarchy of the classes has turned out to be related to the LOCC convertibility:
If a state from a class can be mapped into another one,
then that class can be found higher in the hierarchy.

\setcounter{txtitem}{0}

Now, we list some remarks and open questions.

\txtitem{} The partial separability classification is a 
more fine-grained classification than the Seevinck-Uffink classification \cite{SeevinckUffinkMixSep},
which is a more fine-grained classification than the D{\"u}r-Cirac-Tarrach classification \cite{DurCiracTarrach3QBMixSep},
while it is
more coarse-grained classification than the SLOCC classification \cite{BennettetalEquivalences}, 
which is more coarse-grained classification than the LOCC classification \cite{BennettetalEquivalences,DurVidalCiracSLOCC3QB}.
It considers only the partial separability properties,
but it does this in the fullest detail.
The more coarse-grained 
(Seevinck-Uffink, D{\"u}r-Cirac-Tarrach classification) classifications
and the classifications based on 
$k$-separability \cite{AcinetalMixThreeQB,GuhneTothMultipartite,SeevinckUffinkMixSep} 
and $k$-producibility \cite{SeevinckUffinkMultipartite,GuhneTothMultipartite,TothGuhneMultipartite}
can naturally be described in this framework.

\txtitem{}\label{itm:Summ.Class.ensemble} 
We can elucidate the meaning of the different kinds of state sets
arising in the classification structure in a unified way, 
using the standard ensemble approach of statistical physics.\\
\textit{States in $\mathcal{D}$:} 
We are uncertain about the (pure) state, by which the system is described
(Section \ref{sec:EntMulti.zeroth}).\\
\textit{States in $\mathcal{D}_\alpha$:}
We are uncertain about the (pure) state, by which the system is described,
but we are certain about the split with respect to which the state is separable
(Section \ref{sec:EntMulti.first}).\\
\textit{States in $\mathcal{D}_{\vs{\alpha}}$:}
We are uncertain about the (pure) state, by which the system is described,
and we are also uncertain about the split with respect to which the state is separable,
but we are certain about the possible splits with respect to which the state is separable
(Section \ref{sec:EntMulti.second}).\\
\textit{States in $\mathcal{C}_{\vvs{\alpha}}$:}
We are uncertain about the (pure) state, by which the system is described,
and we are also uncertain about the split with respect to which the state is separable, 
but we are certain about the possible splits with respect to which the state is separable,
and we are also certain about the possible splits with respect to which the state is not separable
(Section \ref{sec:EntMulti.classes}).

\txtitem{}\label{itm:Summ.Class.nonempty} The nonemptiness of the classes was only conjectured (Conjecture \ref{conj:classdef}).
More fully, we could not give necessary and sufficient condition 
for the nonemptiness of the classes in the purely algebraic language of labels.
Probably, methods from geometry or matrix analysis would be needed to solve this puzzle
(Section \ref{sec:EntMulti.classes}).
For a constructive proof,
it would be interesting to construct representative states 
for all classes $\mathcal{C}_{\vvs{\alpha}}$.
For this, it can be helpful to consider not the full state space,
but only some special subsets,
which can be generic enough for intersecting with a sufficient number of different classes,
such as the noisy GHZ-W mixture \cite{SzalaySepCrit},
or GHZ-symmetric states \cite{EltschkaSiewertGHZSymm},
or the magic simplex \cite{UchidaEntEntMagicGHZ},
or mixtures of symmetric Dicke states \cite{WolfeYelinNqbDickeMix}.

\txtitem{} A more challenging issue is
to find utilization for the quantum states of the different classes.
It seems to be promising to find or develop information theoretic tasks, 
such as multipartite secret sharing protocols \cite{LiangetalAnonymous}.

\txtitem{} In close connection with item \ref{itm:Summ.Class.nonempty},
a further geometry-related conjecture could be drafted about the \emph{nonempty classes:}
They are of \emph{nonzero measure.}
It is known in the bipartite case that
the set of separable states is of nonzero measure 
\cite{AcinetalMixThreeQB,BengtssonZyczkowski},
which might motivate this conjecture.

\txtitem{} Note that in the classification,
Levels I and II of the construction are
related to LOCC closedness \eqref{eq:LOCCcloseI} and \eqref{eq:LOCCcloseII}
(state sets $\mathcal{D}_{\vs{\alpha}}$ are closed under LOCC),
while Level III of the construction is 
related to LOCC convertibility \eqref{eq:LOCCconvhierarchy}
(if a state from a class $\mathcal{C}_{\vvs{\beta}}$ can be mapped into $\mathcal{C}_{\vvs{\alpha}}$,
then $\vvs{\beta}\preceq\vvs{\alpha}$).

\txtitem{} In Level III of the construction,
we have the lattice $P_\text{III}$ of class labels \eqref{eq:posetIII}.
We could partially clarify the meaning of the poset $\struct{P_\text{III},\preceq}$;
it is related to the LOCC convertibility 
(see \eqref{eq:LOCCconvhierarchy} in Section \ref{sec:EntMulti.classes}).
On the other hand, being an up-set lattice,
the meet and the join \eqref{eq:posetIII} arise naturally,
however, their meaning is not clear.
On the other hand, we have the poset $\struct{P_{\text{III},\mathcal{C}},\geq_\text{s}}$ of classes \eqref{eq:posetIIIC},
with the ordering $\geq_\text{s}$ related to the (strong) LOCC convertibility \eqref{eq:defLOCCconv.strong}.
\emph{Can a meet and a join also be defined in some motivated way here?}
\emph{Or, what is the meaning of the class 
corresponding to the meet or join of the labels of two classes?}
If Conjecture \ref{conj:LOCCconvhierarchyfull} holds,
then these have the meaning of 
greatest lower and least upper bounds with respect to LOCC convertibility.

\txtitem{} The most important open question is
to prove Conjecture \ref{conj:LOCCconvhierarchyfull} in Level III of the construction,
which would establish a stronger connection between the 
class hierarchy \eqref{eq:posetIII} and the strong LOCC convertibility \eqref{eq:defLOCCconv.strong}.
Based on this, one could compare the different well-defined partial separability properties,
that is, we could say that
states in a given class $\mathcal{C}_{\vvs{\beta}}$
are ``more entangled'' than states in class $\mathcal{C}_{\vvs{\alpha}}$
if $\vvs{\beta}\preceq\vvs{\alpha}$.

\txtitem{} Can a gradation be defined for the lattice $P_\text{III}$?
Because of \eqref{eq:LOCCconvhierarchy},
that would lead to an integer-valued measure of entanglement 
(monotonically decreasing with respect to LOCC).

\txtitem{}\label{itm:Summ.Class.subsys} Note that the partial separability properties
cannot give a full answer for the entanglement \emph{inside} the subsystems.
Of course, if $\varrho\in\mathcal{D}$ is separable with respect to the split $\alpha$,
then $\tr_{\cmpl{K}}\varrho\in\mathcal{D}_K$ is separable with respect to the split
which can be obtained by dropping the elementary subsystems $a\in\cmpl{K}$ from $\alpha$.
(That is, if $\varrho\in\mathcal{D}_\alpha$,
then $\tr_{\cmpl{K}}\varrho\in\mathcal{D}_{\alpha|_K}$, where
$\alpha|_K=\set{K'\cap K\neq\emptyset}{K'\in\alpha}$.)
However, even if we know the class of $\varrho$, 
we cannot give the class of $\tr_{\cmpl{K}}\varrho$.
A well-known example for this is 
the case of the GHZ and W states of three qubits \cite{DurVidalCiracSLOCC3QB},
both of them are tripartite entangled.
The bipartite subsystems of the GHZ state are separable;
that is, if $\varrho=\cket{\text{GHZ}}\bra{\text{GHZ}}
\in\mathcal{C}_{\upset\{\downset\{123\}\}}\subset\mathcal{D}_{\downset\{123\}}$,
then $\tr_3 \varrho\in\mathcal{C}_{\upset\{\downset\{1|2\}\}}=\mathcal{C}_\text{sep}
=\mathcal{D}_{\downset\{1|2\}}$;
while bipartite subsystems of the W state are entangled;
that is, if $\varrho=\cket{\text{W}}\bra{\text{W}}
\in\mathcal{C}_{\upset\{\downset\{123\}\}}\subset\mathcal{D}_{\downset\{123\}}$,
then $\tr_3 \varrho\in\mathcal{C}_{\upset\{\downset\{12\}\}}=\mathcal{C}_\text{ent}
\subset\mathcal{D}_{\downset\{12\}}$.

\txtitem{} The partial separability classification is about the question:
``From which kinds of pure entangled states can a given state be \emph{mixed}?''
Another question \cite{DurCirac3QBMixSep}, 
which is also important from the point of view of quantum information,
but which we have not considered, is:
``Which kinds of pure entangled states can be \emph{distilled out} from a given state?''

\subsection{On the quantification of multipartite entanglement}
\label{sec:Summ.quant}

In the \emph{second part} of the paper,
we have constructed entanglement measures for multipartite quantum systems 
(Section \ref{sec:MeasMulti}).
Besides the usual entanglement monotonicity and discriminance,
we have introduced the multipartite monotonicity,
as a plausible property,
which endows the set of multipartite entanglement measures
with the same hierarchical structure as the partial separability shows.
We have succeeded in constructing a hierarchy of entanglement measures
satisfying these requirements 
(Sections \ref{sec:MeasMulti.second} and \ref{sec:MeasMulti.CnvRoof}),
which is the direct generalization 
of the entanglement entropy for pure states 
and the entanglement of formation for mixed states
(see in \eqref{eq:EntEntII} and \eqref{eq:EntIIOF}).
These measures have information-geometrical meaning,
related to the statistical distinguishability.
A side result is 
another, not multipartite monotonic generalization 
of the entanglement entropy for pure states 
and the entanglement of formation for mixed states
(see in \eqref{eq:IndIImagicg}).

Now, we list some remarks and open questions.

\txtitem{} There are wide-ranging possibilities 
for the generalization of the results,
from which we can conclude
that the entanglement monotonicity together with the discriminance property
does not yield a condition too strong.
The multipartite monotonicity is, however, more demanding.

\txtitem{} The \emph{discriminance,} or \emph{indicator properties} 
\eqref{eq:defIndI} and \eqref{eq:defIndII}, could be omitted, of course,
as is done sometimes in the literature.
This is a part of the construction which can be detached from the 
part dealing with the entanglement monotonicity.
However, we think that the indicator properties are important.
If a quantity is zero for some entangled states,
then it measures not the entanglement, but something else
(which can also be related to entanglement, of course).

\txtitem{} The \emph{multipartite monotonicity properties,} on the other hand,
may or may not be considered plausible enough.
The \emph{weak multipartite monotonicity properties} (vanishing implications) 
\eqref{FirstLabVanish} and \eqref{SecondLabVanish}
are direct consequences of the 
indicator properties \eqref{eq:defIndI} and \eqref{eq:defIndII},
and reflect the hierarchy of multipartite entanglement in a weak sense.
The \emph{multipartite monotonicity properties} \eqref{eq:FirstLabMon} and \eqref{eq:SecondLabMon},
on the other hand, are stronger requirements.

In Level I of the construction, 
the meaning of the multipartite monotonicity seems to be clear:
Entanglement with respect to a coarser partition
cannot be stronger than entanglement with respect to a finer one.
Since, e.g., the tripartite entanglement is considered to be a more powerful resource
than the bipartite entanglement \cite{BorstenGHZW},
one feels that a state can contain a smaller amount of that
than of the bipartite entanglement.

In Level II of the construction, 
the meaning of the multipartite monotonicity is not so clear.
We have, on the one hand, the mathematical analogy with Level I:
entanglement higher in the hierarchy 
cannot be stronger than entanglement lower in there.
(Since the sublattice formed by the principal elements of $P_\text{II}$
is isomorphic to $P_\text{I}$, at least for this sublattice the meaning is clear.)
We have, on the other hand, an interpretation from a statistical approach:
Entanglement, \emph{as a resource} (the same, unified ``notion'' for all splits), is weaker for a given state,
if we are more uncertain about the split with respect to which the state is separable,
and we are certain only about the possible splits with respect to which the state is separable 
(cf.~item \ref{itm:Summ.Class.ensemble}).
Note that if we follow the particular way of construction
based on the statistical distinguishability in information-geometry \eqref{eq:geomcorrII},
then the multipartite monotonicity properties follow automatically.
In this case, we have another interpretation for this,
a third one, coming from information geometry.
Entanglement 
 (the same, unified ``notion'' for all splits),
regarded to be the distinguishability from a subset of states,
is lower for a given state,
if one allows a bigger subset from which the distinguishability is measured.

In summary, using the multipartite monotonicity, we attempt to grasp 
the hierarchy of multipartite entanglement with the measures.
This seems to make the entanglement measures with respect to different splits
(or different down-sets of splits in Level II)
be the manifestations of some ``unified'' notion of entanglement.
This is why we take multipartite monotonicity to be very serious.

\txtitem{} The multipartite monotonicity \eqref{eq:FirstLabMon} and \eqref{eq:SecondLabMon}
means a set of bounds among the measures 
characterizing the multipartite entanglement in the whole system.
(These fall between the global bounds 
\eqref{eq:EntEntIbound},
\eqref{eq:EntEntIIbound}, and
\eqref{eq:EntIIOFbound}.)
It would also be important to obtain bounds among these measures and
the measures characterizing the multipartite entanglement \emph{inside} multipartite subsystems
(cf.~item \ref{itm:Summ.Class.subsys} in Section \ref{sec:Summ.class}),
leading to ``monogamylike'' inequalities \cite{BennettetalMixedStates,CKWThreetangle,EltschkaSiewertMonogamyLorentz}
(not necessarily linear ones).
Finding such bounds is also important 
not only for multipartite entanglement measures,
but also for multipartite quantum correlation measures,
and even for multipartite classical correlation measures.
Knowing these bounds 
would highly improve
our knowledge of correlation and entanglement in quantum systems.
Note that such bounds can follow from known entropic inequalities
\cite{LindenWintervNEntropyIneq,LindenMosonyiWinterREntropyIneq,CadneyetalRankIneq}.

\txtitem{}
A recent approach for the investigation of the structure of multipartite entanglement
is based on the \emph{entropy vector formalism}
 \cite{HuberdeVincenteMultipartEnt,HuberPerarnauLlobetdeVincenteEntropyVector}.
There the convex roof extensions of subsystem entropies are considered.
Using \eqref{eq:cnvroof.sum}, one can obtain
inequalities between the $\alpha$-entanglement of formations and
the sums of the elements of the entropy vector.

\txtitem{} The construction of multipartite entanglement measures is transparent:
The way of that leads parallel to the construction of the partial separability,
so the connection with the partial separability hierarchy is clear.
However, another way has shown up in Section \ref{sec:MeasMulti.second},
different from the one we have followed, based on the $K|\pcmpl{K}$-entanglement measures
(Sections \ref{sec:MeasMulti.first} and \ref{sec:MeasMulti.second}),
which leads to the same result, but shows a deeper motivation; see the next two items.

\txtitem{} Apart from its beautiful properties,
what are the principles making the entropy of the subsystem 
a good choice for measuring the entanglement in \emph{bipartite} pure states?
There are several ways for introducing the entanglement entropy
for measuring bipartite entanglement.

The most fundamental approach is based upon that 
entanglement is considered to be a resource in quantum information theory,
and, related to that, \emph{for bipartite pure states,
the entanglement cost and the distillable entanglement equal to the entropy of entanglement.}
These results are based on coding theory and quantum communication,
which are based on that there is a maximally entangled state 
with respect to LOCC (unique, up to local unitaries) in the bipartite case.
Since in multipartite systems there is no unique maximally entangled state,
this approach cannot be generalized for more than two subsystems.
(A great overview of this can be found in Section 12.5 of \cite{NielsenChuang}.
For some recent results on maximally entangled state sets in the multipartite scenario,
see \cite{CarleetalPurifMaxEnt,SpeedeVincenteKrausRemoteEntPrep,deVincenteSpeeKrausMES,SchwaigerVolumeEntMeas,SpeedeVincenteKrausEntManip}.)

Another approach is that,
since a bipartite state is entangled if and only if its subsystems are mixed \eqref{eq:PsepDecide},
then it is at least plausible to think that
\textit{``the more mixed the marginals, the more entangled is the state.''}
Then a measure of mixedness (entropies) of the subsystem should lead 
to a motivated measure of entanglement of the whole system.
(A slight shortcoming of this reasoning is that not all entropies do the job.
It follows from Theorem \ref{thm:pureEntMon} that only the concave entropies
\eqref{eq:entrConcavity} work for this.)
However, it is not clear how to generalize this approach for the multipartite scenario.
One possibility is that if one consider the entanglement with respect to a split,
then one simply sums up the measures of mixedness of the subsystems with respect to that split,
and one gets Level I of our construction \eqref{eq:IndISumNeumann}.
However, it is not clear how to step further.

There is a third approach, which coincides with the construction
we have carried out (Sections \ref{sec:MeasMulti.first} and \ref{sec:MeasMulti.second}),
but based on different principles;
see the next item.

\txtitem{} We have seen that when we write 
the sum of the von Neumann entropies of the subsystems in a split \eqref{eq:IndISumNeumann},
we actually have a classical correlation measure (with respect to that split)
for pure states \eqref{eq:IndISumNeumannMutInf}.
This reasoning enlightens also the meaning of entanglement itself,
which holds also in the bipartite case \eqref{eq:EntEnt}.
In classical probability theory,
pure states are always uncorrelated,
so if in the quantum case a pure state shows correlation,
then this correlation is considered to be of quantum origin,
and this correlation is \emph{defined to be the entanglement.}
From this point of view, it is plausible to think that
\textit{the quantum versions of classical correlation measures
applied to pure quantum states are pure entanglement measures}
(then they should be extended to mixed states)
in both the bipartite and the multipartite scenarios.
However, entanglement monotonicity should be checked;
it does not seem to be fulfilled automatically.

\txtitem{} As we can see from this reasoning,
one can find other ways for forming the first kind hierarchy of entanglement measures.
Using some distance or divergence $D^\text{gen}$,
one can directly have
 $E^\text{gen}_\alpha(\pi) 
= D^\text{gen}(\pi,\otimes_{K\in\alpha} \pi_K)$,
or one can also follow the more fundamental way
and define a generalized geometric measure of correlation 
(generalized $\alpha$-mutual information)
$I_\alpha^\text{gen}(\varrho)
= \min_{\forall K\in\alpha:\,\omega_K\in\mathcal{D}_K} 
D^\text{gen}(\varrho,\otimes_{K\in\alpha} \omega_K)$,
which leads to the candidate of pure state entanglement measure
$E^\text{gen}_\alpha(\pi) = I_\alpha^\text{gen}(\pi)$.
(The $\alpha$-discriminance \eqref{eq:defIndI} is automatically satisfied by this construction.
On the other hand, it can happen, of course, 
that this construction does not lead to a closed form, in general.)
Here one has several choices again 
for the role of $D^\text{gen}$
(see Sections 12, 13,~and 14~in \cite{BengtssonZyczkowski}).
One can use
the trace-distance, which has also a statistical meaning,
or other distance measures, being less motivated,
or R\'enyi, Tsallis generalizations of the Kullback-Leibler divergence.
We note again that entanglement monotonicity
does not seem to be fulfilled automatically.
This leads to the next question.

\txtitem{} 
\textit{What are the basic properties of correlation measures,
leading to entanglement measures in the construction in the previous item?}

For \emph{entanglement measures,} we have the two main requirements,
the \emph{(entanglement) discriminance,} and the \emph{LOCC monotonicity.}
This latter is composed of two well-understood parts:
\textit{First,} as any correlation, entanglement does not increase locally,
and, 
\textit{second,} while classical correlation does, 
entanglement does not increase by classical communication 
(``classical interaction'') either.

For \emph{correlation measures,} we could also formulate two main requirements,
a \emph{correlation-discriminance} (correlation indicator property):
a correlation measure vanishes exactly for uncorrelated states;
and a (nonincreasing) \emph{monotonicity under LO} (local operations):
a correlation does not increase locally.

Now, \textit{are these properties sufficient
in the construction above?}
That is, if we have an $I_\alpha^\text{gen}(\varrho)$
LO-monotone correlation indicator,
then 
is $E^\text{gen}_\alpha(\pi) = I_\alpha^\text{gen}(\pi)$ an
LOCC-monotone entanglement indicator?

\txtitem{} In the last step of the construction of the entanglement measures,
we have used convex roof extension to step from pure states to mixed states
(see Section \ref{sec:MeasMulti.CnvRoof}).
Convex roof extensions are hard to evaluate.
However, sufficiently motivated
mixed state entanglement measures,
such as entanglement cost, distillation entanglement, or squashed entanglement, 
always seem to be hard to evaluate,
since they always contain an optimization problem
\cite{Horodecki4,EltschkaSiewertEntMeas,HorodeckiEntMeas,PlenioVirmaniEntMeas}.
Among these, the optimization task in the convex roof extension
seem to be the simplest one.
These optimization problems have no solutions in a closed form in general cases.
There are few explicit analytic solutions 
for the convex roof extension \cite{HillWoottersConc,WoottersConc,
LohmayeretalConvRoofTangle,OsterlohetalConvRoofTangle,EltschkaetalConvRoofTangle}.

\txtitem{} An advantage of the convex roof extension
is that it works independently of the dimensions of the subsystems,
so the mixed state entanglement measures by that work for arbitrary dimensions.
However, the numerical optimization depends strongly on the rank of the state,
which can be high if the dimension is high,
resulting in extremely slow convergence, 
which makes the numerical task infeasible in practice, even for small systems.
A recent result is that 
computing a large class of bipartite entanglement measures 
(for example the entanglement of formation) is NP-hard \cite{HuangNPcomplete},
and the same seems to hold for the $\vs{\alpha}$-entanglement of formation.
It is a common belief that some kind of difficult
optimization task cannot be circumvented if one deals with the entanglement of
mixed states.

\txtitem{} It is then an important research direction for practical calculations
to obtain upper and lower bounds for convex roof measures,
the evaluation of which is feasible \cite{EltschkaSiewertGHZSymm,RodriquesDattaLoveBoundingMixMeasures,TothMoroderGuhneEvalConvRoof}.

\txtitem{} Entanglement entropy \eqref{eq:EntEnt} is additive (extensive); 
then so is $\vs{\alpha}$-entanglement entropy \eqref{eq:EntEntII}.
The conjecture about the additivity of the entanglement of formation \eqref{eq:EntOF}
is proven to be false \cite{HastingsNotAdd};
then so is the additivity of the $\vs{\alpha}$-entanglement of formation \eqref{eq:EntIIOF}.

\txtitem{} Since the convex roof extensions of semialgebraic functions
are known to be semialgebraic functions \cite{PetiPriv,BochnakCosteRoy},
it can be useful to use LU-invariant homogeneous polynomials \cite{HWLUA,HWWLUA,PetiLUA1,PetiLUA23,SzDeg6} 
for the role of entanglement measures in Level I of the construction,
which leads to semialgebraic functions in Level II.
This holds, in particular, if one sets out from Tsallis entropy for integer $q\geq2$
in the construction \eqref{eq:fK}-\eqref{eq:IndISum}-\eqref{eq:IndIIqMean}-\eqref{eq:IndIIconvroof}.

\txtitem{} Convex roof extension preserves all the required properties of functions
(see Section \ref{sec:MeasMulti.CnvRoof}).
Are there any other extension methods to step from pure states to mixed states?

\txtitem{} Investigating the correlation and entanglement pattern
in many-body states can be of practical importance
in the properties of strongly correlated systems \cite{AmicoetalEntManybody,GabrielMurgHiesmayrPSMPS} on the one hand
and also in optimizing numerical methods in many-body physics
\cite{LegezaBachBook,SzalayIntJQuantChemReview,BarczaetalEntPatterns} on the other.

\begin{acknowledgments}
Discussions with \emph{Jens Siewert}, \emph{Frank Verstraete}, \emph{Jens Eisert},
\emph{Christian Krumnow}, \emph{Max Pfeffer}, \emph{{\"O}rs Legeza}, \emph{P{\'e}ter Vrana} and \emph{Istv{\'a}n Kov{\'a}cs} 
are gratefully acknowledged.
This project was supported financially by 
the \textit{New Sz\'echenyi Plan of Hungary} (project ID: T\'AMOP-4.2.2.B-10/1–2010-9),
the \textit{Hungarian Scientific Research Fund} (project ID: OTKA-K100908)
and 
the \textit{``Lend\"ulet''} program of the Hungarian Academy of Sciences.
Last but not least,
I thank my wife, \emph{M{\'a}rta}, for her everlasting love, patience and support :).
\end{acknowledgments}

\appendix

\section{On the lattice structure of the classification}
\label{app:lattices}

\subsection{Posets and lattices: basics}
\label{app:lattices.gen}

Here we list some definitions and notations in \emph{order theory},
following \cite{DaveyPriestley} and \cite{Stanley}.

A \emph{partially ordered set}, or \emph{poset}, $\struct{P,\preceq}$
is a set $P$ endowed with a \emph{partial order} $\preceq$,
being a \emph{reflexive}, \emph{antisymmetric} and \emph{transitive} relation,
that is,
for all $x,y,z\in P$,
\begin{subequations}
\label{eq:poax}
\begin{align}
\label{eq:poax.refl}
x&\preceq x, \\
\label{eq:poax.antisymm}
x&\preceq y \;\text{and}\; y\preceq x \quad\Longrightarrow\quad x=y,\\
\label{eq:poax.trans}
x&\preceq y \;\text{and}\; y\preceq z \quad\Longrightarrow\quad x\preceq z.
\end{align}
\end{subequations}

For the posets $\struct{P,\preceq}$ and $\struct{Q,\preceq}$,
a map $\phi:P\to Q$ is an \emph{order isomorphism} when
\begin{equation}
x\preceq y\quad\Longleftrightarrow\quad \phi(x)\preceq\phi(y).
\end{equation}
It follows easily from the reflexivity and antisymmetry of the partial order that
such a map is bijective,
\begin{equation}
\begin{split}
\phi(x)=\phi(y) \quad&\Longleftrightarrow\quad
 \phi(x)\preceq\phi(y) \;\text{and}\; \phi(y)\preceq\phi(x) \\
\quad&\Longleftrightarrow\quad x\preceq y \;\text{and}\; y\preceq x\\
\quad&\Longleftrightarrow\quad x = y.
\end{split}
\end{equation}

A poset $P$ may have a \emph{bottom} and a \emph{top} element,
denoted with $\bot,\top\in P$, if 
\begin{equation}
\forall x\in P:\quad\bot\preceq x\preceq\top.
\end{equation}
(If the bottom and top exist, then they are unique ones,
which is the consequence of the antisymmetry \eqref{eq:poax.antisymm} of the ordering.)

One can define the \emph{minimal} and \emph{maximal} elements of a subset $Q\subseteq P$ as
\begin{subequations}
\begin{align}
\min Q &= \bigset{x\in Q}{(y\in Q\;\text{and}\;y\preceq x)\;\Rightarrow\;y=x},\\
\max Q &= \bigset{x\in Q}{(y\in Q\;\text{and}\;x\preceq y)\;\Rightarrow\;y=x}.
\end{align}
\end{subequations}

A subset $Q\subseteq P$ is a \emph{down-set}, or \emph{order ideal}, if
\begin{subequations}
\begin{equation}
\label{eq:downset}
 \bigl( x\in Q \;\text{and}\; y\preceq x\bigr) \quad \Longrightarrow\quad y\in Q
\end{equation}
(it is ``closed downwards'').
The set of all down-sets of $P$ is denoted with $\mathcal{O}_\downarrow(P)$.
Similarly, a subset $Q\subseteq P$ is an \emph{up-set}, or \emph{order filter}, if
\begin{equation}
\label{eq:upset}
 \bigl( x\in Q \;\text{and}\; x\preceq y\bigr) \quad \Longrightarrow\quad y\in Q
\end{equation}
\end{subequations}
(it is ``closed upwards'').
The set of all up-sets of $P$ is denoted with $\mathcal{O}_\uparrow(P)$.
For a subset $Q\subseteq P$ one can define 
\begin{subequations}
\begin{align}
\label{eq:downQ}
\downset Q &= \bigset{x\in P}{\exists y\in Q: x\preceq y},\\
\label{eq:upQ}
\upset   Q &= \bigset{x\in P}{\exists y\in Q: y\preceq x},
\end{align}
\end{subequations}
which are a down-set and an up-set, respectively.

The \emph{greatest lower bound} or \emph{meet}, $x \wedge y\in P$,
and 
the  \emph{least upper bound} or \emph{join}, $x\vee y\in P$,
of the elements $x$ and $y$ in a poset
are defined, respectively, as
\begin{subequations}
\label{eq:wvGen}
\begin{align}
\label{eq:wvGen.wedge}
\bigl(x\wedge y \preceq x,y\bigr)
\;\text{and}\;
\bigl( z \preceq x,y \;\Rightarrow\; z \preceq x\wedge y \bigr),\\
\label{eq:wvGen.vee}
\bigl(x,y \preceq x\vee y\bigr)
\;\text{and}\;
\bigl( x,y \preceq z \;\Rightarrow\;  x\wedge y \preceq z \bigr).
\end{align}
\end{subequations}
A poset $P$ is called a \emph{lattice},
if, for all $x,y\in P$ pairs, $x \wedge y$ and $x \vee y$ exist.
A poset $P$ is called a \emph{complete lattice},
if, for all $Q\subseteq P$ subsets, $\bigwedge Q$ and $\bigvee Q$ exist.
Every finite lattice is complete.
A finite lattice always has bottom and top elements.
If only the meet or only the join can be defined, then the poset is called
a \emph{meet-semilattice} or \emph{join-semilattice}, respectively.
A finite meet-semilattice always has a bottom element,
a finite join-semilattice always has a top element.

Let $P$ be a finite meet-semilattice having a top element $\top$.
Then the join can be defined as
\begin{subequations}
\begin{equation}
\label{eq:joinconstr}
x\vee y = \bigwedge\upset\{x,y\},
\end{equation}
so $P$ is a lattice
(see Proposition 3.3.1~in \cite{Stanley}).
Dually,
let $P$ be a finite join-semilattice having a bottom element $\bot$.
Then the meet can be defined as
\begin{equation}
\label{eq:meetconstr}
x\wedge y = \bigvee\downset\{x,y\},
\end{equation}
\end{subequations}
so $P$ is a lattice.

Closing this section, we recall some examples \cite{DaveyPriestley,Stanley},
which are used in the constructions in the body of the text.
For a set $X$, its power set $2^X=\{A\subseteq X\}$
is a complete lattice with the ordering $\subseteq$ (inclusion),
the meet $\cap$ (intersection), the join $\cup$ (union),
the bottom element $\bot=\emptyset$, and the top element $\top=X$.
The $\mathcal{O}_\downarrow(P)$ set of all down-sets (ideals) of a poset $P$ 
is a lattice, 
called \emph{down-set lattice},
with the ordering $\subseteq$ (inclusion),
the meet $\cap$ (intersection), the join $\cup$ (union),
the bottom element $\bot=\emptyset$, and the top element $\top=P$.
Dually,
the $\mathcal{O}_\uparrow(P)$ set of all up-sets (filters) of a poset $P$ 
is a lattice, 
called \emph{up-set lattice},
with the ordering $\subseteq$ (inclusion),
the meet $\cap$ (intersection), the join $\cup$ (union),
the bottom element $\bot=\emptyset$, and the top element $\top=P$.
The $\mathcal{O}_\downarrow(P)\setminus\{\emptyset\}$ 
(or $\mathcal{O}_\uparrow(P)\setminus\{\emptyset\}$) set of 
all nonempty down-sets (or nonempty up-sets) of a poset $P$
is not a lattice, in general.
However, if $P$ is a lattice, then there is a bottom (and a top) element $\bot \in P$ ($\top \in P$),
and $\{\bot\}$ (or $\{\top\}$) is the bottom element of $\mathcal{O}_\downarrow(P)\setminus\{\emptyset\}$
(or $\mathcal{O}_\uparrow(P)\setminus\{\emptyset\}$),
which makes it a lattice.

\subsection{Construction of the labels of the first kind by bipartitions}
\label{app:lattices.dualI}

For the proof of \eqref{eq:partitionDual}, we need by \eqref{eq:wvGen.wedge} that
(i) $\alpha\preceq K|\cmpl{K}$ for all $K\in\alpha$, and
(ii) if $\beta \preceq K|\cmpl{K}$ for all $K\in\alpha$ then $\beta\preceq\alpha$.
\textit{For (i)}, we have that
$\alpha=K_1|K_2|\dots|K_{\abs{\alpha}}\preceq K_i|(\cup_{j\neq i}K_j)=K_i|\cmpl{K_i}$,
since every $K_j$ is either identical to $K_i$ or contained in $\cmpl{K_i}$
(the $K_i$ sets are disjoint ones) and \eqref{eq:FirstLabRelDef} holds.
\textit{For (ii)}, let 
$\beta = K'_1|K'_2|\dots|K'_{\abs{\beta}}\preceq K|\cmpl{K}$ for all $K\in\alpha$;
then, by definition  \eqref{eq:FirstLabRelDef},
for all $K'_j\in\beta$ there is a $K_i\in\alpha$ such that $K'_j\subseteq K_i$,
which means that $\beta\preceq\alpha$ by definition \eqref{eq:FirstLabRelDef}.

(Another proof, using the duality principle of order theory, could also be given:
The $K|\cmpl{K}$ bipartitions are the atoms of the dual lattice \cite{DaveyPriestley}.)

\subsection{LOCC closedness of the first kind for mixed states}
\label{app:lattices.LOCCcloseI}

For the proof of \eqref{eq:LOCCcloseI} first we use that every LOCC map $\Lambda$
is also an SO map; that is, it can be written in the form
\begin{subequations}
\begin{equation}
\Lambda(\varrho) = \sum_i \Bigl(\bigotimes_{a\in L} A_{a,i} \Bigr) \varrho 
\Bigl(\bigotimes_{a'\in L} A_{a',i} \Bigr)^\dagger,
\end{equation}
with
\begin{equation}
\sum_i 
\Bigl(\bigotimes_{a'\in L} A_{a',i} \Bigr)^\dagger
\Bigl(\bigotimes_{a \in L} A_{a ,i} \Bigr)
= \Id.
\end{equation}
\end{subequations}
(Note that the reverse is not true.)
By definition \eqref{eq:setID}, an $\alpha$-separable state
can be written in the form
\begin{equation}
\varrho = \sum_j p_j \bigotimes_{K\in\alpha}\pi_{K,j},
\end{equation}
so if $\varrho\in\mathcal{D}_\alpha$, then
\begin{equation*}
\begin{split}
\Lambda(\varrho) 
&= \sum_i \Bigl(\bigotimes_{a\in L} A_{a,i} \Bigr)
\Bigl( \sum_j p_j \bigotimes_{K\in\alpha}\pi_{K,j} \Bigr)
\Bigl(\bigotimes_{a'\in L} A_{a',i} \Bigr)^\dagger\\
&= \sum_j p_j \sum_i \Bigl(\bigotimes_{a\in L} A_{a,i} \Bigr)
\Bigl( \bigotimes_{K\in\alpha}\pi_{K,j} \Bigr)
\Bigl(\bigotimes_{a'\in L} A_{a',i} \Bigr)^\dagger\\
&=\sum_j p_j \underbrace{\sum_i \bigotimes_{K\in\alpha}\Bigl[
\Bigl(\bigotimes_{a\in K}  A_{a ,i} \Bigr) \pi_{K,j}
\Bigl(\bigotimes_{a'\in K} A_{a',i} \Bigr)^\dagger \Bigr]}_{\in\mathcal{D}_\alpha},
\end{split}
\end{equation*}
so $\Lambda(\varrho)\in\Conv\mathcal{D}_\alpha=\mathcal{D}_\alpha$.

\subsection{Order isomorphisms of the first kind}
\label{app:lattices.isomI}

Here we prove \eqref{eq:orderisomIP} and \eqref{eq:orderisomID}.

For the proof of \eqref{eq:orderisomIP},
using the definition \eqref{eq:setIP},
we can reformulate
 $\mathcal{P}_\beta \subseteq \mathcal{P}_\alpha$
as follows:
\begin{equation*}
\begin{split}
\pi \in \mathcal{P}_\beta
\quad&\overset{\eqref{eq:setIP}}{\Longleftrightarrow}\quad
\forall K'\in\beta, \exists \pi'_{K'}\in \mathcal{P}_{K'}:
\pi = \bigotimes_{K'\in\beta} \pi'_{K'}\\
\quad&\Longrightarrow\quad
\forall K \in\alpha, \exists \pi_K   \in \mathcal{P}_K:
\pi = \bigotimes_{K \in\alpha} \pi_K  \\
\quad&\overset{\eqref{eq:setIP}}{\Longleftrightarrow}\quad
\pi \in \mathcal{P}_\alpha.
\end{split}
\end{equation*}
To see the \textit{$\Rightarrow$ implication} in \eqref{eq:orderisomIP},
note that
if $\beta\preceq\alpha$, then, by definition \eqref{eq:FirstLabRelDef},
one can collect every $K'\in\beta$ for which $K'\subseteq K$,
and construct $\pi_K=\bigotimes_{K'\in\beta, K'\subseteq K} \pi'_{K'}\in\mathcal{P}_K$.
This can be done for all $K\in\alpha$, leading to the implication above.
To see the \textit{$\Leftarrow$ implication} in \eqref{eq:orderisomIP},
we prove the contrapositive statement, 
\begin{equation}
\beta\npreceq\alpha \quad\Longrightarrow\quad
 \mathcal{P}_\beta \nsubseteq \mathcal{P}_\alpha.
\end{equation}
For this, we have
\begin{equation}
\label{eq:FirstLabRelDefContra}
\beta\npreceq\alpha\quad\defn\quad \exists K'\in\beta, \forall K\in\alpha : K'\nsubseteq K
\end{equation}
by \eqref{eq:FirstLabRelDef}.
Now, if $\pi\in\mathcal{P}_\beta$,
then it can be written as $\pi = \bigotimes_{K'\in\beta} \pi'_{K'}$
with $\pi'_{K'}\in \mathcal{P}_{K'}$ \eqref{eq:setIP}.
Then
for all $K\in\alpha$,
\begin{equation}
\tr_{\cmpl{K}}\pi
= \tr_{\cmpl{K}}\bigotimes_{K'\in\beta} \pi'_{K'}
\overset{\eqref{eq:ptr}}{=} \bigotimes_{K'\in\beta} \tr_{\cmpl{K}\cap K'} \pi'_{K'}.
\end{equation}
If $\beta\npreceq\alpha$, then by \eqref{eq:FirstLabRelDefContra} we have
that one always finds $K'\in\beta$ and $K\in\alpha$,
for which $K'\cap K\neq\emptyset$, while 
$K'\nsubseteq K$ (equivalently, $K'\setminus K = K'\cap \cmpl{K}\neq\emptyset$).
Then, for such $K'$ and $K$,
if we chose $\pi\in\mathcal{P}$ such that 
$\pi'_{K'}$ is entangled with respect to the $(K'\cap K)|(K'\cap \cmpl{K})$ split,
then $\tr_{\cmpl{K}\cap K'} \pi'_{K'}$ is not pure; 
then $\tr_{\cmpl{K}}\pi\notin\mathcal{P}_K$ by \eqref{eq:PsepDecide}, 
so $\mathcal{P}_\beta \nsubseteq \mathcal{P}_\alpha$.
(Note that this reasoning works only if
$\dim\mathcal{H}_{K'\cap K}>1$, and $\dim\mathcal{H}_{K'\cap \cmpl{K}} >1$,
which follows from that $\dim\mathcal{H}_a>1$, which was a general condition
posed in Section \ref{sec:EntMulti.zeroth}.
Subsystems described with one-dimensional Hilbert spaces
are always uncorrelated from the others,
which makes questions related to correlations ill-defined.
In the multipartite scenario,
on the other hand,
we use one-dimensional Hilbert spaces for representing ``no subsystem,''
being a different notion,
convenient in the tensor algebra.)

For the proof of \eqref{eq:orderisomID}, we have \eqref{eq:orderisomIP},
and we claim
\begin{equation*}
\mathcal{P}_\beta \subseteq \mathcal{P}_\alpha
\quad\Longleftrightarrow\quad
\mathcal{D}_\beta \subseteq \mathcal{D}_\alpha,
\end{equation*}
which comes from the geometry of quantum states.
The \textit{$\Rightarrow$ implication} is obvious from \eqref{eq:setID},
while for the \textit{$\Leftarrow$ implication}
one has 
$\mathcal{P}_\beta=\Extr\mathcal{D}_\beta \subseteq \mathcal{D}_\beta \subseteq \mathcal{D}_\alpha$,
so any $\pi\in\mathcal{P}_\beta$ is an element of $\mathcal{D}_\alpha$;
moreover, it is a pure state, so it is an element also of $\Extr\mathcal{D}_\alpha=\mathcal{P}_\alpha$ \eqref{eq:setIPextr},
which is exactly what we need.

It is general for posets that
if $\alpha\mapsto\mathcal{P}_\alpha$ is an order isomorphism
as in \eqref{eq:orderisomIP}, 
then the two posets are isomorphic; that is,
in our case,
the map $\alpha\mapsto\mathcal{P}_\alpha$ is bijective
(see Appendix \ref{app:lattices.gen}).
The same holds for $\alpha\mapsto\mathcal{D}_\alpha$ based on \eqref{eq:orderisomID}.

\subsection{Meet-semilattice isomorphism of the first kind for pure states}
\label{app:lattices.isomIP}

For the proof of \eqref{eq:meetintersectIPeq},
let
$\pi\in\mathcal{P}_\alpha\cap\mathcal{P}_{\alpha'}$,
where $\alpha=K_1|K_2|\dots|K_{\abs{\alpha}}$
and   $\alpha'=K'_1|K'_2|\dots|K'_{\abs{\alpha'}}$;
then from definition \eqref{eq:setIP} we have
\begin{equation*}
\pi = \bigotimes_{K\in\alpha} \pi_K 
= \bigotimes_{K'\in\alpha'} \pi'_{K'},
\end{equation*}
with $\pi_K\in\mathcal{P}_K$, $\pi'_{K'}\in\mathcal{P}_{K'}$.
For all $K\in\alpha$, we have
\begin{equation*}
\tr_{\cmpl{K}} \pi = \pi_K 
= \bigotimes_{\substack{K'\in\alpha'\\K\cap K'\neq\emptyset}} \tr_{\cmpl{K}}\pi'_{K'}
\overset{\eqref{eq:ptr}}{=} 
  \bigotimes_{\substack{K'\in\alpha'\\K\cap K'\neq\emptyset}} \pi'_{K\cap K'},
\end{equation*}
leading to the decomposition
\begin{equation*}
\pi = \bigotimes_{K\in\alpha} \pi_K
= \bigotimes_{K\in\alpha} 
\bigotimes_{\substack{K'\in\alpha'\\K\cap K'\neq\emptyset}}
\pi'_{K\cap K'};
\end{equation*}
that is, $\pi$ is separable with respect to the split
$\set{K\cap K'\neq\emptyset}{K\in\alpha,K'\in\alpha'}$,
which is just $\alpha\wedge\alpha'$ by \eqref{eq:part.meet},
so we have 
$\mathcal{P}_\alpha\cap\mathcal{P}_{\alpha'}\subseteq\mathcal{P}_{\alpha\wedge\alpha'}$.
The reverse inclusion is the first one in \eqref{eq:meetjoinintersectunionIP},
which completes the proof.

\subsection{Decision of \texorpdfstring{$\alpha$}{alpha}-separability}
\label{app:lattices.IPdecision}

For the proof of \eqref{eq:IPDecide},
note that the \textit{$\Rightarrow$ direction} is obvious 
from the definition \eqref{eq:setIP} of $\alpha$-separable pure states,
while for the \textit{$\Leftarrow$ direction},
we have that if $\tr_{\cmpl{K}}\pi\in\mathcal{P}_K$, then $\pi\in\mathcal{P}_{K|\cmpl{K}}$
by \eqref{eq:PsepDecide}, which holds for all $K\in\alpha$;
then \eqref{eq:IPintersect} leads to the left-hand side.

\subsection{LOCC closedness of the second kind for mixed states}
\label{app:lattices.LOCCcloseII}

For the proof of \eqref{eq:LOCCcloseII}, first we use,
by definition \eqref{eq:setIID}, that every $\varrho\in\mathcal{D}_{\vs{\alpha}}$
can be written in the form $\varrho = \sum_j p_j \pi_j$, where for all $j$,
$\pi_j\in\mathcal{P}_\alpha\subseteq\mathcal{D}_\alpha$ for at least one $\alpha\in\vs{\alpha}$.
Since $\mathcal{D}_\alpha$ is closed under LOCC (see Appendix \ref{app:lattices.LOCCcloseI}),
we have that for a $\varrho\in\mathcal{D}_{\vs{\alpha}}$,
\begin{equation*}
\Lambda(\varrho)=\Lambda\Bigl(\sum_j p_j \pi_j\Bigr)
= \sum_j p_j \underbrace{\Lambda(\pi_j)}_{\substack{\in\mathcal{D}_\alpha,\\ \text{ if $\pi_j\in \mathcal{D}_\alpha$}}},
\end{equation*}
so $\Lambda(\varrho)\in \Conv\cup_{\alpha\in\vs{\alpha}}\mathcal{D}_\alpha = \mathcal{D}_{\vs{\alpha}}$,
because of definitions \eqref{eq:setIIP}, \eqref{eq:setIID}, and \eqref{eq:setIIPextr}.

\subsection{Order isomorphisms of the second kind}
\label{app:lattices.isomII}

Here we prove \eqref{eq:orderisomIIP} and \eqref{eq:orderisomIID}.

For the proof of \eqref{eq:orderisomIIP}, 
using the definitions \eqref{eq:posetII} and \eqref{eq:setIIP},
we have
\begin{equation*}
\begin{split}
\vs{\beta}\preceq\vs{\alpha}
\quad&\overset{\eqref{eq:posetII}}{\Longleftrightarrow}\quad
\vs{\beta}\subseteq\vs{\alpha}\\
\quad&\Longleftrightarrow\quad
\bigset{\mathcal{P}_\beta }{\beta \in\vs{\beta }}\subseteq 
\bigset{\mathcal{P}_\alpha}{\alpha\in\vs{\alpha}}\\
\quad&\Longleftrightarrow\quad
\bigcup_{\beta \in\vs{\beta }}\mathcal{P}_\beta \subseteq
\bigcup_{\alpha\in\vs{\alpha}}\mathcal{P}_\alpha\\
\quad&\overset{\eqref{eq:setIIP}}{\Longleftrightarrow}\quad
\mathcal{P}_{\vs{\beta}} \subseteq \mathcal{P}_{\vs{\alpha}},
\end{split}
\end{equation*}
where the \textit{second implication} is 
because $\alpha\mapsto\mathcal{P}_\alpha$ is bijective
due to \eqref{eq:orderisomIP},
while the \textit{third one} is obvious.

For the proof of \eqref{eq:orderisomIID}, we have \eqref{eq:orderisomIIP},
and we claim
\begin{equation*}
\mathcal{P}_{\vs{\beta}} \subseteq \mathcal{P}_{\vs{\alpha}}
\quad\Longleftrightarrow\quad
\mathcal{D}_{\vs{\beta}} \subseteq \mathcal{D}_{\vs{\alpha}},
\end{equation*}
which can be proven in the same way as 
the parallel result of the first kind in Appendix \ref{app:lattices.isomI}.

Again, it is general for posets that
if $\vs{\alpha}\mapsto\mathcal{P}_{\vs{\alpha}}$ is an order isomorphism
as in \eqref{eq:orderisomIIP},
then the two posets are isomorphic; that is,
in our case,
the map $\vs{\alpha}\mapsto\mathcal{P}_{\vs{\alpha}}$ is bijective
(see Appendix \ref{app:lattices.gen}).
The same holds for $\vs{\alpha}\mapsto\mathcal{D}_{\vs{\alpha}}$ based on \eqref{eq:orderisomIID}.

\subsection{Lattice isomorphism of the second kind for pure states}
\label{app:lattices.isomIIP}

Here we prove \eqref{eq:meetjoinintersectunionIIPeq}.

For the \textit{first part,}
\begin{equation*}
\begin{split}
\mathcal{P}_{\vs{\alpha}}\cap\mathcal{P}_{\vs{\alpha}'}
&\overset{\eqref{eq:setIIP}}{=}
  \bigcup_{\alpha\in\vs{\alpha}} \mathcal{P}_\alpha
  \cap
  \bigcup_{\alpha'\in\vs{\alpha}'} \mathcal{P}_{\alpha'} \\
&\overset{\phantom{\eqref{eq:setIIP}}}{=}
  \bigcup_{\substack{\alpha\in\vs{\alpha}\\\alpha'\in\vs{\alpha}'}}
  \bigl(\mathcal{P}_\alpha \cap \mathcal{P}_{\alpha'}\bigr) 
\overset{\eqref{eq:meetintersectIPeq}}{=}
  \bigcup_{\substack{\alpha\in\vs{\alpha}\\\alpha'\in\vs{\alpha}'}}
  \mathcal{P}_{\alpha\wedge\alpha'} \\
&\overset{\phantom{\eqref{eq:setIIP}}}{\subseteq}
  \bigcup_{\beta\in\vs{\alpha}\cap\vs{\alpha}'}\mathcal{P}_\beta 
\overset{\eqref{eq:posetII}}{=}
  \bigcup_{\beta\in\vs{\alpha}\wedge\vs{\alpha}'}\mathcal{P}_\beta 
\overset{\eqref{eq:setIIP}}{=}
  \mathcal{P}_{\vs{\alpha}\wedge\vs{\alpha}'}
\end{split}
\end{equation*}
where the \textit{first} and \textit{last equations} are by definition \eqref{eq:setIIP},
the \textit{last but one equation} is by definition \eqref{eq:posetII},
the \textit{second equation} is the distributivity of $\cap$ over $\cup$,
and the \textit{third equation} is \eqref{eq:meetintersectIPeq} from Level I.
The \textit{inclusion} is
from that
for all $\alpha\in\vs{\alpha}$ and $\alpha'\in\vs{\alpha}'$,
$\alpha\wedge\alpha'\preceq\alpha \in\vs{\alpha}$ and
$\alpha\wedge\alpha'\preceq\alpha'\in\vs{\alpha}'$
hold, because $\vs{\alpha}$ and $\vs{\alpha}'$ are down-sets \eqref{eq:PII},
so $\alpha\wedge\alpha'\in\vs{\alpha}\cap\vs{\alpha}'$,
so all $\mathcal{P}_{\alpha\wedge\alpha'}$ sets in the left-hand side
appear in the right-hand side as a $\mathcal{P}_\beta$.
The reverse inclusion in the first part of \eqref{eq:meetjoinintersectunionIIPeq}
is the first one in \eqref{eq:meetjoinintersectunionIIP},
which completes the proof.

For the \textit{second part,}
we have
\begin{equation*}
\mathcal{P}_{\vs{\alpha}}\cup\mathcal{P}_{\vs{\alpha}'}
\overset{\eqref{eq:setIIP}}{=}
  \bigcup_{\alpha\in\vs{\alpha}} \mathcal{P}_\alpha
  \cup
  \bigcup_{\alpha'\in\vs{\alpha}'} \mathcal{P}_{\alpha'}
= \bigcup_{\alpha\in\vs{\alpha}\cup\vs{\alpha}'}\mathcal{P}_\alpha
\overset{\eqref{eq:setIIP}}{=}
  \mathcal{P}_{\vs{\alpha}\vee\vs{\alpha}'},
\end{equation*}
where the \textit{first} and \textit{third equations} are by definition \eqref{eq:setIIP};
and the \textit{second one} is obvious using elementary set algebra.

\subsection{Lattice of the labels of the classes}
\label{app:lattices.classes}

For the proof of \eqref{eq:classlabels},
we start with the contrapositive form of \eqref{eq:classemptybyconstruction},
\begin{equation*}
\begin{split}
\mathcal{C}_{\vvs{\alpha}}\neq\emptyset
\quad&\Longrightarrow\quad
\forall \vs{\alpha}\in \vvs{\alpha},\;
\forall \vs{\beta}\notin \vvs{\alpha}\;:\;
\vs{\alpha} \npreceq \vs{\beta}\\
&\Longrightarrow\quad
\forall \vs{\alpha}\in \vvs{\alpha}:
\bigl( \vs{\beta}\notin \vvs{\alpha} \;\Rightarrow\; \vs{\alpha} \npreceq \vs{\beta} \bigr)\\
&\Longleftrightarrow\quad
\forall \vs{\alpha}\in \vvs{\alpha}:
\bigl( \vs{\alpha} \preceq \vs{\beta} \;\Rightarrow\; \vs{\beta}\in \vvs{\alpha}\bigr),
\end{split}
\end{equation*}
where the \textit{last implication} is the
contrapositive reformulation of the parentheses,
leading just to the definition of the up-set lattice $P_\text{III}$ in \eqref{eq:PIII}.

\subsection{Reconstruction of the state sets from classes}
\label{app:lattices.reconstrDII}

For the proof of \eqref{eq:forDbyC2},
we have the equivalent statement
that for all $ \vs{\alpha}\in P_\text{II*}$,
     and all $\vvs{\alpha}\in P_\text{III}$,
\begin{equation}
\label{eq:forDbyC}
\begin{split}
\vs{\alpha}\in\vvs{\alpha}
\quad\Longleftrightarrow\quad
&\vvs{\alpha}\in\upset\bigl\{ \upset\{ \vs{\alpha} \} \bigr\}\\
&\overset{\eqref{eq:upQ}}{=}
\upset\bigl\{  \set{\vs{\beta}\in P_\text{II*}}{\vs{\alpha}\preceq\vs{\beta}} \bigr\}\\
&\overset{\eqref{eq:upQ}}{=}
\bigset{\vvs{\beta}\in P_\text{III}}{ \set{\vs{\beta}\in P_\text{II*}}{\vs{\alpha}\preceq\vs{\beta}}\preceq \vvs{\beta} }.
\end{split}
\end{equation}
To see the \textit{$\Rightarrow$ direction,} note that
if $\vs{\alpha}\in\vvs{\alpha}$, then for all $\vs{\beta}\in P_\text{II*}$ 
for which $\vs{\alpha}\preceq\vs{\beta}$ holds,
also $\vs{\beta}\in\vvs{\alpha}$ holds,
because $\vvs{\alpha}$ is an up-set \eqref{eq:PIII}.
This means that $\upset\{\vs{\alpha}\}\preceq \vvs{\alpha}$ (see \eqref{eq:posetIII}),
so $\vvs{\alpha}\in\upset\{ \upset\{ \vs{\alpha} \} \}$ (see \eqref{eq:upQ}).
To see the \textit{$\Leftarrow$ direction,} note that
if $\vvs{\alpha}\in\bigset{\vvs{\beta}\in P_\text{III}}{ \set{\vs{\beta}\in P_\text{II*}}{\vs{\alpha}\preceq\vs{\beta}}\preceq \vvs{\beta} }$,
then
$\set{\vs{\beta}\in P_\text{II*}}{\vs{\alpha}\preceq\vs{\beta}}\preceq \vvs{\alpha}$;
on the other hand,
$\vs{\alpha}\in\set{\vs{\beta}\in P_\text{II*}}{\vs{\alpha}\preceq\vs{\beta}}$,
so $\vs{\alpha}\in\vvs{\alpha}$ (see \eqref{eq:posetIII}).

\subsection{LOCC convertibility of the classes}
\label{app:lattices.LOCCconv}

For the proof of \eqref{eq:LOCCconvhierarchy},
we prove the contrapositive statement
\begin{equation*}
\begin{split}
\vvs{\beta}\npreceq\vvs{\alpha}
\quad&\overset{\eqref{eq:posetIII}}{\Longleftrightarrow}\quad
\exists \vs{\gamma}\in P_\text{II}:\;
 \vs{\gamma}\in\vvs{\beta}\;\text{and}\;\vs{\gamma}\notin\vvs{\alpha}\\
\quad&\overset{\eqref{eq:CinD}}{\Longleftrightarrow}\quad
 \exists \vs{\gamma}\in P_\text{II}:\;
 \mathcal{C}_{\vvs{\beta}} \subseteq \mathcal{D}_{\vs{\gamma}}
 \;\text{and}\;
 \mathcal{C}_{\vvs{\alpha}} \nsubseteq \mathcal{D}_{\vs{\gamma}}\\
\quad&\overset{\eqref{eq:LOCCcloseII}}{\Longrightarrow}\quad
 \mathcal{C}_{\vvs{\beta}} \overset{\text{LOCC}_\text{w}}{\not\longrightarrow} \mathcal{C}_{\vvs{\alpha}}.
\end{split}
\end{equation*}
The \textit{first implication} is the definition of the level III hierarchy \eqref{eq:posetIII};
the \textit{second one} is \eqref{eq:CinD};
and the \textit{third one} is the LOCC closedness of $\mathcal{D}_{\vs{\gamma}}$ in \eqref{eq:LOCCcloseII}
applied for the weak LOCC convertibility \eqref{eq:defLOCCconv.weak}.

\subsection{Poset of the classes}
\label{app:lattices.classLOCCorder}

For the proof of that 
the strong LOCC convertibility of the classes \eqref{eq:defLOCCconv.strong}
is a partial order,
we have to prove that the properties \eqref{eq:poax} are satisfied.
For the proof of the \textit{reflexivity} \eqref{eq:poax.refl},
note that $\mathcal{C}_{\vvs{\alpha}}\geq_\text{s}\mathcal{C}_{\vvs{\alpha}}$
by choosing the identity channel (which is also a LOCC) for $\Lambda$ in \eqref{eq:defLOCCconv.strong}.
For the proof of the \textit{antisymmetry} \eqref{eq:poax.antisymm},
\begin{equation*}
\begin{split}
\mathcal{C}_{\vvs{\beta}} \geq_\text{s}\mathcal{C}_{\vvs{\alpha}}\;\text{and}\;
\mathcal{C}_{\vvs{\alpha}}\geq_\text{s}\mathcal{C}_{\vvs{\beta}} 
\quad\overset{\eqref{eq:LOCCconvhierarchy}}{\Longrightarrow}\quad
\vvs{\beta}\preceq\vvs{\alpha}\;\text{and}\;
\vvs{\alpha}\preceq\vvs{\beta}& \\
\overset{\eqref{eq:poax.antisymm}}{\Longrightarrow}\quad
\vvs{\beta}=\vvs{\alpha}
\quad\overset{\eqref{eq:classDef}}{\Longrightarrow}\quad
\mathcal{C}_{\vvs{\beta}} = \mathcal{C}_{\vvs{\alpha}}&,
\end{split}
\end{equation*}
where the \textit{first implication} is \eqref{eq:LOCCconvhierarchy},
the \textit{second one} is by the antisymmetry \eqref{eq:poax.antisymm} 
of the partial order \eqref{eq:posetIII},
the \textit{third one} is by definition \eqref{eq:classDef}.
For the proof of the \textit{transitivity} \eqref{eq:poax.trans}, if 
$\mathcal{C}_{\vvs{\gamma}} \geq_\text{s}\mathcal{C}_{\vvs{\beta}}$  and
$\mathcal{C}_{\vvs{\beta}}  \geq_\text{s}\mathcal{C}_{\vvs{\alpha}}$, then, by definition \eqref{eq:defLOCCconv.strong},
for all $\varrho\in\mathcal{C}_{\vvs{\gamma}}$ there is a $\Lambda$ LOCC
such that $\Lambda(\varrho)\in\mathcal{C}_{\vvs{\beta}}$, 
and for this, by definition \eqref{eq:defLOCCconv.strong},
there is a $\Lambda'$ LOCC such that $\Lambda'(\Lambda(\varrho))\in\mathcal{C}_{\vvs{\alpha}}$,
and the composition of LOCC operations is also an LOCC operation.
(Note that this is the point where 
the weak LOCC convertibility \eqref{eq:defLOCCconv.weak} would not be sufficient.)

\section{On entanglement measures}
\label{app:EntMon}

\subsection{Convexity and concavity of operator functions}
\label{app:EntMon.cnvcnc}

Here we recall a useful result in the theory of trace functions
from Section 2.2 of \cite{CarlenIneqs}.

Let $f:\field{R}\to\field{R}$ be continuous
and $F = \tr\circ f: \Lin_\text{SA}\mathcal{H}\to\field{R}$
the associated \emph{trace function};
then
if $f$ is monotonically increasing (decreasing), then $F$ is monotonically increasing (decreasing),
and
if $f$ is convex (concave), then $F$ is convex (concave).

Based on these, 
the \eqref{eq:entrConcavity.Neumann} and 
\eqref{eq:entrConcavity.Tsallis} concavity of the 
von Neumann \eqref{eq:NeumannEntr} and Tsallis \eqref{eq:TsallisEntr} entropies 
can be proven
using the functions $f(x) = -x\ln x$ and $f(x) = \frac1{1-q} (x^q-x)$.
For the \eqref{eq:entrConcavity.Renyi} concavity of the 
R{\'e}nyi entropy \eqref{eq:RenyiEntr},
we have that $f(x) = x^q$ is concave if and only if $q\leq1$,
then so is $F(\varrho) = \tr\varrho^q$,
while $\ln(x)$ is concave and monotonically increasing, so $\ln\tr\varrho^q$ is also concave.
However, for $q\geq1$, $F(\varrho) = \tr\varrho^q$ is convex,
while $-\ln(x)$ is convex and monotonically decreasing, so the concavity or convexity
cannot be decided using this method.

\subsection{Uncorrelated state with minimal relative entropy}
\label{app:EntMon.geom}

Here we recall the proof of \eqref{eq:geomCorrIArgmin} from \cite{Modi-2010}.
For a $\varrho\in\mathcal{D}$, and a partition $\alpha$,
we use the notation $\varrho_K:=\tr_{\cmpl{K}}\varrho\in\mathcal{D}_K$ for all $K\in\alpha$,
and let $\omega_K\in\mathcal{D}_K$ for all $K\in\alpha$.
Suppose that \eqref{eq:geomCorrImin} takes its minimum at $\{\omega_K\}$.
Then 
\begin{align*}
0&\;\leq
   D^\text{KL}\Bigl(\varrho\;\Big\Vert\bigotimes_{K\in\alpha}\varrho_K\Bigr)
  -D^\text{KL}\Bigl(\varrho\;\Big\Vert\bigotimes_{K\in\alpha}\omega_K\Bigr)\\
&\overset{\eqref{eq:KullbackLeiblerDiv}}{=}
  -\tr\Bigl(\varrho\ln\bigotimes_{K\in\alpha}\varrho_K\Bigr)
  +\tr\Bigl(\varrho\ln\bigotimes_{K\in\alpha}\omega_K\Bigr)\\
&\;=-\sum_{K\in\alpha}\tr(\varrho_K\ln\varrho_K)
  +\sum_{K\in\alpha}\tr(\varrho_K\ln\omega_K)\\
&\;=-\tr\Bigl(\bigotimes_{K\in\alpha}\varrho_K\ln\bigotimes_{K\in\alpha}\varrho_K\Bigr)
  +\tr\Bigl(\bigotimes_{K\in\alpha}\varrho_K\ln\bigotimes_{K\in\alpha}\omega_K\Bigr)\\
&\overset{\eqref{eq:KullbackLeiblerDiv}}{=}
  -D^\text{KL}\Bigl(\bigotimes_{K\in\alpha}\varrho_K\;\Big\Vert\bigotimes_{K\in\alpha}\omega_K\Bigr),
\end{align*}
where the \textit{first and last equalities}
are the \eqref{eq:KullbackLeiblerDiv} definition of the relative entropy
and the \textit{second and third equalities} are from the linearity of the trace
and the additivity of the logarithm,
$\ln(\omega_1\otimes\omega_2\otimes\dots)
 =\ln(\omega_1\otimes\Id\otimes\dots)
     (\Id\otimes\omega_2\otimes\dots)\dots
 =\ln(\omega_1\otimes\Id\otimes\dots)
 +\ln(\Id\otimes\omega_2\otimes\dots)
 +\dots$.
Since the relative entropy is non-negative \eqref{eq:RelEntrNNeg},
then the above is zero, 
which leads to that $\omega_K=\varrho_K$ because of \eqref{eq:RelEntrDiscr}.

\subsection{Pure entanglement measures}
\label{app:EntMon.pure}

Here we recall Horodecki's proof \cite{HorodeckiEntMeas}
for Theorem \ref{thm:pureEntMon} given in Section \ref{sec:MeasBasics}.

A function is \emph{symmetric} in its arguments 
if it does not change its value for the permutation of its arguments,
and it is \emph{expansible} 
if it takes the same values for arguments $(x_1,\dots,x_d)$ and $(x_1,\dots,x_d,0)$.
So, first of all, from (i) it follows that
\begin{equation}
\label{eq:same}
F(\tr_{\cmpl{K}}\pi) = F(\tr_K\pi),
\end{equation}
since the two arguments have the same spectrum, apart from the multiplicity of the zero eigenvalue.
Let us decompose the LOCC $\Lambda$ into the \emph{pure} operations $\Lambda_i$ consisting of single Kraus operators each, 
$\Lambda_i(\cdot)=A_i(\cdot)A_i^\dagger$.
These operations are separable ones, that is, 
$A_i=A_{1,i}\otimes A_{2,i}\otimes\dots\otimes A_{n,i}$,
and they can further be decomposed into the composition of $\Lambda_{a,i}(\cdot)$ acting nontrivially on the $a$th subsystem only.
Applying these to the initial (pure) state $\pi$ results in the ensemble of \emph{pure} states $\pi_i'=\frac1{p_i}\Lambda_{a,i}(\pi)$
(with probabilities $p_i=\tr\Lambda_{a,i}(\pi)$).
The resulting mixed states are then $\pi'=\sum_i p_i\pi_i'$.

Take a $\Lambda_{a,i}$, and let $K_*=K$ if $a$ is not contained in $K$, and $K_*=\cmpl{K}$ otherwise.
Then $\Lambda_{a,i}$ leaves subsystem $K_*$ invariant, $\tr_{\cmpl{K_*}}\pi'=\tr_{\cmpl{K_*}}\pi$,
which leads to
\begin{equation}
\label{eq:invarK}
\tr_{\cmpl{K_*}}\pi = \tr_{\cmpl{K_*}}\pi' = \sum_i p_i \tr_{\cmpl{K_*}}\pi_i' \quad\text{for $a\notin K_*$.}
\end{equation}
Now we can write
\begin{equation*}
\begin{split}
f_K(\pi)
\overset{\substack{\eqref{eq:pureVidal}\\\eqref{eq:same}}}{=}
F(\tr_{\cmpl{K_*}}\pi)
\overset{\eqref{eq:invarK}}{=}
F\Bigl(\sum_i p_i \tr_{\cmpl{K_*}}\pi_i'\Bigr)\\
\geq
\sum_i p_i F(\tr_{\cmpl{K_*}}\pi_i')
\overset{\substack{\eqref{eq:pureVidal}\\\eqref{eq:same}}}{=}
\sum_i p_i f_K(\pi_i'),
\end{split}
\end{equation*}
where the \textit{first and last equalities} are that both subsystems can be used \eqref{eq:same} in construction \eqref{eq:pureVidal},
the \textit{second one} is \eqref{eq:invarK}, and the inequality is the concavity (ii) of Theorem \ref{thm:pureEntMon}.

\subsection{Convex roof extension preserves entanglement monotonicity}
\label{app:EntMon.cnvRoofs}
Here we recall Horodecki's proof \cite{HorodeckiEntMeas}
for Theorem \ref{thm:averageConvRoof} given in Section \ref{sec:MeasBasics}.

Let $f:\mathcal{P}\to\field{R}$ be a function satisfying \eqref{eq:averagePure}, that is,
\begin{equation}
\label{eq:averagePureagain}
\sum_i p_i f(\pi_i') \leq f(\pi)
\end{equation}
for all $\pi\mapsto\{(p_i,\pi_i')\}$ ensembles
resulting from a LOCC $\Lambda$ consisting of the $\Lambda_i$ \emph{pure} operations.
Take an $f$-optimal pure decomposition $\{(q_j,\pi_j)\}$ of $\varrho$, that is, 
$\varrho=\sum_jq_j\pi_j$,
and $\sum_jq_jf(\pi_j)$ 
in the right-hand side of \eqref{eq:cnvroofext} takes its minimum,
so
\begin{equation}
\label{eq:optimaldec}
\cnvroof{f}(\varrho)= \sum_jq_jf(\pi_j).
\end{equation}
Applying the $\Lambda_i$ pure operators to the pure states $\pi_j$ of this ensemble results in 
the ensembles of \emph{pure} states $\pi_{j,i}'=\frac1{p_{j,i}}\Lambda_i(\pi_j)$
(with probabilities $p_{j,i}=\tr\Lambda_i(\pi_j)$).
Applying the $\Lambda_i$ pure operators to the mixed state $\varrho$ results in
\begin{equation}
\label{eq:rhoitraf}
\varrho_i'=\frac1{p_i}\Lambda_i(\varrho)=\frac1{p_i}\sum_jq_j\Lambda_i(\pi_j)=\frac1{p_i}\sum_jq_jp_{j,i}\pi_{j,i}'
\end{equation}
(with probability $p_i=\tr\Lambda_i(\varrho)=\sum_jq_j\tr\Lambda_i(\pi_j)=\sum_jq_jp_{j,i}$).
With these, we can write
\begin{equation*}
\begin{split}
\cnvroof{f}(\varrho) 
&\overset{\eqref{eq:optimaldec}}{=}
  \sum_jq_j f(\pi_j)\\ 
&\overset{\eqref{eq:averagePureagain}}{\geq}
  \sum_jq_j \sum_i p_{j,i}f(\pi_{j,i}')\\
&\overset{\eqref{eq:cnvroofpure}}{=}
  \sum_jq_j \sum_i p_{j,i}\cnvroof{f}(\pi_{j,i}')\\
&\overset{\phantom{\eqref{eq:cnvroofpure}}}{=}
  \sum_i p_i\frac1{p_i}\sum_jq_jp_{j,i}\cnvroof{f}(\pi_{j,i}')\\
&\overset{\eqref{eq:cnvroofcnv}}{\geq}
  \sum_i p_i \cnvroof{f}\Bigl(\frac1{p_i}\sum_jq_jp_{j,i}\pi_{j,i}'\Bigr)
\overset{\eqref{eq:rhoitraf}}{=} \sum_i p_i \cnvroof{f}(\varrho_i'),
\end{split}
\end{equation*}
where 
the \textit{first equality} is the optimality \eqref{eq:optimaldec},
the \textit{first inequality} is due to the entanglement monotonicity of $f$ 
on pure states \eqref{eq:averagePureagain},
the \textit{second equality} is \eqref{eq:cnvroofpure},
the \textit{second inequality} is the convexity of the convex roof extension \eqref{eq:cnvroofcnv},
and the \textit{last equality} is \eqref{eq:rhoitraf}.

Note that this reasoning does not depend on 
whether $\{(q_j,\pi_j)\}$ is a result of a LOCC or some other class of operations,
as far as \eqref{eq:averagePureagain} holds.

\subsection{Convex roof extension preserves discriminance}
\label{app:EntMon.cnvRoofsDisc}

For the proof of \eqref{eq:cnvroofDisc}, 
let $\mathcal{P}_*\subseteq\mathcal{P}$
and $\mathcal{D}_*=\Conv\mathcal{P}_*\subseteq\mathcal{D}=\Conv\mathcal{P}$,
and
let $f:\mathcal{P}\to[0,\infty)$; then 
\begin{equation*}
\begin{split}
\varrho\in\mathcal{D}_*
\quad&\Longleftrightarrow\quad \varrho\in\Conv\mathcal{P}_*\\
\quad&\Longleftrightarrow\quad \varrho=\sum_ip_i\pi_i \;\text{with}\; \pi_i\in\mathcal{P}_*\\
\quad&\overset{\eqref{eq:measdiscr}}{\Longleftrightarrow}\quad 
                               \varrho=\sum_ip_i\pi_i \;\text{with}\; f(\pi_i)=0 \\
\quad&\Longleftrightarrow\quad \cnvroof{f}(\varrho)=0,
\end{split}
\end{equation*}
where 
the \textit{first and second implications} are from the assumptions above
and the \textit{third implication} is the assumption in \eqref{eq:cnvroofDisc},
which is the  discriminance property \eqref{eq:measdiscr} for the pure state function.
The condition $f\geq0$ is necessary for the \textit{last implication:}
To see the \textit{$\Rightarrow$ direction,} note that 
the minimum in the convex roof extension \eqref{eq:cnvroofext} of a non-negative function is zero \eqref{eq:cnvroofbound},
which is attained if the left-hand side holds,
and 
to see the \textit{$\Leftarrow$ direction,} note that
if the convex roof extension \eqref{eq:cnvroofext} of a non-negative function vanishes,
then there exists a decomposition for pure states for which the function vanishes.

\subsection{Convex roof extension preserves the invariance}
\label{app:EntMon.cnvRoofsInv}

For the proof of \eqref{eq:cnvroofinv}, note that
the \textit{$\Leftarrow$ direction} is obvious from \eqref{eq:cnvroofpure}, while
for the \textit{$\Rightarrow$ direction} we have
\begin{equation*}
\begin{split}
\cnvroof{f}(G\varrho G^\dagger) 
&\overset{\eqref{eq:cnvroofext}}{=}
  \min_{\sum_i p_i \pi_i=G\varrho G^\dagger} \sum_i p_i f(\pi_i)\\
&\;=\min_{\sum_i p_i \pi'_i=\varrho} \sum_i p_i f(G\pi'_i G^\dagger)\\
&\;=\min_{\sum_i p_i \pi'_i=\varrho} \sum_i p_i f(\pi'_i) 
\overset{\eqref{eq:cnvroofext}}{=} \cnvroof{f}(\varrho),
\end{split}
\end{equation*}
where the \textit{first and last equalities} are 
the definition of the convex roof extension \eqref{eq:cnvroofext},
the \textit{second one} is by using the notation $\pi'_i=G^{-1} \pi_i (G^\dagger)^{-1}$,
and the \textit{third one} is the condition in the left-hand side of \eqref{eq:cnvroofinv}.

\subsection{Convex roof extension is monotonic}
\label{app:EntMon.cnvRoofsMon}

For the proof of \eqref{eq:cnvroofMon}, note that
the \textit{$\Leftarrow$ direction} is obvious from \eqref{eq:cnvroofpure}, while
for the \textit{$\Rightarrow$ direction}
take a $g$-optimal decomposition $\{(p_i,\pi_i)\}$ of $\varrho$,
that is, $\varrho = \sum_ip_i\pi_i$ for which the next equality holds
in the convex roof minimization \eqref{eq:cnvroofext},
and
\begin{equation*}
\begin{split}
\cnvroof{g}(\varrho) 
&=    \sum_i p_i g(\pi_i) 
 \geq \sum_i p_i f(\pi_i) \\
&\geq \min_{\sum_{i'} p'_{i'} \pi'_{i'}=\varrho} \sum_{i'} p'_{i'} f(\pi'_{i'})
\overset{\eqref{eq:cnvroofext}}{=}\cnvroof{f}(\varrho),
\end{split}
\end{equation*}
where the
\textit{first inequality} is the left-hand side in \eqref{eq:cnvroofMon},
the \textit{second one} is because a $g$-optimal decomposition is not necessarily $f$-optimal, 
and the \textit{last equality} is the definition \eqref{eq:cnvroofext} of the convex roof extension.

\subsection{Some other properties of convex roof extension}
\label{app:EntMon.cnvRoofsEtc}

The proof of \eqref{eq:cnvroof.c} is obvious 
from the definition \eqref{eq:cnvroofext} of the convex roof extension.

For the proof of \eqref{eq:cnvroof.sum}, 
take an $(f+g)$-optimal decomposition $\{(p_i,\pi_i)\}$ of $\varrho$,
that is, $\varrho = \sum_ip_i\pi_i$ for which the next equality holds
in the convex roof minimization \eqref{eq:cnvroofext},
and
\begin{equation*}
\begin{split}
&\cnvroof{(f+g)}(\varrho) 
=    \sum_i p_i \bigl( f(\pi_i)+g(\pi_i) \bigr)\\
&\;= \Bigl( \sum_i p_i f(\pi_i) \Bigr) + \Bigl( \sum_i p_i g(\pi_i) \Bigr)\\
&\;\geq \min_{\sum_{i'} p'_{i'} \pi'_{i'}=\varrho} \sum_{i'} p'_{i'} f(\pi'_{i'})
+    \min_{\sum_{i'} p'_{i'} \pi'_{i'}=\varrho} \sum_{i'} p'_{i'} g(\pi'_{i'})\\
&\overset{\eqref{eq:cnvroofext}}{=}
\cnvroof{f}(\varrho)+\cnvroof{g}(\varrho),
\end{split}
\end{equation*}
where the \textit{inequality} is  because an $(f+g)$-optimal decomposition
is not necessarily $f$-optimal or $g$-optimal,
and the \textit{last equality} is the definition \eqref{eq:cnvroofext} of the convex roof extension.

For the proof of \eqref{eq:cnvroof.min}, note that
$\min\{f,g\} \leq f,g$, so 
$\cnvroof{(\min\{f,g\})}\leq \cnvroof{f} , \cnvroof{g}$ by \eqref{eq:cnvroofMon},
from which
$\cnvroof{(\min\{f,g\})}\leq\min\{\cnvroof{f},\cnvroof{g}\}$ follows.

\section{On the properties of sums and means}
\label{app:Means}

\subsection{Definitions  and properties of \texorpdfstring{$q$}{q}-sums and \texorpdfstring{$q$}{q}-means}
\label{app:Means.Power}

Let $\ve{x}=(x_1,\dots,x_m)\in \field{R}^m$, $\ve{x}>0$. 
(The latter is meant elementwisely, $x_j>0$, $j=1,\dots,m$.)

The \emph{$q$-sums} of $\ve{x}$ are defined for nonzero $q\in\field{R}$ parameters as
\begin{subequations}
\label{eq:qSums}
\begin{equation}
\label{eq:qSums.q}
N_q(\ve{x}) := \Bigl[\sum_{j=1}^m x_j^q\Bigr]^{1/q}\quad\text{for $q\neq 0$}.
\end{equation}
It can be defined for positive and negative infinities by its limits, leading to
\begin{align}
\label{eq:qSums.max}
N_{+\infty}(\ve{x}) &:= \lim_{q\to+\infty}N_q(\ve{x}) = \max_j(x_j),\\
\label{eq:qSums.min}
N_{-\infty}(\ve{x}) &:= \lim_{q\to-\infty}N_q(\ve{x}) = \min_j(x_j).
\end{align}
Since $\lim_{q\to 0^+}N_q(\ve{x})=\infty$ and
      $\lim_{q\to 0^-}N_q(\ve{x})=0$,
$N_q$ cannot be made continuous in $q=0$.
\end{subequations}
We have the usual \emph{sum} $N_1$,
the \emph{harmonic sum} $N_{-1}$,
and the \emph{quadratic sum} $N_2$.

The \emph{$q$-means} or \emph{power-means} of $\ve{x}$ are defined for nonzero $q\in\field{R}$ parameters as
\begin{subequations}
\label{eq:qMeans}
\begin{equation}
\label{eq:qMeans.q}
M_q(\ve{x}) := \Bigl[\frac1m\sum_{j=1}^m x_j^q\Bigr]^{1/q}\quad\text{for $q\neq 0$}.
\end{equation}
It can be defined for parameter zero and for positive and negative infinities by its limits, leading to
\begin{align}
\label{eq:qMeans.max}
M_{+\infty}(\ve{x}) &:= \lim_{q\to+\infty}M_q(\ve{x}) = \max_j(x_j),\\
\label{eq:qMeans.min}
M_{-\infty}(\ve{x}) &:= \lim_{q\to-\infty}M_q(\ve{x}) = \min_j(x_j),\\
\label{eq:qMeans.0}
M_0(\ve{x})         &:= \lim_{q\to0}M_q(\ve{x}) = \Bigl[\prod_j x_j \Bigr]^{1/m}.
\end{align}
\end{subequations}
We have the \emph{geometric mean} $M_0$,
the usual \emph{arithmetic mean} $M_1$,
the \emph{harmonic mean} $M_{-1}$,
and the \emph{quadratic mean} $M_2$.

The $q$-sums and $q$-means above are defined for strictly positive $x_j$'s;
however, we would like to use them for non-negative values, too.
For $q>0$, the definitions \eqref{eq:qSums} and \eqref{eq:qMeans} 
work well for $x_j=0$ values.
For $q<0$, notice that 
\begin{equation*}
M_q(\ve{x}) = M_{-\abs{q}}(\ve{x}) = \biggl[m \frac1{\frac1{x_1^\abs{q}}+\dots+\frac1{x_m^\abs{q}}} \biggr]^{1/\abs{q}},
\end{equation*}
showing that $\lim_{x_j\to 0^+}M_q(\ve{x})=0$,
and the same holds for $N_q$.
So defining for $q<0$ and $q\leq0$ and for any $x_j=0$ the $q$-sum and $q$-mean by their limit, 
$N_q(\ve{x})=M_q(\ve{x})=0$,
allows us to use $q$-sum and $q$-mean of $\ve{x}\geq0$ non-negative numbers.

Let us see the most important properties of $q$-sums and $q$-means.
$N_q$ is continuous for $0\neq q\in\field{R}$, $N_q\geq0$,
$M_q$ is continuous for $q\in\field{R}$, $M_q\geq0$,
and they have the following \emph{vanishing properties:}
\begin{subequations}
\label{eq:qSumVanish}
\begin{align}
\label{eq:qSumVanish.and}
\text{if $q>0$:}\quad
N_q(\ve{x}) = 0\quad&\Longleftrightarrow\quad \forall j:\; x_j=0,\\
\label{eq:qSumVanish.or}
\text{if $q<0$:}\quad
N_q(\ve{x}) = 0\quad&\Longleftrightarrow\quad \exists j:\; x_j=0,
\end{align}
\end{subequations}
and
\begin{subequations}
\label{eq:qMeanVanish}
\begin{align}
\label{eq:qMeanVanish.and}
\text{if $q>0$:}\quad
M_q(\ve{x}) = 0\quad&\Longleftrightarrow\quad \forall j:\; x_j=0,\\
\label{eq:qMeanVanish.or}
\text{if $q\leq0$:}\quad
M_q(\ve{x}) = 0\quad&\Longleftrightarrow\quad \exists j:\; x_j=0.
\end{align}
\end{subequations}
They are \emph{homogeneous functions;} that is, 
\begin{equation}
\text{for $c\geq0$:}\quad
N_q(c\ve{x})=cN_q(\ve{x}),\quad
M_q(c\ve{x})=cM_q(\ve{x}).
\end{equation}
For all $0\neq q\in\field{R}$, $N_q(\ve{x})$ 
and for all $q\in\field{R}$,  $M_q(\ve{x})$ are
\emph{monotonically increasing} for all arguments $x_j$ 
(see Appendix \ref{app:qMean.mon}),
and their \emph{convexity/concavity} properties are
\begin{subequations}
\label{eq:qSumCon}
\begin{align}
\label{eq:qSumCon.vex}
N_q\Bigl(\sum_i p_i \ve{x}_i\Bigr) \leq \sum_i p_i N_q(\ve{x}_i)
  \quad&\Longleftrightarrow\quad q\geq1,\\
\label{eq:qSumCon.cave}
N_q\Bigl(\sum_i p_i \ve{x}_i\Bigr) \geq \sum_i p_i N_q(\ve{x}_i)
 \quad&\Longleftrightarrow\quad 0\neq q\leq1,
\end{align}
\end{subequations}
and
\begin{subequations}
\label{eq:qMeanCon}
\begin{align}
\label{eq:qMeanCon.vex}
M_q\Bigl(\sum_i p_i \ve{x}_i\Bigr) \leq \sum_i p_i M_q(\ve{x}_i)
  \quad&\Longleftrightarrow\quad q\geq1,\\
\label{eq:qMeanCon.cave}
M_q\Bigl(\sum_i p_i \ve{x}_i\Bigr) \geq \sum_i p_i M_q(\ve{x}_i)
 \quad&\Longleftrightarrow\quad q\leq1
\end{align}
\end{subequations}
(see Appendix \ref{app:qMean.con}).
On the other hand, for all $\ve{x}\geq0$,
$N_q(\ve{x})$ is monotonically \emph{decreasing} for the parameter $q$,
and it has a discontinuity in $q=0$, as
\begin{subequations}
\label{eq:qSumIneq}
\begin{align}
0<q< q'\quad\Longrightarrow\quad M_q(\ve{x}) \geq M_{q'}(\ve{x}),\\
q< q'<0\quad\Longrightarrow\quad M_q(\ve{x}) \geq M_{q'}(\ve{x}),\\
q<0<q' \quad\Longrightarrow\quad M_q(\ve{x}) \leq M_{q'}(\ve{x}).
\end{align}
\end{subequations}
On the other hand, for all $\ve{x}\geq0$, 
$M_q(\ve{x})$ is monotonically \emph{increasing} for the parameter $q$,
which is the \emph{$q$-mean inequality,}
\begin{equation}
\label{eq:qMeanIneq}
q< q'\quad\Longrightarrow\quad M_q(\ve{x}) \leq M_{q'}(\ve{x}),
\end{equation}
with equality if and only if $x_1=\dots=x_m$.
From this, it immediately follows that the $q$-mean of numbers
is between the minimal \eqref{eq:qMeans.min} and the maximal \eqref{eq:qMeans.max} ones.
On the other hand, the inequality between the arithmetic and geometric means
$M_0(\ve{x}) \leq M_1(\ve{x})$ is a particular case of this.

\subsection{Definitions and properties of quasi-sums and quasi-arithmetic means}
\label{app:Means.Quasi}

Let $\ve{x}=(x_1,\dots,x_m)\in \field{R}^m$, $\ve{x}\geq0$. 
For a continuous strictly monotonic function $h:\field{R}\to\field{R}$,
let the \emph{quasi-sum} of $\ve{x}$ be defined as
\begin{equation}
\label{eq:SumQuasi}
N_h(\ve{x}) := h^{-1}\Bigl(\sum_{j=1}^m h(x_j)\Bigr),
\end{equation}
and
the \emph{quasi-arithmetic mean} of $\ve{x}$ is defined as
\cite{Kolmogorov-1930,Marichal-2000}
\begin{equation}
\label{eq:MeanQuasiArithm}
M_h(\ve{x}) := h^{-1}\Bigl(\frac1m\sum_{j=1}^m h(x_j)\Bigr).
\end{equation}
Note that
in the two cases, $h$ and the linearly, respectively, affinly transformed functions
$ah$, respectively, $ah+b$
(with $a,b\in\field{R}$, $a\neq0$)
lead to the same functions,
$N_h=N_{ah}$, and
$M_h=M_{ah+b}$.
The choice $h(x)=x^q$ gives back the $q$-sum and $q$-mean for $q\neq0$,
while $h(x)=\ln(x)$ gives back the geometric mean.
For the quite general function class of the quasi-arithmetic means,
only few of the properties can be known, in general 
\cite{Marichal-2000,Micic-2012,BenTalEntrMeans}, 
most of which are true also for the quasi-sums with minor modifications.

\subsection{Monotonicity}
\label{app:qMean.mon}

For the monotonicity of the $q$-sums \eqref{eq:qSums.q}
and $q$-means \eqref{eq:qMeans.q} for $\ve{x}>0$,
we have for the latter one for $q\neq0$ that
\begin{equation}
\label{eq:qMeanqDiv}
\frac{\partial M_q(\ve{x})}{\partial x_i} = \frac1{m^{1/q}}\Bigl(\sum_k x_k^q\Bigr)^{1/q-1} x_i^{q-1} \geq 0.
\end{equation}
The same holds for $N_q$.
We have for the geometric mean $M_0$ \eqref{eq:qMeans.0},
\begin{equation}
\label{eq:qMean0Div}
\frac{\partial M_0(\ve{x})}{\partial x_i} = 
\frac1m\Bigl(\prod_k x_k\Bigr)^{1/m} x_i^{-1} \geq 0.
\end{equation}

\subsection{Convexity, concavity}
\label{app:qMean.con}

For an $f:\field{R}^m\to\field{R}$,
if $\dom f$ is op	en and the $\frac{\partial^2 f(\ve{x})}{\partial x_j\partial x_i}$ Hessian exists,
then
$f$ is convex (concave) if and only if $\dom f$ is convex and 
$\frac{\partial^2 f(\ve{x})}{\partial x_j\partial x_i}$ is positive (negative) semidefinite
\cite{BoydVandenbergheConvOpt}.

For the convexity/concavity of the $q$-sums \eqref{eq:qSums.q}
and $q$-means \eqref{eq:qMeans.q} for $\ve{x}>0$,
we have for the latter one for $q\neq0$ that
the Hessian from \eqref{eq:qMeanqDiv} is as follows,
\begin{widetext} 
\begin{equation}
\frac{\partial^2 M_q(\ve{x})}{\partial x_j\partial x_i} =  
 \frac1{m^{1/q}}(1-q)\Bigl(\sum_k x_k^q\Bigr)^{1/q-2}x_i^{q-1}x_j^{q-1} 
+\frac1{m^{1/q}}(q-1)\Bigl(\sum_k x_k^q\Bigr)^{1/q-1}\delta_{i,j}x_i^{q-2}.
\end{equation}
Then, taking an $\ve{u}\in\field{R}^m$,
we are interested in the sign of
\begin{equation}
\begin{split}
\sum_{j,i} u_j \frac{\partial^2 M_q(\ve{x})}{\partial x_j\partial x_i} u_i 
&= \frac1{m^{1/q}}(q-1) \Bigl(\sum_k x_k^q\Bigr)^{1/q-2}
\Bigl[\Bigl(\sum_i x_i^q\Bigr)\Bigl(\sum_j u_j^2 x_j^{q-2}\Bigr) - \Bigl(\sum_iu_ix_i^{q-1}\Bigr)^2  \Bigr] \\
&=\frac1{m^{1/q}}(q-1) \Bigl(\sum_k x_k^q\Bigr)^{1/q-2}
\Bigl[ \sum_i (x_i^{q/2})^2 \sum_j(u_jx_j^{q/2-1})^2 - \Bigl(\sum_i(u_ix_i^{q/2-1})(x_i^{q/2})\Bigr)^2\Bigr].
\end{split}
\end{equation}
The content of the square bracket $[\;]$ is non-negative, 
which is the Cauchy-Bunyakovsky-Schwarz inequality 
for the vectors of components $u_ix_i^{q/2-1}$ and $x_i^{q/2}$.
So we have that
$M_q$ is convex for $q\geq1$
and 
$M_q$ is concave for $q\leq1$.
The same holds for $N_q$.

For $q=0$,
the Hessian from \eqref{eq:qMean0Div} is as follows,
\begin{equation}
\frac{\partial^2 M_0(\ve{x})}{\partial x_j\partial x_i} =  
 \frac1{m^2}\Bigl(\prod_k x_k\Bigr)^{1/m} x_i^{-1} x_j^{-1}
-\frac1m    \Bigl(\prod_k x_k\Bigr)^{1/m} \delta_{i,j}x_i^{-2}
\end{equation}
Then, taking an $\ve{u}\in\field{R}^m$,
we are interested in the sign of
\begin{equation}
\begin{split}
\sum_{j,i} u_j \frac{\partial^2 M_0(\ve{x})}{\partial x_j\partial x_i} u_i 
&= \frac1{m^2}\Bigl(\prod_k x_k\Bigr)^{1/m} 
\Bigl[\Bigl(\sum_i u_ix_i^{-1}\Bigr)\Bigl(\sum_j u_jx_j^{-1}\Bigr)
- m\sum_i u_i^2x_i^{-2}  \Bigr]\\
&=\frac1{m^2}\Bigl(\prod_k x_k\Bigr)^{1/m}
\Bigl[
\Bigl(\sum_i (1)(u_ix_i^{-1})\Bigr)^2
- \Bigl(\sum_i (1)^2\Bigr)\Bigl(\sum_i (u_ix_i^{-1})^2\Bigr)
\Bigr].
\end{split}
\end{equation}
The content of the square bracket $[\;]$ is nonpositive,
which is the Cauchy-Bunyakovsky-Schwarz inequality
for the vectors of components $1$ and $u_ix_i^{-1}$,
so we have that $M_0$ is concave.

\section{On constructions of entanglement monotones}
\label{app:MeanMon}

\subsection{By concavity}
\label{app:MeanMon.lemGMeasure}

For the proof of Lemma \ref{lem:GMeasure},
let $\Lambda=\sum_i\Lambda_i$ a LOCC 
with the decomposition into pure suboperations $\Lambda_i$,
and
$\pi_i'=\frac1{p_i}\Lambda_i(\pi)$ 
with probability $p_i = \tr \Lambda_i(\pi)$.
Then we can write
\begin{equation*}
\begin{split}
\sum_ip_iG(f_1,\dots,f_m)(\pi'_i)
&=\sum_ip_iG\bigl(f_1(\pi'_i),\dots,f_m(\pi'_i)\bigr)
\leq G\Bigl( \sum_ip_if_1(\pi'_i),\dots,\sum_ip_if_m(\pi'_i)\Bigr)\\
&\leq G\bigl(f_1(\pi),\dots,f_m(\pi)\bigr)
=G(f_1,\dots,f_m)(\pi)
\end{split}
\end{equation*}
where the \textit{first inequality} is the concavity of $G$
and the \textit{second inequality} is the assumption \eqref{eq:averagePure} 
together with the monotonicity $G$.

\subsection{A new entanglement monotone}
\label{app:MeanMon.magic}

For the proof of that $M_{\ln\circ g}(E_{\alpha_1},\dots,E_{\alpha_{\abs{\vs{\alpha}}}})$ 
in \eqref{eq:IndIImagicg} is an entanglement monotone,
first we claim
the monotonicity and concavity of the quasi-arithmetic mean
$M_{\ln\circ g} = g^{-1} \circ M_0 \circ g$,
with $g(x)=1-\ee^{-x}$, written out as
\begin{equation*}
M_{\ln\circ g}(\ve{x}) = -\ln\Bigl[1-\Bigl(\prod_{k=1}^m(1-\ee^{-x_k})\Bigr)^{1/m}\Bigr] 
= -\ln\Bigl[1-\Pi^{1/m}\Bigr],
\end{equation*}
where we use the shorthand notation $\Pi:=\prod_{k=1}^m(1-\ee^{-x_k})$.
Let $\ve{x}>0$, then we have that
\begin{equation*}
\frac{\partial\Pi}{\partial x_i} = \frac{\ee^{-x_i}}{1-\ee^{-x_i}}\Pi.
\end{equation*}
First, let us see the first partial derivatives,
\begin{equation}
\frac{\partial M_{\ln\circ g}(\ve{x})}{\partial x_i}
= \frac1{1-\Pi^{1/m}} \frac1m \Pi^{1/m-1} \frac{\partial\Pi}{\partial x_i} 
= \frac1{1-\Pi^{1/m}} \frac1m \Pi^{1/m} \frac{\ee^{-x_i}}{1-\ee^{-x_i}}\geq0,
\end{equation}
so the function is monotonically increasing in all arguments.
Then the Hessian is
\begin{equation}
\begin{split}
&\frac{\partial^2 M_{\ln\circ g}(\ve{x})}{\partial x_j\partial x_i} \\
&\quad=\Bigl[\frac{\partial}{\partial x_j} \frac1{1-\Pi^{1/m}}\Bigr]\frac1m \Pi^{1/m} \frac{\ee^{-x_i}}{1-\ee^{-x_i}}
+\frac1{1-\Pi^{1/m}} \frac1m \Bigl[\frac{\partial}{\partial x_j} \Pi^{1/m} \Bigr] \frac{\ee^{-x_i}}{1-\ee^{-x_i}}
+\frac1{1-\Pi^{1/m}} \frac1m \Pi^{1/m}\Bigl[\frac{\partial}{\partial x_j}\frac{\ee^{-x_i}}{1-\ee^{-x_i}}\Bigr]\\
&\quad= \Bigl[\frac1m\frac{\ee^{-x_j}}{1-\ee^{-x_j}}\frac{\Pi^{1/m}}{(1-\Pi^{1/m})^2} \Bigr]\frac1m \Pi^{1/m} \frac{\ee^{-x_i}}{1-\ee^{-x_i}}
+\frac1{1-\Pi^{1/m}} \frac1m \Bigl[\frac1m\frac{\ee^{-x_j}}{1-\ee^{-x_j}}\Pi^{1/m}\Bigr]\frac{\ee^{-x_i}}{1-\ee^{-x_i}}\\
&\quad +\frac1{1-\Pi^{1/m}} \frac1m \Pi^{1/m}\Bigl[\frac{-\delta_{i,j}\ee^{-x_i}}{(1-\ee^{-x_i})^2}\Bigr]\\
&\quad=\frac1{m^2}\frac{\Pi^{1/m}}{(1-\Pi^{1/m})^2} \Bigl[
\Pi^{1/m} \frac{\ee^{-x_j}}{1-\ee^{-x_j}} \frac{\ee^{-x_i}}{1-\ee^{-x_i}}
+(1-\Pi^{1/m})\frac{\ee^{-x_j}}{1-\ee^{-x_j}} \frac{\ee^{-x_i}}{1-\ee^{-x_i}}
-m(1-\Pi^{1/m}) \frac{\delta_{i,j}\ee^{-x_i}}{(1-\ee^{-x_i})^2}
\Bigr] \\
&\quad=\frac1{m^2}\frac{\Pi^{1/m}}{(1-\Pi^{1/m})^2} \Bigl[
\frac{\ee^{-x_j}}{1-\ee^{-x_j}} \frac{\ee^{-x_i}}{1-\ee^{-x_i}} 
-m(1-\Pi^{1/m}) \frac{\delta_{i,j}\ee^{-x_i}}{(1-\ee^{-x_i})^2}
\Bigr].
\end{split}
\end{equation}
Then, taking an $\ve{u}\in\field{R}^m$,
we are interested in the sign of
\begin{equation}
\begin{split}
\sum_{j,i} u_j \frac{\partial^2 M_{\ln\circ g}(\ve{x})}{\partial x_j\partial x_i} u_i =
\frac1{m^2} \frac{\Pi^{1/m}}{(1-\Pi^{1/m})^2}\Bigl[
\Bigl(\sum_i \frac{u_i\ee^{-x_i}}{1-\ee^{-x_i}}\Bigr)^2
-m(1-\Pi^{1/m})\sum_i\frac{u_i^2\ee^{-x_i}}{(1-\ee^{-x_i})^2}
\Bigr].
\end{split}
\end{equation}
We claim that the content of the square bracket $[\;]$ is nonpositive,
which follows from
\begin{equation*}
\Bigl(\sum_i \frac{u_i\ee^{-x_i}}{1-\ee^{-x_i}}\Bigr)^2
\leq
\sum_i\Bigl( \frac{u_i\sqrt{\ee^{-x_i}}}{1-\ee^{-x_i}}\Bigr)^2
\sum_j\Bigl( \sqrt{\ee^{-x_j}}\Bigr)^2
\leq
\sum_i\Bigl( \frac{u_i\sqrt{\ee^{-x_i}}}{1-\ee^{-x_i}}\Bigr)^2
m(1-\Pi^{1/m}),
\end{equation*}
where the first inequality is the Cauchy-Bunyakovsky-Schwarz inequality
for the vectors of components $\frac{u_i\sqrt{\ee^{-x_i}}}{1-\ee^{-x_i}}$ and $\sqrt{\ee^{-x_i}}$,
and the second inequality is $\sum_j\ee^{-x_j}\leq m(1-\Pi^{1/m})$,
which is rearranged as $\Pi^{1/m}\leq 1-\frac1m\sum_j\ee^{-x_j}$,
\begin{equation*}
\Pi^{1/m}
= \Bigl(\prod_j(1-\ee^{-x_j})\Bigr)^{1/m}
\leq \frac1m\sum_j(1-\ee^{-x_j})
= 1-\frac1m\sum_j\ee^{-x_j}, 
\end{equation*}
which is the inequality between the geometric and arithmetic means
(see equation \eqref{eq:qMeanIneq} for $q=0$ and $q'=1$).
So we can conclude that the function $M_{\ln\circ g}:\field{R}^m\to\field{R}$ is concave.
Now, $M_{\ln\circ g}(E_{\alpha_1},\dots,E_{\alpha_{\abs{\vs{\alpha}}}})$ 
in \eqref{eq:IndIImagicg} is an entanglement monotone,
since Lemma \ref{lem:GMeasure} holds for that,
which completes the proof.

\end{widetext} 

\bibliography{multipartmeasures}{}

\end{document}